\documentclass[review]{elsarticle}

\usepackage{amssymb,amsfonts,amsmath}
\usepackage[english]{babel}
\usepackage{lineno}
\modulolinenumbers[5]
\usepackage{graphicx}
\usepackage{epsfig}
\usepackage{float}
\usepackage{hyperref}
\usepackage{sidecap}
\usepackage[dvipsnames]{xcolor}
\usepackage{mathtools}
\usepackage[labelfont=bf]{caption} 
\usepackage[top=0.75in,bottom=0.75in,left=0.75in,right=0.5in]{geometry} 
\usepackage{outlines}
\usepackage{enumitem}
\usepackage{todonotes}
\usepackage{hyperref}

\usepackage{palatino} 
\usepackage[T1]{fontenc}
\usepackage[utf8]{inputenc}

\journal{Communications in Nonlinear Science and Numerical Simulation}

\begin{document}
\title{Homoclinic puzzles and chaos in a nonlinear laser model}

\author{K. Pusuluri \fnref{myfootnote}\corref{mycorrespondingauthor}}
\ead{pusuluri.krishna@gmail.com}
\address{837 NeurDS lab, Neuroscience Institute, Georgia State University, Petit Science Center, 100 Piedmont Av., Atlanta, GA 30303, USA}
\author{H.G.E. Meijer}
\ead{h.g.e.meijer@utwente.nl}
\address{Department of Applied Mathematics, University of Twente, PO Box 217, 7500 AE Enschede, Netherlands}
\author{A.L. Shilnikov} 
\ead{ashilnikov@gsu.edu}
\address{Neuroscience Institute and Department of Mathematics and Statistics, Georgia State University, Petit Science Center, 100 Piedmont Av., Atlanta, GA 30303, USA}

\begin{abstract}
We present a case study elaborating on the multiplicity and self-similarity of homoclinic and heteroclinic bifurcation structures in the 2D and 3D parameter spaces of a nonlinear laser model with a Lorenz-like chaotic attractor. In a symbiotic approach combining the traditional parameter continuation methods using MatCont and a newly developed technique called the Deterministic Chaos Prospector (DCP) utilizing symbolic dynamics on fast parallel computing hardware with graphics processing units (GPUs), we exhibit how specific codimension-two bifurcations originate and pattern regions of chaotic and simple dynamics in this classical model. We show detailed computational reconstructions of key bifurcation structures such as Bykov T-point spirals and inclination flips in 2D parameter space, as well as the spatial organization and 3D embedding of bifurcation surfaces, parametric saddles, and isolated closed curves (isolas).
\end{abstract}

\maketitle
\section{Introduction}

it is evident today the incorporation of complex mathematical elements in combination with the latest computational breakthroughs is the key drive that can further stimulate significant advances in fundamental sciences and cutting-edge engineering. The development of algorithmically simple and generalizable mathematical methods exploiting fast and comprehensive massively parallel simulations using graphics processing units (GPU) ensures better quantitative and, more importantly, qualitative understanding of the nature of complex nonlinear dynamics occurring in various multi-parametric models originating in applications of physical and living systems. The goal of this paper is to deepen our understanding of homoclinic bifurcation theory by demonstrating the universality of the causes and the rules underlying deterministic chaotic dynamics \cite{ABS:1977,AS:1983,GW79,SSTC:1998,Shilnikov:2001,Homburg:2010} using a particular application from nonlinear optics -- a reduced laser model \cite{Moloney:1987,Forysiak:1991}. 

While basic transition mechanisms of oscillations emerging from stable equilibria through (homoclinic) saddle-node and Andronov-Hopf bifurcations are well presented in diverse applications, their higher dimensional analogues such as various homoclinic and torus bifurcations still remain poorly understood. One of the limiting factors behind such inadequacy is that expert tools for ODE models, i.e., the numerical continuation packages AUTO \cite{auto07p,homcont} and MatCont \cite{Matcont}, require advanced knowledge on a user's part to be able to handle those bifurcations. These tools can detect and analyze the essential homoclinic and heteroclinic structure and bifurcations in the phase and parametric spaces of a system. This allows to identify the primary and codimension-two (cod-2) bifurcations, both local and non-local, and the corresponding curves to assess the skeleton of the bifurcation diagram for the system. The main shortcoming of this computational approach is that it requires particular skills and patience to perform a strenuous reconstruction of the ``complete'' bifurcation unfolding in the 2D parameter plane, by individually continuing a few dozens of principal bifurcation curves, one after another \cite{ASHIL93}, as seen in Figs. \ref{fig:sigma15Cont},\ref{fig:sigma20}b,\ref{fig:sigma100Cont} in this paper. The recent MatCont releases have a strong built-in engine and support for numerical detection and analysis of typical low-codimension homoclinic bifurcations of saddle equilibria in autonomous systems \cite{DeWitte:2012}. Nevertheless, such studies involving even basic homoclinic structures remain state-of-the-art. One reason is that the built-in algorithms of a parameter-continuation software package must catch up with the rigorously developed theoretical results that are often rooted in or inspired by diverse applications with interesting dynamics. An alternative computational approach widely used in nonlinear dynamics is based on evaluating the maximal or several Lyapunov exponents \cite{Wolf:1985}. Such a largest Lyapunov exponent approach computes some average rate of change, convergence or/and divergence, of the distance between two close, long-term transients. By definition, negative, zero and positive Lyapunov exponents are associated with stable equilibria, stable periodic orbits, and chaotic dynamics in the system, respectively. With a brute-force approach for sweeping 1D parametric pathways or 2D bi-parametric planes of the system, one can detect regions of regular/simple and complex/chaotic dynamics, as well as multistability, if it exists, by employing multiple initial conditions. While this brute-force approach based on the evaluation of Lyapunov exponents can effectively locate stability windows within regions of chaos \cite{Gallas:2010,Barrio:2011}, it falls short in disclosing a variety of essential bifurcation structures in the parameter space that play a pivotal role to understand complex dynamics and their origin. Moreover, very long simulations are needed to reliably estimate the Lyapunov exponents. Given the current state-of-the-art, there is still a pressing need for tools and detailed evaluations of the global bifurcation unfolding of any system in question, to elucidate the contributions and multiplicity of the local and non-local bifurcations detected, and to determine how they shape regions of simple and complex dynamics in the corresponding 2D or even 3D parameter spaces. The objective of this paper is to somewhat remedy the situation by showcasing our know-how with an open source GPU-based computational toolkit based on symbolic dynamics, that should be accessible and practical for the nonlinear dynamics community.

Over the last several years, we developed a basis for new theoretical and computational approaches to explore the origin of complex dynamics and the characteristic bi-parametric patterns for so-called {\em Lorenz-like} systems \cite{Barrio:2012,Xing:2014}. Our motivation is to extend the existing theory of homoclinic bifurcations of low-codimensions \cite{ABS:1977,Shilnikov:2001, Bykov:1980,AGLST:2014,AD02,Belyakov1,GW79,ALS91,Shilnikov:1965,LP67,Shilnikov:2012,GOST05,Bykov:1998} and to make it accessible and practical for the nonlinear dynamics community. Recently, we have advanced an approach called the ``Deterministic Chaos Prospector'' (DCP) that utilizes symbolic representations of simple and chaotic dynamics \cite{Pusuluri:2017,Pusuluri:2018, pusuluri2019symbolic, Pusuluri2020}, based on our earlier works \cite{Xing:2014,xwbs}. This allows for fast and effective identification of bifurcation structures underlying and governing deterministic chaos in various systems. The ideas of this computational research trend are inspired by and based on the classical results of L.P.~Shilnikov on homoclinic bifurcations \cite{SSTC:1998, Shilnikov:1965,Shilnikov:1968}. We introduce and discuss the basic elements of homoclinic bifurcations in the text later, as well as how short-term symbolic dynamics can be introduced to disclose a stunning array of homoclinic structures and their organizing centers in all Lorenz-like systems, including the optically pumped laser (OPL) model under consideration in this paper. Some pilot results on the use of symbolic dynamics for the OPL model can be found in \cite{Barrio:2012, Pusuluri:2018}. In addition to simple dynamics associated with stable equilibria and periodic orbits, this system reveals a broad range of bifurcation structures that are typical for many ODE models from nonlinear optics and ones with the Lorenz attractor \cite{Xing:2014,Pusuluri:2017,xwbs,Barrio:2014,xwzs2015}. These include homoclinic orbits and heteroclinic connections between saddle equilibria that are the key building blocks of deterministic chaos in most systems. Their bifurcation curves with characteristic spirals around T(terminal)-points, along with other codimension-2 points, are the organizing centers that shape regions of complex and simple dynamics in the parameter space of such systems. Moreover, the latest advances in GPU and parallel computing techniques have empowered us to achieve a tremendous degree of parallelization to reconstruct highly detailed bi-parametric sweeps. A remarkable wealth of homoclinic bifurcations are disclosed in such a system, in a matter of just a few seconds on a desktop workstation powered by an Nvidia Tesla K40 GPU. It is fair to say that currently there is no other computational technique that is able to reveal the ordered intricacy in these bifurcation diagrams with such a stunning clarity and speed. Next, one can corroborate the type of homoclinic bifurcations using numerical continuation with AUTO or MatCont. While we do so in this paper, we underline that DCP performs impressively at just a fraction of the time taken for traditional continuation and other serial computational approaches. We also demonstrate how DCP can reveal intricate parametric regions of complex -- structurally-unstable dynamics against those with simple -- stable dynamics for a system, by exploiting the sensitivity of deterministic chaos for arbitrary long-term solutions whose symbolic descriptions are processed using periodicity-detection and Lempel-Ziv-complexity \cite{Pusuluri:2018, Lempel:1976} algorithms.

The multifarious set of complex dynamics exhibited by the OPL system makes it a prime candidate for this study, and enables us to demonstrate the versatility of our approach. In this paper, we employ the computational toolkit DCP and the numerical continuation package MatCont to disclose the remarkable features of the parameter space of the OPL model, due to various homoclinic and heteroclinic bifurcations that originate and underlie its complex, Lorenz-like dynamics. Specifically, using DCP we achieve the following objectives:\\
\textbf{(1)} We disclose the universal, self-similar fine organization of homoclinic and heteroclinic bifurcation structures in bi-parametric sweeps using short-term symbolic sequences. The key ones such as T-point spirals, saddles, inclination flips, and other codimension-2 bifurcation points, are also validated by MatCont-based numerical continuation, which provides all the necessary details, coordinates, eigenvalues etc.\\
\textbf{(2)} We elaborate on the spatial organization and embedding of the identified bifurcation structures in 3D parameter space, including two different types of parametric saddles and two different types of isolated closed curves (isolas for short). To our knowledge, this is the first elaborate 3D computational reconstruction of such bifurcation surfaces. \\
\textbf{(3)} We detect regions of simple and chaotic dynamics in the bi-parametric sweeps using a novel approach based on long-term symbolic sequences, combined with the periodicity-detection algorithms and the notion of Lempel-Ziv-complexity.

This paper is organized as follows. First, we introduce the OPL model, and present its key dynamical and bifurcational features. Next, we give a brief overview of the codimension-2 homoclinic inclination flip bifurcation, as well as a heteroclinic bifurcation involving all the three saddle equilibria in the phase space that corresponds to the so-called T(terminal)-point in the parameter plane of the model. This is followed by details of the numerical methods and the results of this study, which are presented in the order of (i) transient dynamics for global bifurcation structure, (ii) long term simple vs. complex dynamics, (iii) detailed examination of transient and long term behaviors near the organizing centers, and (iiii) 3D organization of special structures in the parametric space including branching and bridging saddles as well as annular and semi-annular isolas. 

\section{Optically pumped laser (OPL) model}\label{OPL}

\begin{figure}[t!]
\centering
\includegraphics[width=.35\textwidth]{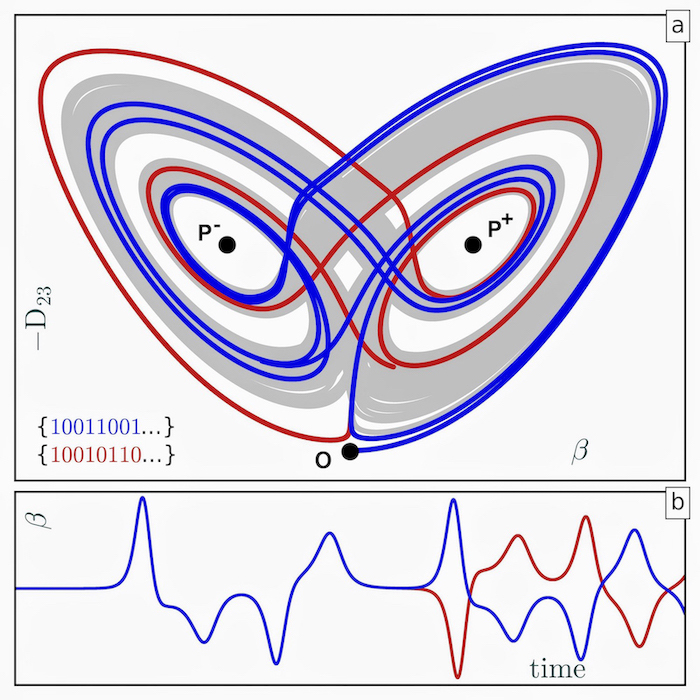}
\caption{\textbf{(a)} The Lorenz attractor in the OPL-model: ($-D_{23},\,\beta$)-projection of the strange attractor (gray background) with the lacuna (associated with a white figure-8 interior). Two close trajectories (red and blue) that diverge when they come back to the saddle equilibrium state (black dot), their (color matched) symbolic representations, and the time-progression of their corresponding $\beta$-coordinates \textbf{(b)} are presented. Parameter values: $\sigma=1.5$, $a=3.8$, $b=0.43858$ (red) or $b=0.43855$ (blue).}\label{fig:symbolicLorenzLacunae}
\end{figure}

\begin{figure}[t!]
\centering
\includegraphics[width=.5\textwidth]{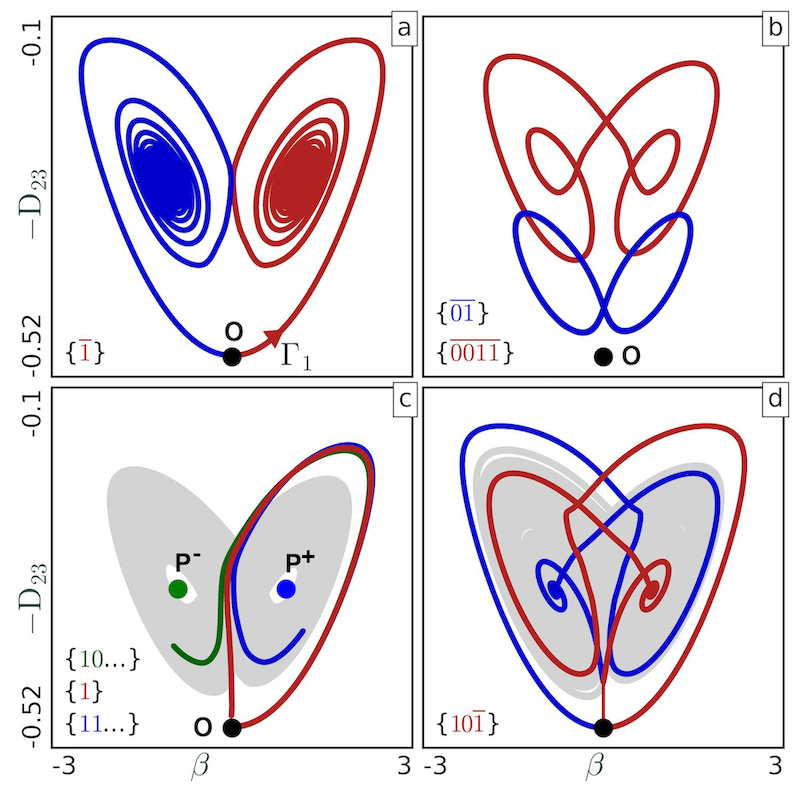}
\caption{\textbf{(a)} Phase-space projections of the symmetric unstable separatrices ($\Gamma_1/\Gamma_2$) of the saddle equilibrium (black dot) converging to the stable equilibria $P^+$ and $P^-$ at $(a,b)=(3.7,0.52)$; $\Gamma_1$ (red) generates the binary sequence $\{1111...\}$ or $\{\overline{1}\}$. \textbf{(b)} Two stable periodic orbits coded as $\{\overline{01}\}$ (blue) and $\{\overline{0011}\}$ (red) at parameters $(4.2, 0.583)$ and $(3.37326, 0.313333)$, respectively. \textbf{(c)} Primary homoclinic orbit, coded as $\{1\}$, making a single turn around $P^+$ at $(3.827, 0.51903)$, with a Lorenz-like attractor in the background. After splitting, when it misses the saddle, $\Gamma_1$ will make another turn around $P^+$ (blue, $\{11...\}$) at $(3.827, 0.54)$ or around $P^-$ (green, $\{10...\}$) at $(3.827, 0.50)$. \textbf{(d)} Heteroclinic connections between the saddle and the two saddle-foci  occurring at the T-point $T_1$ (see the bifurcation diagram in Fig.~\ref{fig:sigma15_IFs_T1T2}b) are shown. Red curve ($\{10\overline{1}\}$) connects $\Gamma_1$ and $P^+$, while the symmetric connection between $\Gamma_2$ and $P^-$ is shown in blue. The initially red red trajectory is further continued to generate the chaotic Lorenz attractor on the  background for $(a,b)=(3.68199, 0.3517)$; Here $\sigma=1.5$.}\label{fig:traces}
\end{figure}

The universality of characteristic chaotic oscillations that occur in various models with the Lorenz attractor, originating from hydrodynamics, magnetodynamics and nonlinear optics, was first shown by H.~Haken in 1970s, followed by experimental demonstrations in $Xe$, $He-Ne$, and $NH_3$ infrared gas lasers \cite{Haken:1975,Haken:1985,Casperson:1978,Weiss:1982,Weiss:1985,Weiss:1986}. Optically pumped, far infrared lasers are known to demonstrate a variety of nonlinear dynamic behaviors, including Lorenz-like chaos \cite{Weiss:1995}. An acclaimed example of the modeling studies for chaos in nonlinear optics is the two-level laser model proposed by \cite{Haken:1975}, to which the Lorenz equation can be reduced. The validity of Lorenz dynamics inherently persisting in three-level laser models was widely questioned back then. The reduced 6D model of a resonant three-level optically pumped laser (OPL) introduced in \cite{Moloney:1987,Forysiak:1991,Moloney:1989}, and described by equations~(\ref{lasermodel}) below, was shown to possess a variety of dynamical and structural features of Lorenz-like systems, including stationary and periodic behaviors emerging through $\mathbb{Z}_2$-pitchfork and Andronov-Hopf bifurcations. Specifically, the model demonstrates Lorenz-like chaotic behavior (see Fig.~\ref{fig:symbolicLorenzLacunae}), with all the quintessential organizing structures, pivotal to understand its complex dynamics. These include various homoclinic and heteroclinic bifurcation structures of co-dimensions one and two, the so-called  Bykov T-points with the associated spirals, as well as parametric saddles for switching branches in the parameter plane of this model, that are also seen in the classical Lorenz and Shimizu-Morioka models \cite{ABS:1977,ASHIL93,Barrio:2011,Barrio:2012,Xing:2014,ALS91,xwbs,xwzs2015,Bykov:1993}. Similar structures were also discovered in another non-linear optics model describing a laser with a saturable absorber, which can be locally reduced to the Shimizu–Morioka model near a steady-state solution with triple zero Lyapunov exponents \cite{Vladimirov:1993,Shilnikov:1993}. 

The OPL model \cite{Forysiak:1991,Moloney:1989} is given by the following six ODEs:
\begin{eqnarray}\label{lasermodel}
\dot{\beta}~~ &=& -\sigma \beta + g p_{23}, \nonumber \\
\dot{p}_{21} &=& -p_{21} - \beta p_{31} + a D_{21}, \nonumber\\
\dot{p}_{23} &=& -p_{23} + \beta D_{23} - a p_{31}, \nonumber\\
\dot{p}_{31} &=& -p_{31} + \beta p_{21} + a p_{23}, \\
\dot{D}_{21} &=& -b(D_{21}-D_{21}^0)-4 a p_{21} - 2 \beta p_{23}, \nonumber\\
\dot{D}_{23} &=& -b(D_{23}-D_{23}^0)-2 a p_{21} - 4 \beta p_{23}, \nonumber
\end{eqnarray}
where $a$ and $\beta$ are the Rabi flopping quantities representing the electric field amplitudes at pump and emission frequencies, respectively; $b$ is the ratio between population and polarization decay rates; $\sigma$ represents the cavity loss parameter; $g$ is the unsaturated gain; $p_{ij}$'s are the normalized density matrix elements for the transitions between levels $i$ and $j$; and $D_{ij}$'s are the population differences between levels $i$ and $j$. The system parameters $a$, $b$ and $\sigma$ can be manipulated experimentally. Furthermore, we set $g=50,D^{0}_{21}=1,$ and $D^{0}_{23}=0$. In this study, we treat $a$ and $\beta$ as the key bifurcation parameters for constructing biparametric scans and parameter continuations, at several fixed values of $\sigma$. Several such 2D scans may also be put together for 3D reconstructions of the parameter space of the model. Note that Eqs.~(\ref{lasermodel}) are $\mathbf{Z}_2$--symmetric under the coordinate transformation $(\beta,p_{21},p_{23},p_{31},D_{21},D_{23}) \leftrightarrow$ $(-\beta,p_{21},-p_{23},-p_{31},D_{21},D_{23})$, similar to other Lorenz-like systems. 

\subsection{Simple dynamics in the OPL-model} 

\begin{figure*}[t!]
\centering
\includegraphics[width=0.5\textwidth]{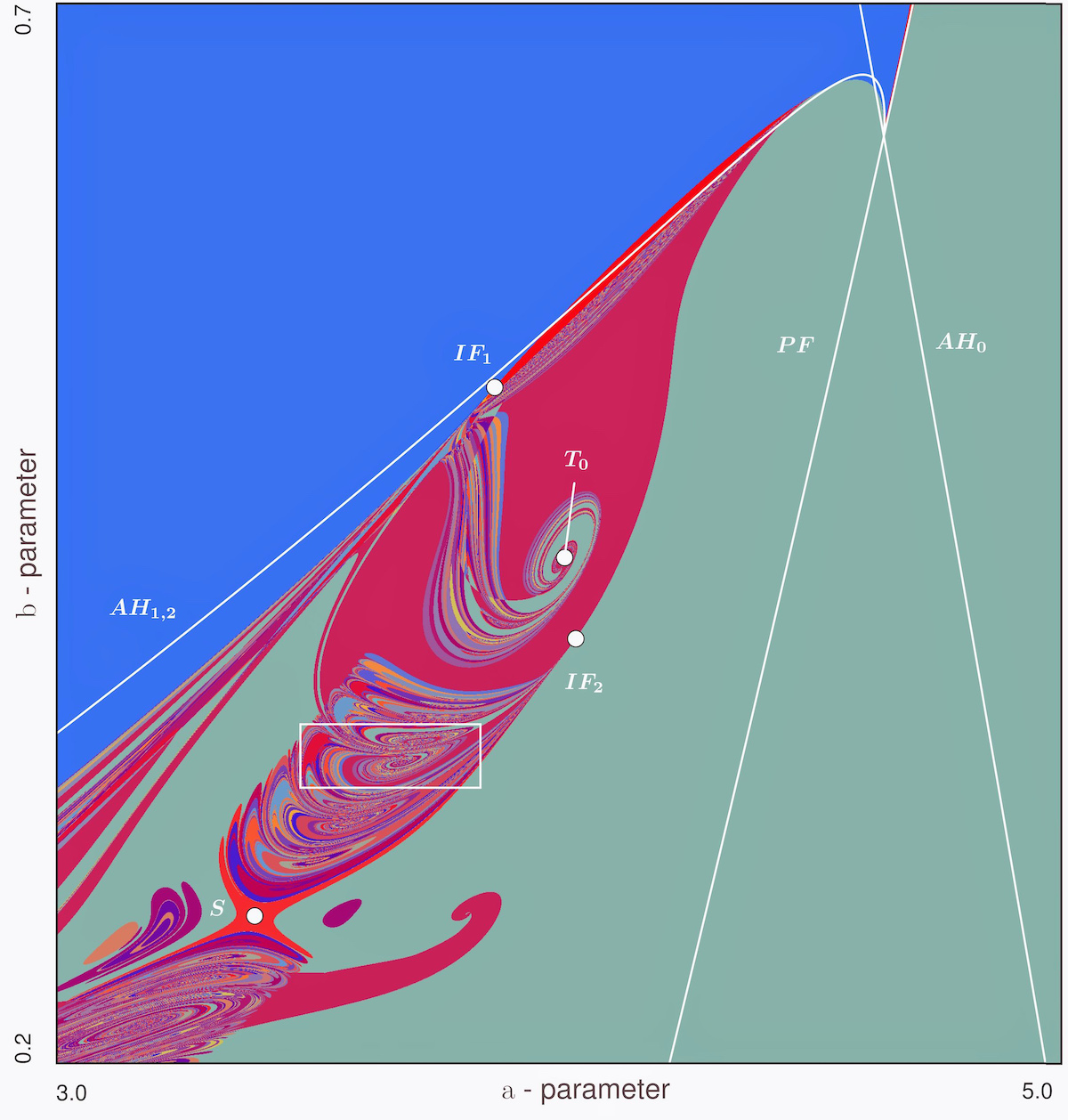}
\caption{The $(a,b)$-parametric sweep of the OPL-model at $\sigma=1.5$ using short symbolic sequences $\{k_i{\}}_{i=5}^{12}$ to detect homoclinic and heteroclinic bifurcation structures (compare Fig.~\ref{fig:sigma15Cont}). Regions of solid colors correspond to topologically identical dynamics while their boundaries represent homoclinic bifurcation curves. Superimposed white lines $PF$, $AH_0$ and $AH_{1,2}$ correspond to pitchfork and supercritical Andronov-Hopf bifurcations of equilibria $O$ and $P^\pm$, respectively. Points labeled by $IF_1$, $IF_2$ correspond to cod-2 inclination flip bifurcations, $T_0$ is the primary T-point $\{1\overline{0}\}$, and $S$ is a bridging saddle (see Fig.~\ref{fig:saddleMerge},\ref{fig:saddle3D}). White box marks the region including $T_{1,2}$ that is magnified in Fig.~\ref{fig:sigma15_IFs_T1T2}b.}\label{fig:sigma15}
\bigskip
\includegraphics[width=0.5\textwidth]{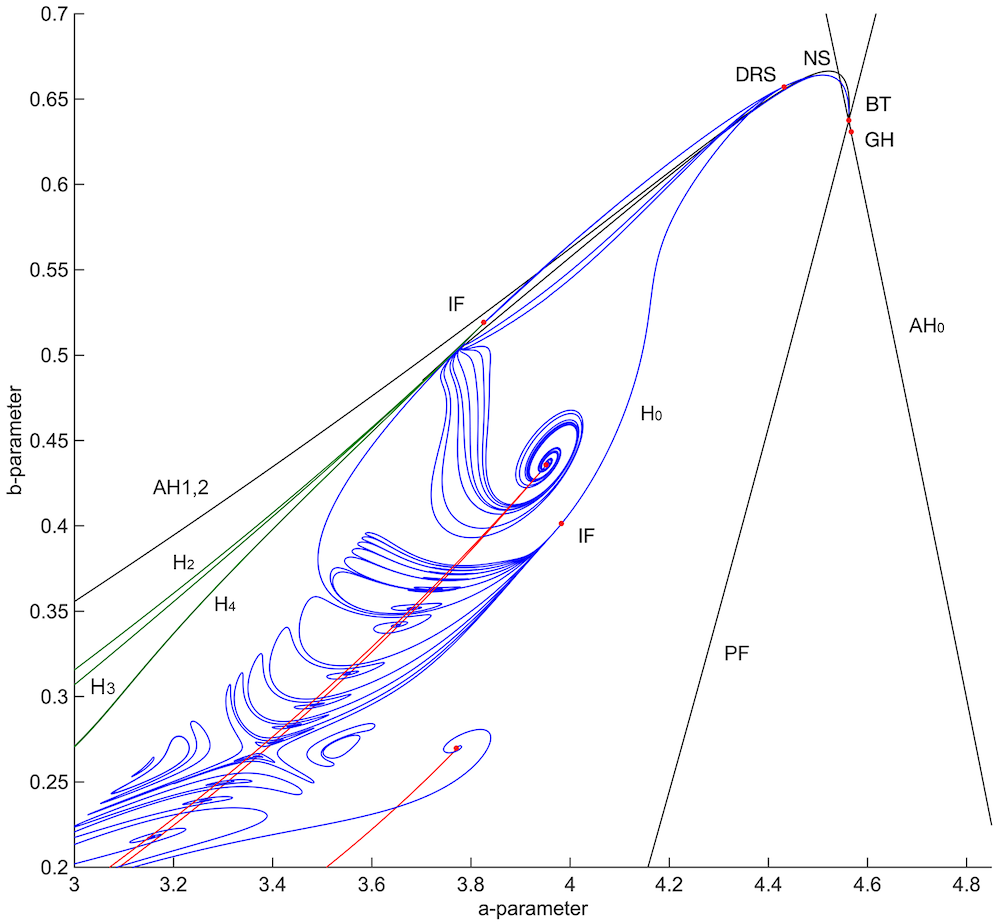}
\caption{Reverse reconstruction of the sweep presented in Fig.~\ref{fig:sigma15} using the continuation approach with MatCont. Blue and red curves originating from $IF_1$, $IF_2$ and $T_0$ correspond to homoclinic orbits of the saddle $O$ and the heteroclinic connections between $O$ and the saddle-foci $P^{\pm}$, respectively. $PF$, $AH_0$ and $AH_{1,2}$ stand for pitchfork and Andronov-Hopf bifurcation curves crossing at the ${\mathbf Z_2}$-symmetric cod-2 Bogdanov/Khorozov-Takens (BT) point. Other cod-2 points on the primary homoclinic bifurcation curve $H_0$ include $NS$ for a neutral saddle, $DRS$ for a change of the stable leading eigenvalue, and the point $GH$ stands for the cod-2 Bautin bifurcation of the change in super-criticality on $AH$-curve. Green lines labeled by $H_{2,3,4}$ originating from $IF_1$ correspond to single-sided homoclinic orbits coded as $\{11\}$, $\{111\}$ and so forth, see Fig.~\ref{fig:sigma15_IF_3D}d. Note that secondary T-points are located within the wedge bounded by two red lines corresponding to homo- and heteroclinic bifurcations of the saddle-foci.}\label{fig:sigma15Cont}
\end{figure*}

The OPL system (\ref{lasermodel}) has either a single non-lasing steady state $O$ or an extra pair of equilibria $P^\pm$ that emerge following the loss of stability of $O$ through a pitchfork bifurcation. Some elements of the simple, Morse-Smale dynamics including stable equilibria and various periodic orbits are sampled in Fig.~\ref{fig:traces} at $\sigma=1.5$. It was shown in the original papers \cite{Forysiak:1991,Moloney:1989}, as well as recent ones \cite{Barrio:2012, Pusuluri:2018}, that the OPL-model is quite rich in bifurcations -- the list includes two super-critical Andronov-Hopf and pitchfork bifurcations, occurring respectively on the curves $AH_0$ (for $O$), $AH_{1,2}$ (for $P^\pm$), and $PF$ in the bifurcation diagrams presented in Figs.~\ref{fig:sigma15},\ref{fig:sigma15Cont}. The description of simple dynamics starts off with the cod-2 point labeled as $BT$ which stands for the Bogdanov-Takens bifurcation of an equilibrium state with two zero Lyapunov characteristic exponents. Given that the OPL-model is $\mathbf{Z}_2$-symmetric, and so is the local central manifold containing such an equilibrium state, this should be referred to as a Khorozov-Takens bifurcation \cite{Shilnikov:2001}. Its unfolding includes four bifurcation curves originating from the $BT$-point in the parameter plane, $AH_0$, $PF$, $AH_{12}$ and $H_0$, which we indeed retrieve all computationally using Matcont.
 
It is convenient to describe the bifurcation unfolding of simple solutions of the OPL model first, starting off with the trivial equilibrium state $O$, where $\beta=p_{23}=p_{31}=0$. This equilibrium remains stable in the parameter space to the right of the bifurcation curve labeled $AH_0$ in Figs.~\ref{fig:sigma15},\ref{fig:sigma15Cont}. The curve $AH_0$ corresponds to a supercritical Andronov-Hopf bifurcation that gives rise to a stable figure-8 periodic orbit shown in Fig.~\ref{fig:traces}b (blue), existing for parameter values to the left of $AH_0$. The curve labeled $PF$ corresponds to a pitchfork bifurcation that gives rise to a couple of additional equilibrium states $P^{\pm}$ with $\beta\neq 0$ (see Fig.~\ref{fig:traces}c blue and green dots) emerging from $O$, which then becomes a saddle. This saddle is of (5,1)-type, i.e., $O$ has a pair of 1D unstable separatrices $\Gamma_{1,2}$ due to a single positive eigenvalue and a 5D stable manifold due to five eigenvalues with negative real parts (including a complex conjugate pair). The equilibrium states $P^{\pm}$ undergo another super-critical Andronov-Hopf bifurcation occurring on the curve $AH_{12}$ (Figs.\ref{fig:sigma15},\ref{fig:sigma15Cont}) so that a pair of stable periodic orbits emerge from $P^{\pm}$. In addition, the unfolding of the symmetric codimension-2 $BT$-point includes another bifurcation curve labeled $H_0$. It corresponds to the occurrence of a homoclinic figure-8 pattern made up of two separatrix orbits $\Gamma_{1,2}$ of $O$ that each turn once around $P^{\pm}$, respectively (see Fig.~\ref{fig:traces}c red for $\Gamma_{1}$). This is a so-called ``gluing'' bifurcation: after the stable periodic orbits on either side become the separatrix loops of the saddle $O$, they get ``glued'' to produce a stable figure-8 periodic orbit. Note that the saddle $O$ included into the homoclinic figure-8, has two leading eigenvalues, the positive and the largest negative (closest to the imaginary axis), whose sum, known as the saddle-value, remains negative near the $BT$-point. This condition assures the stability of the periodic orbits. 
 
\subsection{Cod-2 bifurcations: homoclinic and heteroclinic skeleton }

The system~(1) undergoes a homoclinic bifurcation when both unstable separatrices $\Gamma_{1,2}$ of the saddle equilibrium $O$ come back to itself, along the stable leading directions on the 5D stable manifold (Fig.~\ref{fig:traces}c). The primary homoclinic bifurcation occurs on the curve $H_0$ in the 2D parameter diagrams represented in Figs.~\ref{fig:sigma15},\ref{fig:sigma15Cont}. The goal of this paper is to demonstrate the pivotal role of such bifurcations for Lorenz-like systems. Furthermore, we want to disclose the structure of global homoclinic unfoldings in the 3D parameter space of the OPL-model by illustrating different ways in which the homoclinic curves arrange themselves, morph, and switch branches across saddle points, as well as how such parametric saddles emerge/disappear as pairs of T-points merge together.

In his Ph.D. thesis, L.P. Shilnikov generalized the theory of homoclinic bifurcations of saddle and saddle-node equilibrium states, which would lead to the emergence of a single stable periodic orbit in $\mathbb R^n$, $n \ge 3$ \cite{LP62,LP63}. Later, in 1968 he published another paper proving the existence and uniqueness conditions of a saddle periodic orbit emerging from a homoclinic loop of a plain saddle with a positive saddle value in $\mathbb R^{3}$ and higher dimensions \cite{Shilnikov:1968}. In it, he pointed out at three specific  conditions giving rise to codimension-2 bifurcations code-named {\em a zero saddle value}, {\em a zero separatrix value} and {\em the change of the leading direction} at the saddle, that are, respectively, and alternatively known today as a resonant or neutral saddle, an inclination-flip and an orbit-switch (see Fig.~\ref{fig:sigma15_IF_3D}). Furthermore, it was shown by L.P. Shilnikov that upon the fulfillment of certain conditions, these bifurcations can lead to the onset of complex dynamics in $\mathbf{Z_2}$-symmetric systems, specifically, to the emergence of the Lorenz attractor in the phase space, right next to the homoclinic butterfly occurring near these cod-2 points on the bifurcation curves in the parameter diagram of such systems  \cite{LP81}, see also \cite{ALS91,Rob89,Ry90,Shilnikov:1993}. These   results became a scientific folklore long time ago, i.e., lacking the proper acknowledgment of the original author, unlike his famous Shilnikov saddle-focus \cite{Shilnikov:1965,LP67,LP70} and a lesser known Shilnikov saddle-node or saddle-saddle \cite{LP69,Shilnikov:2007,AGLST:2014}.

A salient feature of the OPL-model is the inclination-flip bifurcations ($IF$-points in Figs.~\ref{fig:sigma15} and \ref{fig:sigma15Cont}) that occur on the primary homoclinic butterfly unlike the case of the Shimizu-Morioka (SM) model where it occurs on the double homoclinic loops; however, the SM-model shows the presence of a resonant homoclinic butterfly of the saddle with zero saddle value. Recall that the saddle value is the sum of the characteristic exponents (or their real values in the complex case) that are the closest (or leading) to the origin in the complex plane. Its sign determines whether a stable or unstable periodic orbit bifurcates from the homoclinic loop, or in the symmetric case, whether the homoclinic butterfly bifurcation glues two stable orbits into the stable figure-8 pattern, or can lead to the emergence of the Lorenz-like attractor near the resonant saddle in the parameter space.

\begin{figure}[t!]
\centering
\includegraphics[width=.6\textwidth]{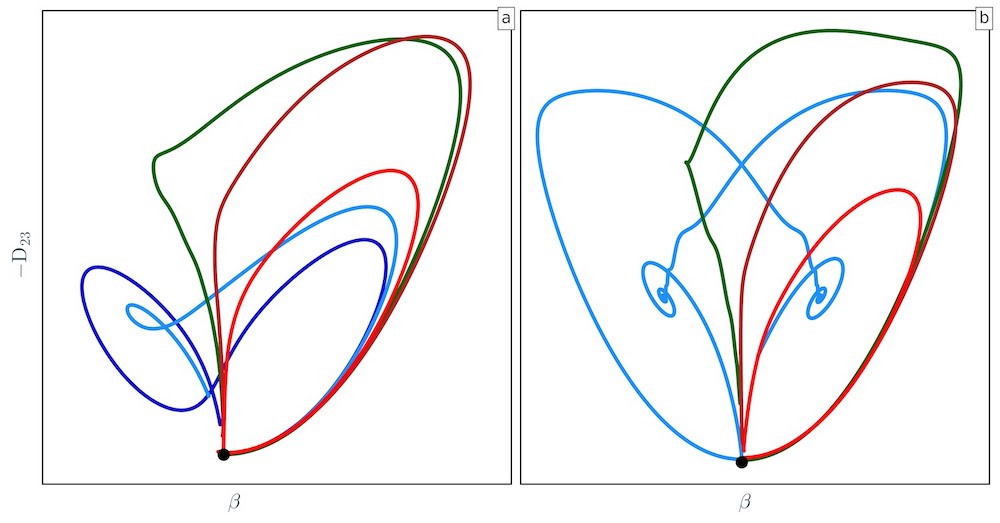}
\caption{\textbf(a) The primary homoclinic orbit, coding $\{1\}$, transforming into a double loop, coding $\{10\}$, as the bifurcation parameters are sampled along the closed corresponding curve $H_0$ in Figs.\ref{fig:sigma15} and \ref{fig:sigma15Cont} at $\sigma=1.5$. Between $DRS$ and $IF_1$ points, $\Gamma_1$ tracing out an oriented single loop [light red, $(4.124910, 0.595354)$] comes back to $O$, right-tangent to $D_{23}$-direction on the right; Beyond $IF_1$, the homoclinic orbit [dark red, $(3.827, 0.51903)$] becomes twisted (non-oriented) as $\Gamma_1$ comes back left-tangent to the leading direction. Between $IF_1$ and $IF_2$, the twisted single loop [green, $(3.76, 0.354)$] turns around the 1D stable manifold of the saddle-focus $P^-$ (see Fig.~\ref{fig:sigma15_IF_3D}); Beyond $IF_2$, the double loop [light blue, $(4.125120, 0.479283)$] becomes oriented and morphs into the homoclinic butterfly [blue, $(4.241910, 0.598092)$] as we approach the $NS$ point. \textbf(b) Transformation stages of the primary homoclinic orbit $\{1\}$ into the heteroclinic connection $\{1\bar 0\}$ [light red $(3.701, 0.8734)$ $\rightarrow$ dark red $(3.321, 0.69885) \rightarrow$ green $(3.03, 0.39331)\rightarrow$ light blue $(3.276, 0.5186)$] along the bifurcation curve $H_0$ spiraling onto the primary T-point at $\sigma=2.0$, see Fig.~\ref{fig:sigma20}. In this case, as we sample the $H_0$-curve beyond $IF_2$, the twisted loop makes more turns around the stable manifold of $P^-$, until it becomes the heteroclinic connection with $P^-$ at the T-point.}\label{fig:sigma15_tracesHomoclinic}
\end{figure}

\begin{table}
\centering 
\begin{tabular}{|c|c|c|}
\hline
$\sigma$ & a & b \\
\hline
1.50 & 4.4309 & 0.6572\\
1.55 & 4.3394 & 0.6973\\
1.60 & 4.2382 & 0.7258\\ 
1.65 & 4.1333 & 0.7453\\ 
1.70 & 4.0291 & 0.7584\\
1.75 & 3.9285 & 0.7671\\
1.80 & 3.8325 & 0.7728\\
\hline
\end{tabular}
\caption{Table of the coordinates of the orbit-switch (DRS) points in the 3D $(a,b,\sigma)$-parameter space of the OPL-model corresponding to double-real stable eigenvalues of the saddle $O$.}\label{tab:DRS}
\end{table}

Let us go over the sequence of cod-2 bifurcations that occur along the primary homoclinic curve $H_0$ in the parameter space. For reference, see Figs.~\ref{fig:sigma15},\ref{fig:sigma15Cont} at $\sigma=1.5$ (as well as Figs.\ref{fig:sigma20},\ref{fig:sigma100} corresponding to different cuts at $\sigma=2$ and $10$) in the parameter space of the OPL-model. The saddle $O$ is of (5,1)-type, i.e., has the eigenvalues that can be ordered as follows: $\dots<\lambda_3 <\lambda_2< 0<\lambda_1$. For such homoclinic saddles we need to evaluate the values of the following quantities: the saddle value $\Sigma=\lambda_2 + \lambda_1$ whose sign determines the stability of a periodic orbit emerging from a homoclinic loop -- negative/positive $\Sigma$ values imply stable/saddle orbits; alternatively, one can use the first saddle index $\nu= |\lambda_2|/\lambda_1>1$ or $<1$. Whenever $\Sigma=0$, one needs to evaluate the second saddle index - the smallest quantity $\mu= |2\lambda_2|/\lambda_1$ or $\mu= |\lambda_3|/\lambda_1$. This holds true for the inclination flip case as well, when $\nu<1$.

The homoclinic figure-8 connection stands for the case where both the unstable separatrices $\Gamma_{1,2}$ of the saddle $O$ come back to itself, opposite to each other along the leading $\mathbf{Z_2}$-symmetric direction on the stable manifold, whereas in the homoclinic butterfly case, both $\Gamma_{1,2}$ come back tangent to each other and to the asymmetric leading direction -- the vertical axis in Figs.~\ref{fig:traces} and \ref{fig:sigma15_tracesHomoclinic}. Near the cod-2 point $BT$, the homoclinic connection is initially of the figure-8 case, with a stable gluing bifurcation ($\Sigma<0$) occurring in a plane. As the bifurcation curve $H_0$ is continued further away from $BT$, Matcont detects the following sequence of singular homoclinic bifurcations depending on the $\sigma$-cut. In case of $\sigma=1.5$ in Fig.~\ref{fig:sigma15}, they are labeled as follows: $NS$ corresponds to a resonant saddle with a neutral or zero saddle value $\Sigma=\lambda_1+\lambda_2=0$ or with a saddle index $\nu=|\lambda_2|/\lambda_1=1$. After that the homoclinic connection is continued with $\Sigma>0$, and hence gives rise to unstable/saddle periodic orbits. This should explain the reason for the presence of a Bautin bifurcation ($GH$) point in this bifurcation puzzle, near the Bogdanov-Takens phenomenon. Recall that its unfolding includes the saddle-node bifurcation through which stable and unstable periodic orbits merge and annihilate. The following homoclinic cod-2 bifurcation detected by MatCont is the homoclinic orbit-switch ($DRS$) due to the change of the leading direction. In $\mathbf{Z_2}$-symmetric systems like the OPL-model, this corresponds to the transition from the homoclinic figure-8 to the homoclinic butterfly with with $\Sigma>0$. It follows from theory that a thin Lorenz-like attractor, hardly indistinguishable from the homoclinic butterfly without significant magnification, can already emerge near this point in the parameter space. One can see a loci of bifurcation curves emerging from the $DRS$-point, in a vicinity of the $BT$-point, in Figs.~\ref{fig:sigma15Cont} and \ref{fig:sigma20}, which will be further described later. Table~\ref{tab:DRS} lists the coordinates of the cod-2 homoclinic orbit-switch points in the $(a,b,\sigma)$-parameter space of the OPL-model. Fig.~\ref{fig:sigma15_tracesHomoclinic}a shows how the primary homoclinic loop morphs into a double loop as we move along the primary homoclinic bifurcation curve $H_0$ in Fig.~\ref{fig:sigma15Cont} at $\sigma=1.5$. For a different parametric cut in Fig.~\ref{fig:sigma20} at $\sigma=2.0$, it forms a heteroclinic connection with the saddle foci as shown in Fig.~\ref{fig:sigma15_tracesHomoclinic}b.

\subsubsection{Inclination-flip homoclinic bifurcation}

\begin{figure}[ht!]
\centering
\includegraphics[width=0.6\textwidth]{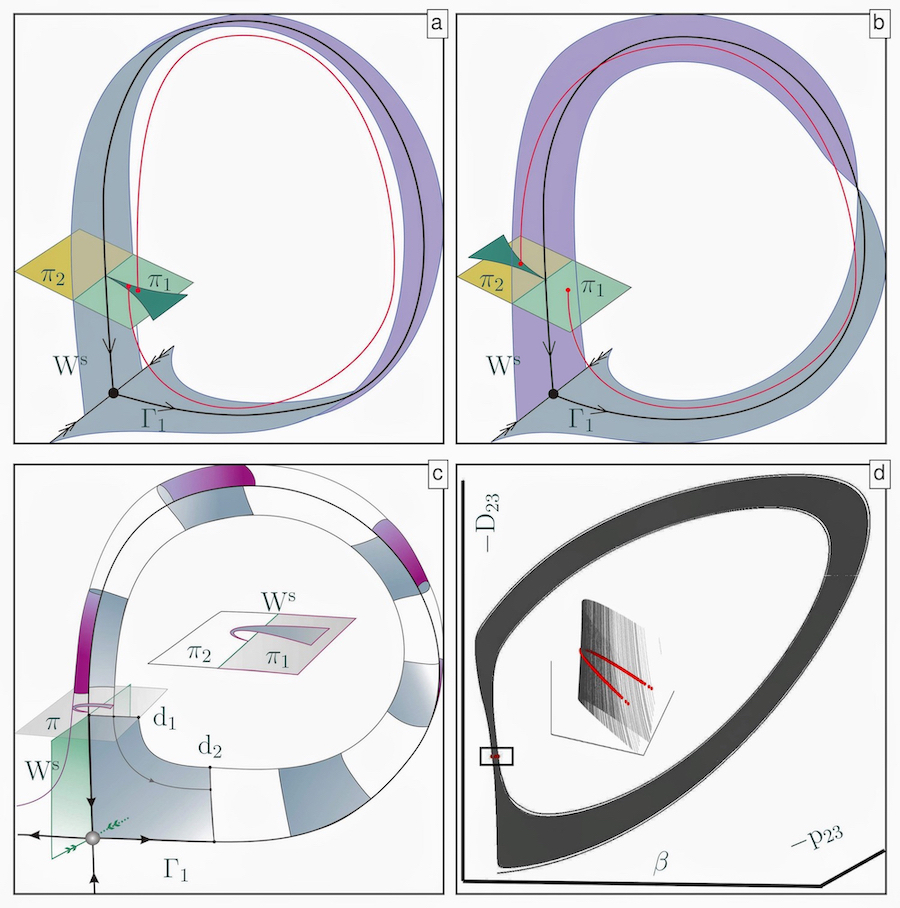}
\caption{Geometry and consequences of the cod-2 {\it inclination flip} homoclinic bifurcation in a 3D phase space. \textbf{(a)} A section $\pi_1$ of a 2D cross-section transverse to the stable manifold $W^s$ of the saddle with a saddle index $\nu<1$ is taken back along the orientable primary homoclinic loop $\bar \Gamma_1$ (in black) and a close trajectory (red) so that its image (darker wedge) is mapped back onto the original $\pi_1$, whereas in the non-orientable or twisted case \textbf{(b)}, it is mapped onto the opposite section $\pi_2$ (dark yellow). This makes the surface ``spanned'' by the unstable and non-leading stable eigenvectors look like a M\"obius band.
\textbf(c) After the IF bifurcation, the flow bends the surface of the local manifold spanned by the unstable and leading stable eigenvectors, so that its image has the shape of a hook or a Smale horse-shoe.
\textbf(d) This gives rise the onset of one of two ``one-sided'' chaotic attractors  in the OPL-model near each primary homoclinic loop (at $a=3.7, b=0.48415447, \sigma=1.5$) with a distinct bent horseshoe (red dots) on a cross-section near the saddle in this 3D phase-space projection. 
}\label{fig:sigma15_IF_3D}
\end{figure}

\begin{figure}[ht!]
\centering
\includegraphics[width=.4\textwidth]{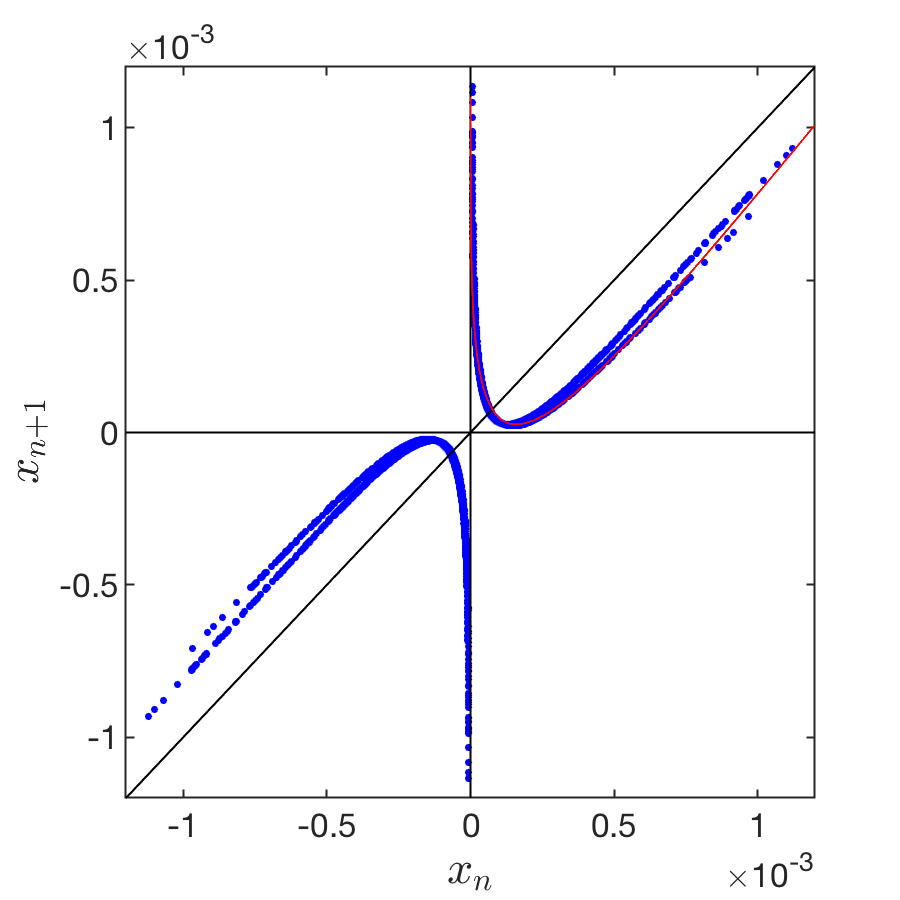}
\caption{Unimodal shape of a symmetric 1D  map $x_n \to x_{n+1}$ derived from the simulated 2D return map in Fig.~\ref{fig:sigma15_IF_3D}d to illustrate the dynamics due to period-doubling and homoclinic bifurcations occurring in the fitted model (red line) given by Eq.~(\ref{map}) near the primary inclination-flip bifurcation. As $\mu$ is decreased further, both ``one-sided'' non-oriented attractors (Fig.~\ref{fig:sigma15_IF_3D}d) merge into one symmetric after the folds of the map cross over the $x_n$-axis.}\label{fig:1Dmap}
\end{figure}


Of special interest in this paper is the role(s) of cod-2 inclination flip bifurcations (labeled as $IF$) in the parameter space and how they shape and organize the global bifurcation unfolding of the OPL-model, see Figs.~\ref{fig:sigma15},\ref{fig:sigma15Cont},\ref{fig:sigma15_IF1_short_long}. The IF bifurcation occurs when an orientable stable or unstable manifold transitions to a non-orientable one, when followed along the homoclinic orbit. Unlike the Shimizu-Morioka model \cite{Xing:2014}, this bifurcation in the OPL model involves the primary homoclinic butterfly and not the double homoclinic loops. As mentioned previously, L.P. Shilnikov pointed out the conditions under which the IF bifurcation gives rise to the onset of the Lorenz attractor in $\mathbf{Z_2}$-symmetric systems. These conditions helped determine and verify the regions of existence of the Lorenz attractor, as well as enabled computer assisted proofs of chaotic dynamics in the canonical Lorenz model, without stable orbits and homoclinic tangencies \cite{ABS:1983,Bykov:1992,Tucker:1999,Viana:2000}. One such condition showed that the inclination-switch bifurcation can also lead to the onset of stable orbits nearby in the phase space, even in the case of an expanding saddle whose saddle index satisfies the condition $1/2<\nu<1$. Below, we will highlight the essence of the inclination-switch bifurcation. Its in-depth analysis is given in \cite{SSTC:1998, Shilnikov:2001}.

3D images in Figure~\ref{fig:sigma15_IF_3D} illustrate the concept of an inclination flip bifurcation that transforms an orientable homoclinic loop of the saddle (Figure~\ref{fig:sigma15_IF_3D}a) into a non-orientable one (Figure~\ref{fig:sigma15_IF_3D}b). This is depicted by a single twist of the surface traced out by a small, local section of the strongly stable direction, due to the smaller eigenvalue $\lambda_3$ ($\dots<\lambda_3<\lambda_2<0 <\lambda_1$) of the saddle. Let us explore the global return map $\mathbf{T}$ that takes a 2D cross-section, $\pi=\pi_1 \cup \pi_2$, transverse to the stable manifold $W^s$ of the saddle with a saddle index $\nu<1$, onto itself along the homoclinic loop. The 1D outgoing separatrix of the saddle ( here $\Gamma_1$) comes back along the leading direction (vertical, due to $\lambda_2$). Before the inclination-flip (Figure~\ref{fig:sigma15_IF_3D}a), the section $\pi_1$ (the right half, relative to $W^s$) is mapped back on to the original $\pi_1$ (darker wedge), along the orientable primary homoclinic loop $\Gamma_1$ (in black). A nearby trajectory (red) is also shown. Post the inclination flip (Figure~\ref{fig:sigma15_IF_3D}b), in the non-orientable or twisted case, the section $\pi_1$ is mapped onto the opposite section $\pi_2$ (dark yellow). This makes the surface ``spanned'' by the unstable (due to $\lambda_1$) and the non-leading stable (due to $\lambda_3$) eigenvectors look like a M\"obius band.

Figure~\ref{fig:sigma15_IF_3D}c depicts the action near an IF bifurcation using a 2D local Poincar\'e map. The map takes a small interval $d_1 \ll 1$ on $\pi$ into $d_2 \sim d_1^\nu > d_1$. For $\nu<1$, the local map near the saddle is typically an expansion, stretching phase areas or volumes. Let us consider a small portion of the local manifold $\mathfrak{M}$ tangent to the span of the leading stable (due to $\lambda_2$) and the leading unstable eigenvectors of the saddle $O$. Near an IF bifurcation, as we take $\mathfrak{M}$ along the outgoing separatrix loop in a succession, it bends and hits the cross-section $\pi$ as a hook, narrowly squeezed due to the strongly stable eigenvalue $\lambda_3$. This hook $\mathbf{T}\pi_1$ of the pre-image $\pi_1$ looks like a Smale horseshoe on the cross-section (see Fig.~\ref{fig:sigma15_IF_3D}c inset). This bending makes the image of $d_2$ shorter than the original $d_1$, making the global map $\mathbf{T}$ a contraction ($\mathbf{T}d_1 < d_1$) as it overcomes the local expansion near the saddle ($d_2 > d_1$). Note that the global map $\mathbf{T}$ is an expansion prior to the IF bifurcation. Figure~\ref{fig:sigma15_IF_3D}d illustrates the onset of one of two ``one-sided'' chaotic attractors in the OPL-model near the inclination flip point on the primary homoclinic loop (at $a=3.7, b=0.48415447, \sigma=1.5$, next to the $IF_1$-point in  Fig.~\ref{fig:sigma15}) due to the presence of such a hook/Smale horseshoe (red), filled in with the points of intersection of transverse segments of the outgoing separatrix $\Gamma_1$ of the saddle $O$ on a nearby cross-section. The inset depicts a simulated analogue of the bending manifold $\mathfrak{M}$ sketched in Fig.~\ref{fig:sigma15_IF_3D}c.

The geometric model of the Lorenz attractor (presented in \cite{ABS:1983}) can be elaborated using a 2D return map near the primary homoclinic butterfly of the two separatrix loops of the saddle. For a system to possess a genuine chaotic attractor devoid of homoclinic tangencies or stable orbits, this map should satisfy a few analytical conditions. The boundary of its existence region is marked by a violation of these criteria. The 2D map near the IF bifurcations can be further reduced to a simplified 1D map (Fig.~\ref{fig:1Dmap}) in the following form \cite{Shilnikov:2001}:

\begin{equation}\label{map}
x_{n+1} = \left [ \mu + A |x_{n}|^{\nu} +o(|x_{n}|^{2\nu}) \right ] \cdot \mbox{sign}(x_n), 
\end{equation}
where $1/2<\nu={|\lambda_2|}/{\lambda_1}<1$ is the saddle index, $A$ is the separatrix value, and $\mu$ is the splitting parameter of the homoclinic loop \cite{ASHIL93}. Near the IF bifurcations when $0 \sim |A| \ll 1$, the term $o(|x_{n}|^{2\nu})$ is no longer negligible. Fig.~\ref{fig:1Dmap} demonstrates the bent shape of such a return map, derived from the simulated 2D map, shown in Fig.~\ref{fig:sigma15_IF_3D}d, for the ``one-sided'' chaotic attractor. As $\mu$ is decreased further, both ``one-sided'' (left/right) non-oriented attractors merge into a symmetric one. The map illustrates possible dynamics near the primary  IF bifurcation due to period-doubling bifurcations in the 1D Poincar\'e map, leading to the emergence of multiple secondary homoclinic curves from this cod-2 point, as well as multiple stable periodic orbits (Figs.~\ref{fig:sigma15_IFs_T1T2},\ref{fig:sigma12_short_long},\ref{fig:sigma15_IF1_short_long},\ref{fig:sigma100_IF2_short_long}) when the return map starts bending again and again to resemble the return map of a Shilnikov saddle-focus but with a finite number of Smale horse-shoes \cite{Xing:2014}.

\subsubsection{Bykov ``terminal'' T-points}

\begin{figure}[ht!]
\centering
\includegraphics[width=0.35\textwidth]{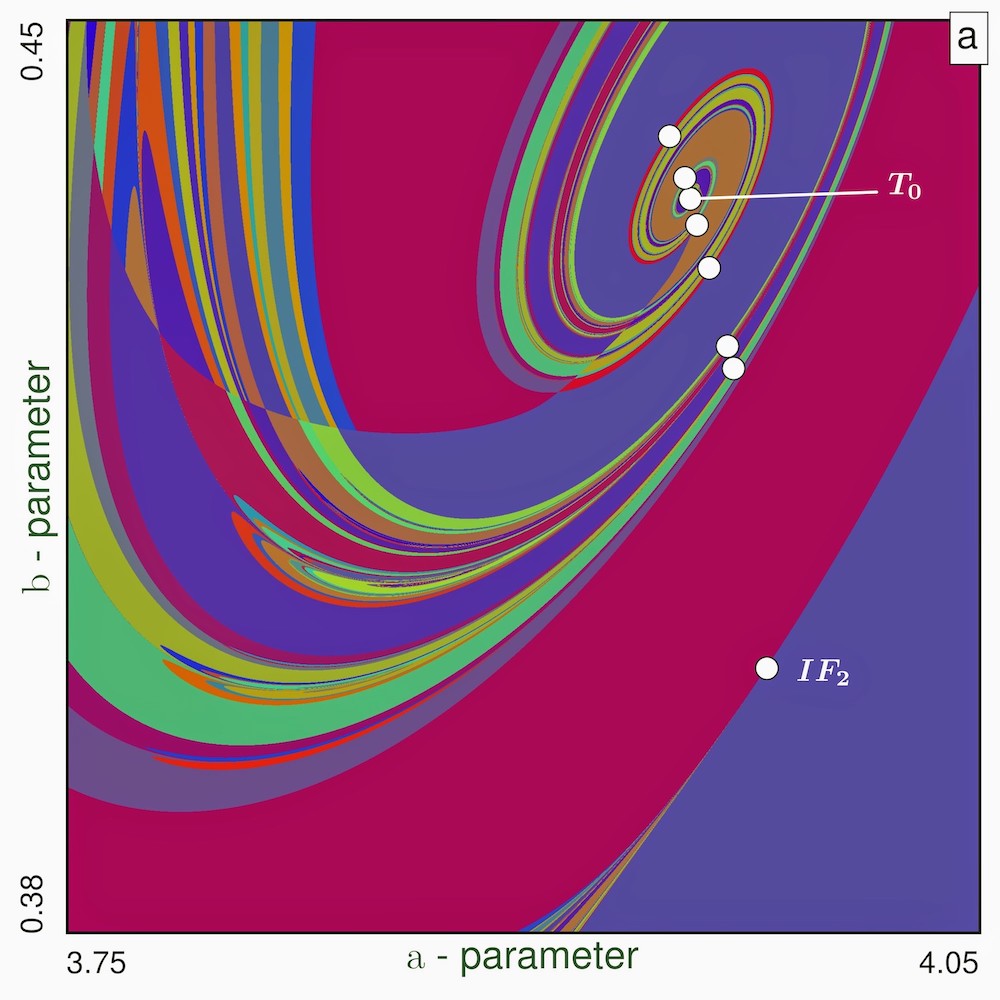}
\includegraphics[width=0.35\textwidth]{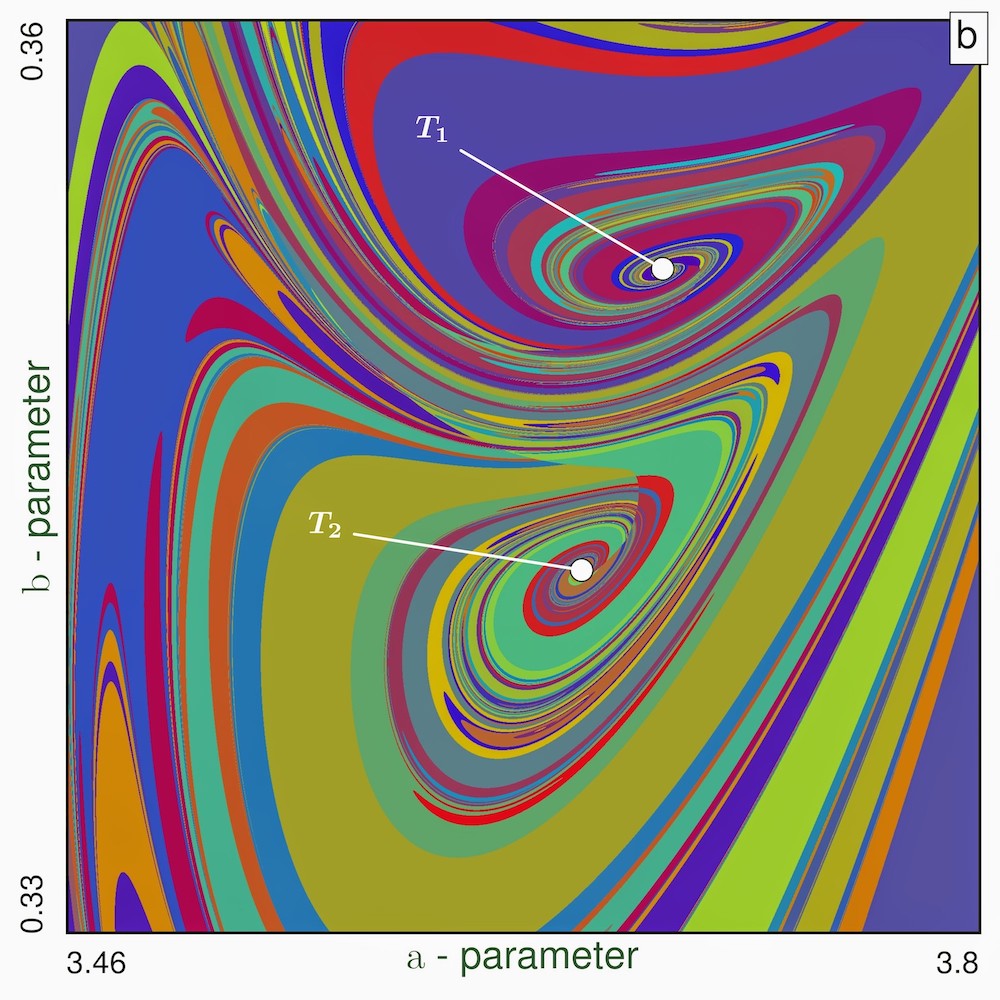}
\caption{\textbf{(a)} Magnification of the vicinity of the primary T-point $T_0$ from Fig.~\ref{fig:sigma15}, using $\{k_i{\}}_{i=5}^{12}$. Multiple secondary inclination-flip points (marked with white dots on critical points of homoclinic spirals) originating from $IF_2$ and passing throughout $T_0$ give rise to other  homoclinic curves spiraling onto  secondary T-points nearby. \textbf{(b)} Magnification of the white box region in Fig.~\ref{fig:sigma15}, using $\{k_i{\}}_{i=3}^{10}$, showing symmetric T-points $T_1$ $\{10\overline{1}\}$ and $T_2$ $\{11\overline{0}\}$, above and below the primary homoclinic bifurcation curve $H_0$, respectively. The corresponding phase trajectory at $T_1$ is shown in Fig.~\ref{fig:traces}d.
}\label{fig:sigma15_IFs_T1T2}
\end{figure}

\begin{figure}[ht!]
\centering 
\includegraphics[width=0.35\textwidth]{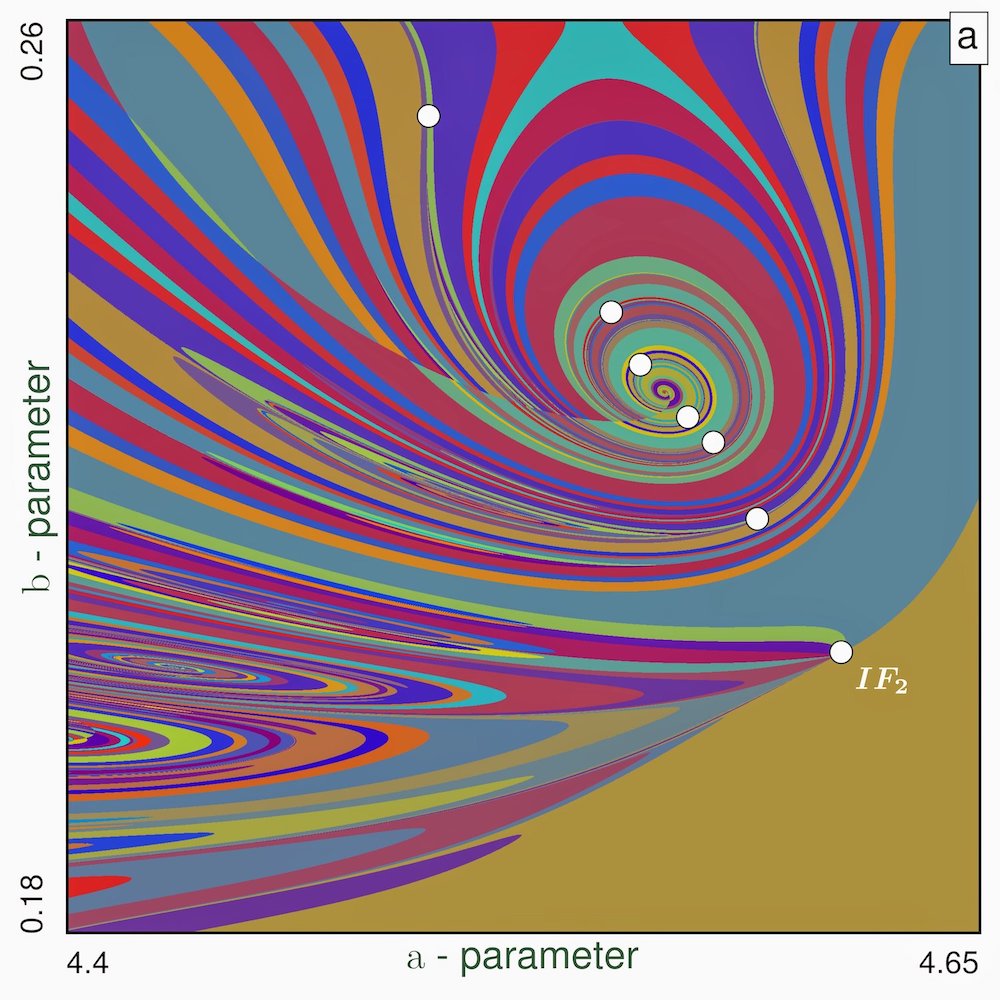}
\includegraphics[width=0.35\textwidth]{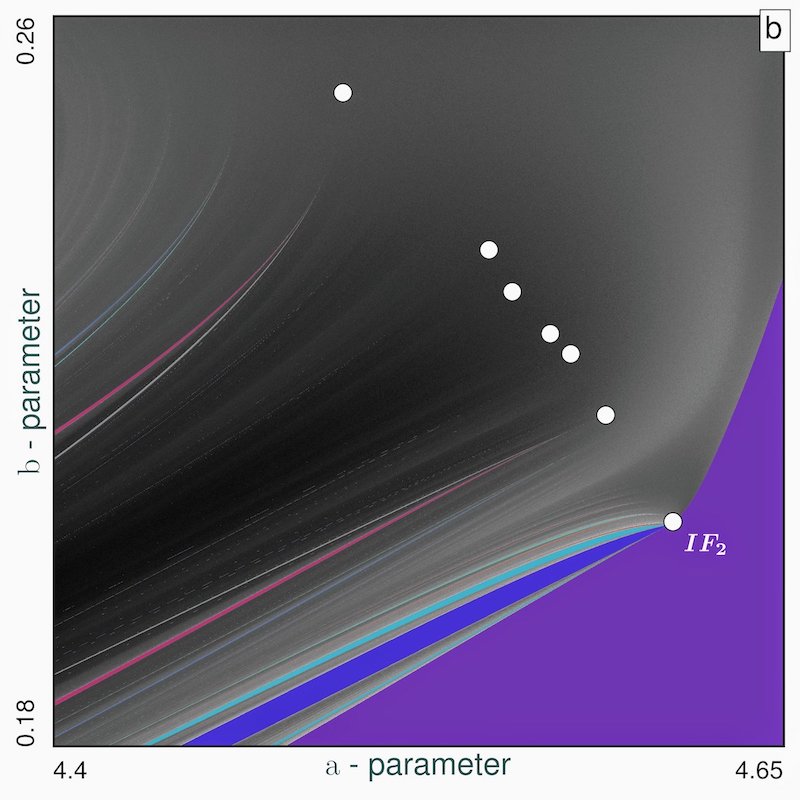}
\caption{Short-term and long-term sweeps to disclose the multiplicity of basic inclination-flip bifurcation points (white dots) at $\sigma=1.2$. \textbf{(a)} $\{k_i{\}}_{i=2}^{9}$ sweep illustrates  a locus of  homoclinic curves converging towards the primary and secondary T-points to the inclination-flip points $IF_2$ \textbf{(b)} Long $\{k_i{\}}_{i=1000}^{1999}$ DCP-based sweep reveals a variety of large and narrow stability windows, also known as the {\em Shilnikov flames}, originating below subsequent inclination flip points located on the boundary (not shown here) separating the region of the Lorenz attractor (above it) from that of quasi-attractors coexisting with stable periodic orbits.}\label{fig:sigma12_short_long}
\end{figure}

A typical signature for Lorenz-like systems is the complex universal and self-similar characteristic spirals in the parameter plane, organized around a central point called a Bykov terminal point (T-point) as seen in Figs.~\ref{fig:sigma15},\ref{fig:sigma15Cont} and detailed magnifications are presented in Fig.~\ref{fig:sigma15_IFs_T1T2} \cite{Pusuluri:2018}. Each characteristic spiral around a T-point in the parameter plane corresponds to a homoclinic loop of the saddle $O$ in the phase space, such that with each turn of the spiral approaching the T-point, the outgoing separatrix of saddle makes increasing number of loops around a saddle-focus $P^\pm$, before finally forming a closed heteroclinic connection between the saddle and the saddle focus at the T-point. Fig.~\ref{fig:traces}d shows a one-way heteroclinic connection between the saddle and the saddle focus for parameter values near the T-point $T_1$ in Fig.~\ref{fig:sigma15_IFs_T1T2}b, with $(a,b)\sim(3.68179,0.3517)$. Here, the unstable separatrix $\Gamma_1$ (red) of the saddle $O$ makes one loop around $P^+$, followed by another loop around $P^-$, then merges with the incoming separatrix of $P^+$, thus effectively making infinite loops around $P^+$ before emerging out. The symmetric heteroclinic connection along $\Gamma_2$ from $O$ to $P^-$, is shown in blue.

Near a T-point, there exist countably many secondary T-points with increasing complexity of heteroclinic connections in the phase space and with similar bifurcation structures as the central T-point in the parameter plane, although on a smaller scale (see Figs.~\ref{fig:sigma15_IFs_T1T2}, \ref{fig:sigma12_short_long}). Multiple inclination flip (IF) homoclinic bifurcations of the saddle occurring along the characteristic spirals of the T-points (detected with MatCont and shown as white dots in Figs.~\ref{fig:sigma15_IFs_T1T2}a,\ref{fig:sigma12_short_long}) give rise to saddle-node and period-doubling bifurcations of periodic orbits \cite{Shilnikov:1993,Shilnikov:2001}. Fine organization of the structure of the chaotic regions and stability windows near the T-point and surrounding IF points is revealed in greater detail in \ref{fig:sigma12_short_long}b. In addition, the unfolding of a T-point also includes other curves corresponding to the homoclinic connections of the saddle-focus satisfying the Shilnikov condition \cite{Shilnikov:1965,LP70,Shilnikov:2007} that give rise to a denumerable set of saddle periodic orbits nearby \cite{LP67}, as well as those corresponding to heteroclinic connections between both saddle-foci \cite{Bykov:1980,Bykov:1993,Glendinning:1986}.

\section{Numerical methods} 

\subsection{Parametric continuation with MatCont}

We constructed bifurcation diagrams such as Fig.~\ref{fig:sigma15Cont} using numerical continuation with MatCont \cite{Matcont}, including homotopy to initialize continuation of homoclinic and heteroclinic orbits, shown as blue and red curves, respectively \cite{DeWitte:2012}. First, we locate the local bifurcation curves of the symmetric and asymmetric equilibria. This also involves an Andronov-Hopf bifurcation. Following the limit cycle, we encounter a homoclinic bifurcation where the period explodes to infinity. In the region where the inclination flip occurs, $O$ has a single unstable direction. To determine these curves, we use homotopy to locate the connecting orbits, where we set the distances from the equilibrium to be $10^{-3}$ at most, but typically $10^{-4}$. Once a connecting orbit has been located, we continue it with $a,b$ as free parameters and the ``period'' as auxiliary continuation parameter. During continuation, we monitor the test functions for various codimension-two bifurcations, including inclination flip and neutral homoclinic bifurcations.

While results have only been proven for three-dimensional vector fields, see \cite{Homburg:2010,Bykov:1993,Glendinning:1986}, we follow these descriptions. We consider the following leading eigenvalues $...<\lambda_3<\lambda_2<0<\lambda_1$. To characterize the type of inclination flip, we define the quantities 
\begin{equation}
\nu_1=-\lambda_{2}/\lambda_{1},\qquad\qquad \nu_2=-\lambda_{3}/\lambda_{1}.
\end{equation}
If $\nu_1<\frac{1}{2}$ or $\nu_2<1$, then there are infinitely many $N$-homoclinic orbits emerging from this cod-2 bifurcation for each fixed $N\geq 2$.

\subsection{Symbolic dynamics: Deterministic Chaos Prospector (DCP)}

Our idea of symbolic dynamics is similar to the so-called kneadings \cite{Milnor:1988,GS93,GH96}, originally introduced for 1D continuous unimodal (with a single critical point) maps of an interval, and the binary description of their complex dynamics. A kneading invariant (see Eqs.~(\ref{kneadingsMain},\ref{kneadingsSum})) was meant to serve as a modulus to topologically conjugate such maps and ODE systems generating such maps, see \cite{GW79,Shilnikov:2001,Xing:2014,Barrio:2014,R78,Malkin91,TW93,GH90}. In order to apply such a symbolic technique to a generic Lorenz-like system of higher dimensions, only wave forms of a symmetric variable progressing in time, that consistently start from the same initial condition near the saddle are required.

The system exhibits the presence of a Lorenz-like attractor which is both dynamically and structurally unstable (see Fig.~\ref{fig:traces}c gray background), due to an abundance of homoclinic bifurcations of the saddle equilibrium $O$. Both the outgoing separatrices of $O$, $\Gamma_1$ and $\Gamma_2$, densely fill out the two spatially symmetric wings of the Lorenz attractor in an unpredictable flip-flop pattern. In addition to this, as we vary the parameters of the system, the attractor itself undergoes bifurcations and reorganizes the flip-flop patterns, resulting in structural instability. Such changes are marked by homoclinic bifurcations of the saddle $O$, where the outgoing separatrices, $\Gamma_1$ and $\Gamma_2$, after a few flip-flops, merge with the stable manifold of $O$ (see Fig.~\ref{fig:traces}c red curve). We make use of this property of homoclinic bifurcations to identify regions of topologically identical dynamics in parameter space by following the positive unstable separatrix $\Gamma_1$ of $O$ and converting the flip-flop patterns of the trajectory around $P^\pm$ into a binary sequence \(\{k_n\}\) defined as:

\begin{equation}\label{kneadingsMain}
k_n = \begin{dcases*}
1, & when $\Gamma_1$ loops around the ``right'' saddle-focus \(P^+\), \\
0, & when $\Gamma_1$ loops around the~ `left''~ saddle-focus \(P^-\). 
\end{dcases*}
\end{equation}

This can be simplified by just considering complete or partial loops of the trajectory on the positive or negative side of $\beta$, to be represented by the symbols 1 or 0, respectively. Therefore, the primary homoclinic bifurcation of Fig.~\ref{fig:traces}c (red) is symbolically represented as $\{1\}$ since $\Gamma_1$ makes one loop around $P^+$ before returning to $O$, while the blue and green curves are represented as $\{11...\}$ and $\{10...\}$, respectively. Repetitive sequences are represented with an over-bar, as shown in Fig.~\ref{fig:traces}a,b,d, showing convergence to a fixed point, periodic orbit, or heteroclinic connection, respectively. For example, the blue periodic orbit of Fig.~\ref{fig:traces}b (transient omitted) generates the infinite sequence ($\{010101010101...\}$), which is shortly represented as $\{\overline{01}\}$. We use the notation $\{k_i{\}}_{i=p}^q$, to represent a sequence starting from the $p^{th}$ symbol up-to the $q^{th}$ symbol, with $i,p,q$ in $\mathbb{Z}^+$ and $0 \leq p \leq q$. As we always initiate trajectories along $\Gamma_1$, we consider $k_0=1$.

Since a single homoclinic loop, upon change of a parameter, can undergo transmutation into a double loop via an inclination flip, without an underlying homoclinic bifurcation (see Fig.~\ref{fig:sigma15_tracesHomoclinic}), the above approach can give rise to artifacts near the IF bifurcations, e.g. Fig.~\ref{fig:sigma15_IFs_T1T2}a, where a single homoclinic loop starts to twist and then makes an extra loop on the left. We can avoid some of those artifacts by a slightly modified approach, given by: 

\begin{equation}\label{kneadingsModified}
k_n = \begin{dcases*}
1, & whenever $\Gamma_1$ reaches $\beta$-maxima and $\beta > 0$ and $\displaystyle  \frac{dD_{23}}{dt} < 0$,\\
0, & whenever $\Gamma_1$ reaches $\beta$-minima and $\beta < 0$ and $\displaystyle \frac{dD_{23}}{dt} > 0$.
\end{dcases*}
\end{equation}

With the former approach (\ref{kneadingsMain}), the artifacts are confined to a small region near the IF bifurcations, while the modified approach (\ref{kneadingsModified}) confines the artifacts to regions away from IF bifurcations. Therefore, we use approach (\ref{kneadingsMain}) for most of the biparametric scans (Fig.~\ref{fig:sigma15},\ref{fig:sigma15_IFs_T1T2} etc.), except for close zoom-ins near IF points (Fig.~\ref{fig:sigma15_IF1_short_long}), for which we use approach (\ref{kneadingsModified}).

\subsubsection{2D/3D parametric scans}

Topologically identical regions in the parametric plane can be identified by defining a formal convergent power series P \cite{Pusuluri:2017,Milnor:1988} for a sequence $\{k_i{\}}_{i=p}^q$ defined as:
\begin{equation}\label{kneadingsSum}
P = \sum_{i=p}^{q} \frac{k_i}{2^{q+1-i}}. 
\end{equation}
For example, the value of this power series for the sequence $\{10100101...\}$ with $p=0$ and $q=7$ can be computed as : 
$
P = \dfrac{1}{2^8}+\dfrac{0}{2^7}+\dfrac{1}{2^6}+\dfrac{0}{2^5}+\dfrac{0}{2^4}+\dfrac{1}{2^3}+\dfrac{0}{2^2}+\dfrac{1}{2} = 0.64453125. 
$ 
The value of the sum $P$ lies between 0 (for the sequence $\{\overline{0}\}$) and 1 (for the sequence $\{\overline{1}\}$ in the limit as $(q-p) \rightarrow \infty$). Two different sets of parameter values that show topologically identical behavior in their trajectories, result in identical sequences \(\{k_i\}\), and thus, have the same $P$ values. Therefore, $P$ serves as a dynamic invariant, which we can use in the construction of biparametric scans such as Fig.~\ref{fig:sigma15}, as follows. The parameter $\sigma$ is kept constant while we vary the parameters $a$ and $b$ (at a resolution of 2000x2000). For each set of values of these parameters, we follow the right unstable separatrix $\Gamma_1$ of $O$ to obtain the sequence \(\{k_i\}\) and the power series $P$. We then use a colormap to project the $P$ values onto the biparametric plane, where topological similarity between regions is identified by their equivalent colors, and the boundaries between adjacent regions represent homoclinic bifurcation curves. The colormap is constructed by discretizing the range of $P$, i.e., [0,1], into $2^{8}$ bins, and assigning them RGB-color values from 0 through 1, for each of Red, Green and Blue colors in decreasing, random, and increasing fashion, respectively. Greater weight is assigned to the symbols towards the end of the sequences \(\{k_i\}\) in the computation of $P$, to maximize the color contrast between neighboring regions in a scan, that differ in the last symbol, separated by a homoclinic  curve at their boundary. Numerical integration is performed using the classic Runge-Kutta method (RK4) with fixed step size $dt=0.01$. The computation of these trajectories is massively parallelized by running on separate GPU threads using CUDA. This allows for the construction of bi-parametric scans such as Fig.~\ref{fig:sigma100}, in as little as 8 seconds on an Nvidia Tesla K40 GPU. Visualizations are done in Python. In order to construct complex bifurcation structures in the 3D parametric space, such as Figs.~\ref{fig:sigma15_sigma20_transition3D}a,\ref{fig:sigma15_sigma20_transition3D_zooms},\ref{fig:saddle3D}, we obtain a large number of biparametric scans in the $(a,b)$ parametric plane, as we continuously vary the third parameter $\sigma$. Such arrays of scans are then rendered together using the open source volume exploration tool Drishti \cite{Limaye:2012}, which performed the best with our huge datasets, compared to other available tools for 3D rendering. 

\subsubsection{Long term behavior}
The significance of our DCP tool is that not only can it reveal the short term transient dynamics and the underlying homoclinic, heteroclinic, saddle, and T-point spiral structures, but it can also be employed to reveal the long term behavior of the system (Fig.~\ref{fig:sigmaVariations_longTerm}). When looking at the long term behavior of a trajectory after omitting a long transient, one should be aware of the shift-symmetry of periodic orbits. Depending on the length of the initial transient omitted, the same symmetric 8-shaped periodic orbit (Fig.~\ref{fig:traces}b - blue) can be represented by either $\{\overline{01}\}$) or $\{\overline{10}\}$), and hence, needs to be normalized. In addition, if there is an asymmetric periodic orbit such as $\{\overline{011}\}$), the symmetry of the system implies the existence of another periodic orbit which is its symmetric counterpart, $\{\overline{100}\}$). This implies that all six of the periodic sequences $\{\overline{011}\}$), $\{\overline{110}\}$), $\{\overline{101}\}$), $\{\overline{100}\}$), $\{\overline{001}\}$) and $\{\overline{010}\}$) need to be normalized. We accomplish this by means of an algorithm we call Periodicity Correction (PC), where we identify the periodic structure within a sequence, along with its symmetric counterpart, and then normalize the sequence to the smallest valued circular permutation. Thus, both $\{\overline{01}\}$) and $\{\overline{10}\}$) are normalized to $\{\overline{01}\}$) and all six of the asymmetric periodic orbits described previously are normalized to the lowest valued sequence $\{\overline{001}\}$).
PC helps in identifying regions of simple, Morse-Smale dynamics of stable fixed points and periodic orbits by eliminating artifacts arising from symmetries of the system or the symbolic sequence. Structurally unstable, chaotic regions are marked by the lack of periodicity in their symbolic representations. Alternatively, we can also employ a compression algorithm, like Lempel-Ziv-76 (LZ) \cite{Lempel:1976}, to find the complexity of a binary sequence. LZ scans a sequence from left to right, and adds a new word to the vocabulary every time a previously unencountered sub-string is detected. Since the circular permutations of a periodic orbit will have identical complexity, this approach can also detect windows of stability amidst structurally unstable chaotic regions. 

While PC can detect stable periodic orbits efficiently even at short sequence lengths, LZ requires very long sequences and is not as effective. Within chaotic regions, on the other hand, PC cannot distinguish between different sequences while LZ separates very complex strings from not-so-complex strings, akin to Lyapunov exponents. To harness the best of both worlds, we therefore combine both PC and LZ. For a long term biparametric sweep, we first run the symbolic sequences through PC to detect the existence of, and to normalize, periodic orbits. The normalized sequences are then used to compute the power series sum $P$ and colored with the colormap as previously described. We then run the aperiodic sequences through LZ to detect their complexity. We color the LZ-complexity values in shades of gray, with darker gray representing greater complexity. Together, this results in long term bi-parametric sweeps such as Figs.~\ref{fig:sigma12_short_long},\ref{fig:sigmaVariations_longTerm},\ref{fig:sigma15_IF1_short_long}b,\ref{fig:sigma100_IF2_short_long}b,, where darker shades of gray represent greater structural instability and chaotic dynamics, while all other solid colors represent simple Morse-Smale dynamics of stable fixed points or periodic orbits. On the whole, we coin this collection of symbolic tools the `Deterministic Chaos Prospector (DCP)'. The open source software developed is available at \url{https://bitbucket.org/pusuluri_krishna/deterministicchaosprospector/}.

\section{Results}

Results are presented in the order of global transient dynamics, global long term dynamics, detailed magnifications near the organizing centers for both transient and long term behavior, followed by parametric saddles, isolas, and their 3D embedding.
 
\subsection{Transient dynamics and global bifurcation structure}

\subsubsection{Case: $\sigma=1.5$}

\begin{table}[ht!]
\centering 
\begin{tabular}{|c|c|c|ccl|l|}
\hline
$\sigma$ & a & b & $\lambda_{1}$ & $\lambda_{2}$& ~~~~$\lambda_{3}$ & $\nu$ \\
\hline
1.50 & 3.8264 & 0.5189 & 1.8954 & -.5189 &-.7594$\pm$ 7.6491i & 0.4 \\
   & 3.9821 & 0.4011 & 1.5603 & -.4011 & -.6614 & 0.42 \\
\hline
1.55 & 3.7616 & 0.5436 & 1.9390 & -.5436 & -.7718$\pm$7.5199i & 0.4 \\ 
   & 3.8950 & 0.4160 & 1.6461 & -.4160 & -.7080$\pm$7.7845i & 0.43 \\
\hline
1.60 & 3.7011 & 0.5663 & 1.9754 & -.5663 & -.7831$\pm$7.3990i & 0.4 \\ 
   & 3.8127 & 0.4284 & 1.7197 & -.4284 & -.7142$\pm$7.6200i & 0.36 \\
\hline
1.65 & 3.6443 & 0.5872 & 2.0059 & -.5872 & -.7936$\pm$7.2856i & 0.4 \\
   & 3.7345 & 0.4386 & 1.7833 & -.4386 & -.6552 & 0.36 \\
\hline
1.70 & 3.5907 & 0.6065 & 2.0313 & -.6065 & -.8033$\pm$7.1788i & 0.4\\
   & 3.6601 & 0.4470 & 1.8386 & -.4470 & -.6522 & 0.35 \\
\hline
1.75 & 3.5401 & 0.6246 & 2.0524 & -.6246 & -.8123$\pm$7.0778i & 0.4 \\
   & 3.5891 & 0.4536 & 1.8872 & -.4536 & -.6491 & 0.34 \\
\hline
1.80 & 3.4922 & 0.6414 & 2.0699 & -.6414 & -.8208$\pm$6.9820i & 0.4\\
   & 3.5210 & 0.4588 & 1.9298 & -.4588 & -.6460 & 0.33 \\
\hline
1.90 & 3.4029 & 0.6719 & 2.0959 & -.6719 & -.8337 & 0.39 \\ 
   & 3.3928 & 0.4653 & 2.0008 & -.4653 & -.6401 & 0.32 \\
\hline
2.00 & 3.5705 & 0.8120 & 1.8373 & -.7615 & -.8120 & 0.44\\
   & 3.3209 & 0.6988 & 2.1127 & -.6988 & -.8182 & 0.38 \\
   & 3.5705 & 0.8120 & 1.8373 & -.7615 & -.8120 & 0.44 \\
\hline
10.0 & 1.2609 & 0.4956 & 0.5175 & -.4884 & -.4956 & 0.96 \\
   & 1.1570 & 0.1936 & 0.5710 & -.1936 & -.4444 & 0.77\\
   & 1.1247 & 0.1941 & 0.6225 & -.1941 & -.4987 & 0.79\\
   & 1.0681 & 0.1896 & 0.6969 & -.1896 & -.5765 & 0.82\\
\hline
\end{tabular}
\caption{Table listing codim-2 inclination-flip ($IF$) points and the three leading eigenvalues $\lambda_3< \lambda_2<0<\lambda_1$ of the saddle $O$. }
\label{tab:IF}
\end{table}

We start this section with $\sigma=1.5$. Fig.~\ref{fig:sigma15} shows the bifurcation diagram in $(a,b)$-parametric plane at $\sigma=1.5$ using symbolic sequences $\{k_i{\}}_{i=5}^{12}$. Two sets of parameters that result in topologically identical dynamics for the symbolic sequence under consideration, are colored identically in the bi-parametric sweep. Boundaries between such regions represent a shift in the dynamics due to homoclinic bifurcations. The corresponding bifurcation diagram obtained using continuation is shown in Fig.~\ref{fig:sigma15Cont}. Blue curves represent homoclinic bifurcation curves while heteroclinic bifurcation curves are shown in red. 

A homoclinic bifurcation curve of the trivial equilibrium $O$ emerges from $BT$, which will be of primary interest as we vary $\sigma$. As we follow this primary curve we encounter a codim 2 point $DRS$ with two real stable eigenvalues and the leading stable direction changes. Here, a double, twisted homoclinic loop emerges. As we move along the primary homoclinic bifurcation curve $H_0$ from $DRS$ towards $IF_1$ in the parametric plane (Fig.~\ref{fig:sigma15Cont}), in the phase space (Fig.~\ref{fig:sigma15_tracesHomoclinic}a) the primary homoclinic loop is oriented and approaches $O$ from the right (light red curve). Moving on along $H_0$, we find a first inclination flip bifurcation $IF_1$ of complex type, where the primary homoclinic orbit approaches $O$ along its stable manifold (Fig.~\ref{fig:sigma15_tracesHomoclinic}a - dark red). As we move further along $H_0$, a twisted single loop approaches $O$ from the left (green), and $\nu_1$ and $\nu_2$ increase again when we find a second inclination flip point $IF_2$. Beyond $IF_2$, the homoclinic orbit changes its shape from a single loop around one non-trivial equilibrium to a double loop configuration around both non-trivial equilibria, that gradually grows bigger (light blue $\rightarrow$ dark blue). Finally the primary curve gets close to the DRS point where the continuation terminates. This corroborates theoretical expectations although verification that the exact end point of this primary homoclinic curve is at the DRS point, is still open. 

From the eigenvalues (see Table \ref{tab:IF}), we find that the inclination flip at $IF_1$ and other cases are all of complex type so that infinitely many homoclinic bifurcation curves emerge from this point $IF_1$. The homoclinic bifurcation curves all spiral to the T-points such as $T_0$ (marked by the heteroclinic connection $\{1\overline{0}\}$). From the T-point $T_0$, two heteroclinic bifurcation curves also emerge connecting the saddle-foci $P^{\pm}$. These curves are rather straight and it may be appreciated that the homoclinic spirals to secondary T-points end up close to these heteroclinic curves. The inclination flip points, T-points, saddles($S$) and closed loops are further explored in Sections \ref{section_T_IF},\ref{section_Saddle_CL}.

\begin{figure*}[hb!]
\centering
\includegraphics[width=0.47\textwidth]{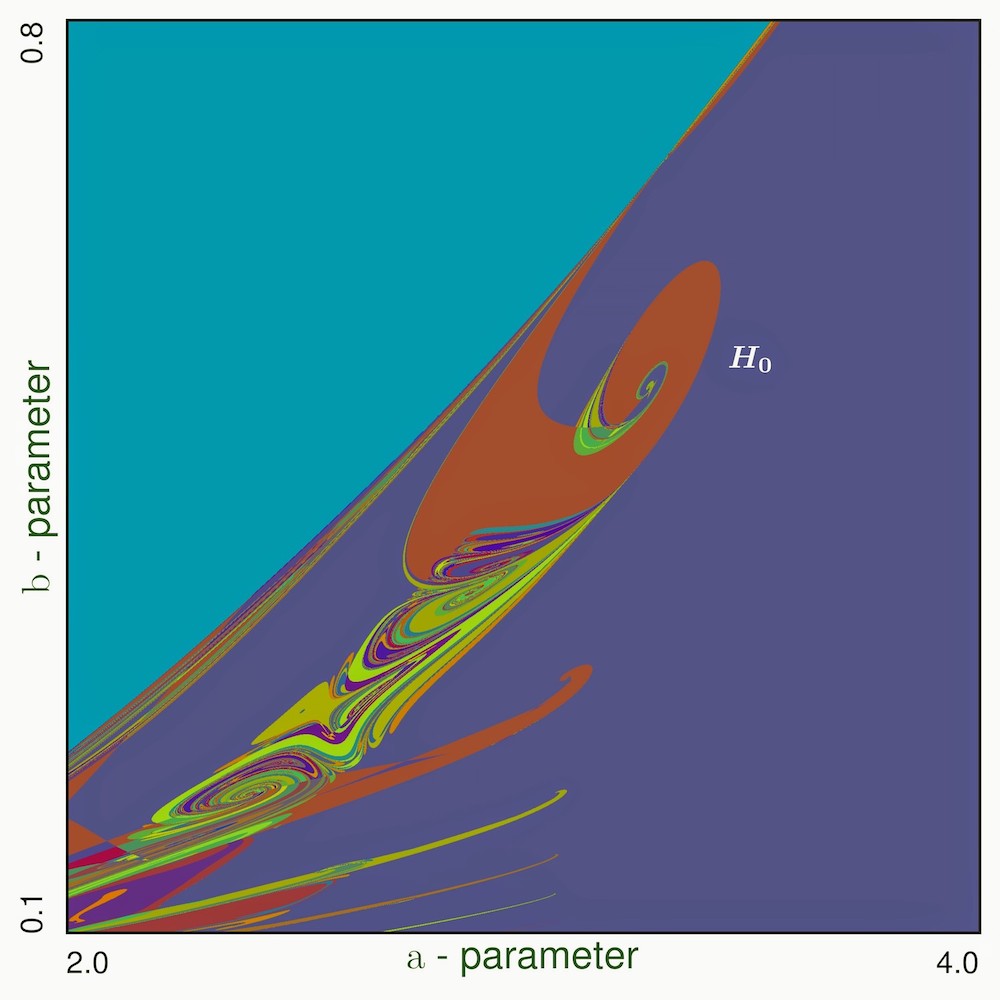}
\includegraphics[width=0.48\textwidth, height=0.46\textwidth]{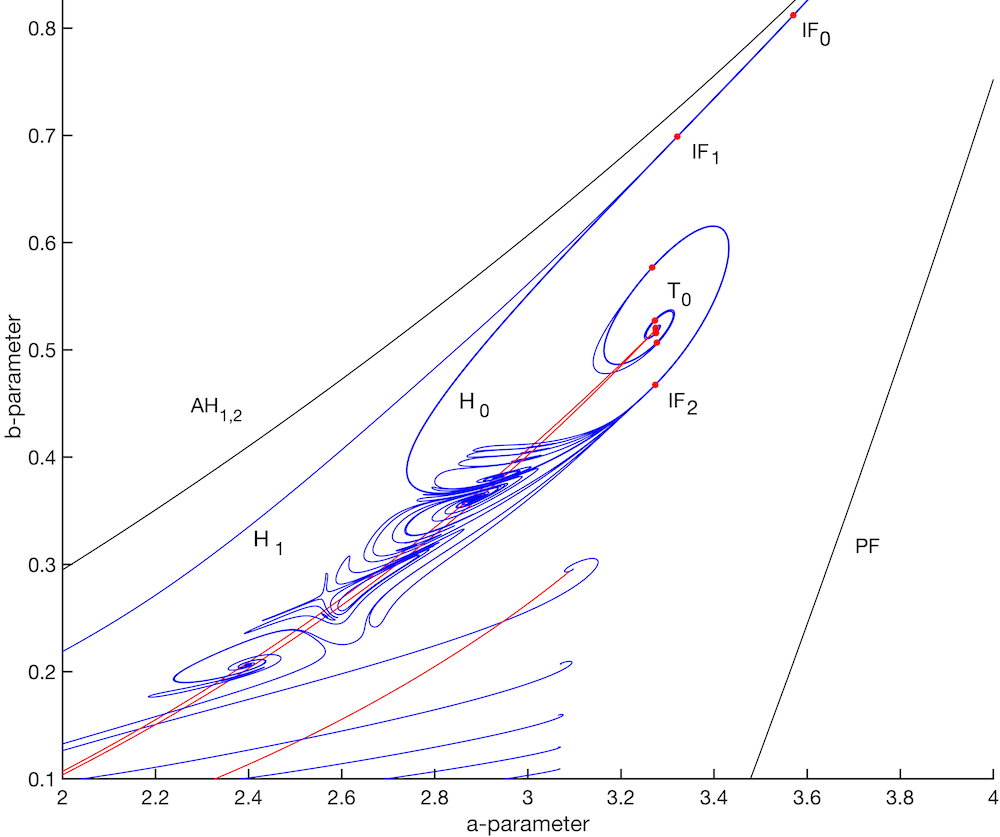}
\caption{{\bf (a)} The $(a,b)$-parametric sweep of the OPL-model at $\sigma=2.0$ using short symbolic sequences $\{k_i{\}}_{i=3}^{10}$ to detect principal homoclinic and heteroclinic bifurcations of the saddle $O$. {\bf (b)} MatCont reconstruction depicting the key bifurcations of equilibrium states, multiple cod-2 points (with their labels as previously described) on the primary homoclinic bifurcation curve $H_0$ spiraling onto the T-point $T_0$, rather then going back to the cod-2 $NS$-point as before in Fig.~4. Shown in blue are the curves corresponding to various homoclinic and heteroclinic connections of the saddle. Note that secondary T-points are located within the wedge bounded by two red lines corresponding to homo- and heteroclinic bifurcations of the saddle-foci.}\label{fig:sigma20}
\end{figure*}

\subsubsection{Case: $\sigma=2.0$}

The $(a,b)$-parametric sweep for $\sigma=2.0$ using symbolic sequences $\{k_i{\}}_{i=3}^{10}$ and continuation is shown in Fig.~\ref{fig:sigma20}. We find results are globally quite similar for the pitchfork and AH curves. The primary homoclinic bifurcation curve $H_0$ has some marked differences though. First, the codimension-two neutral saddle point has disappeared simultaneously altering the nature of the BT point. Second, the primary homoclinic bifurcation curve no longer ends at the DRS point but spirals towards the T-point. In this case, as we sample parameters beyond $IF_2$, in the phase space the twisted loop starts making more and more revolutions around the stable manifold of $P^-$, until it forms a heteroclinic connection to $P^-$ at the T-point. As shown in Fig.~\ref{fig:sigma15_tracesHomoclinic}b, this transfiguration of the primary homoclinic orbit into a heteroclinic connection to $P^-$ happens through the orbits light red $\rightarrow$ dark red $\rightarrow$ green $\rightarrow$ light blue as we move along $H_0$. Detailed transition of the behavior of $H_0$ between $\sigma=1.5$ and $\sigma=2.0$ is explored further in Section.\ref{section_Saddle_CL}. We also note that the second leading stable eigenvalue has changed from complex at $\sigma=1.5$ to real for $\sigma=2.0$. This transition occurs at $\sigma\approx 1.90$.

\subsubsection{Case: $\sigma=10.0$}

\begin{figure*}[t!]
\centering

\includegraphics[width=0.55\textwidth]{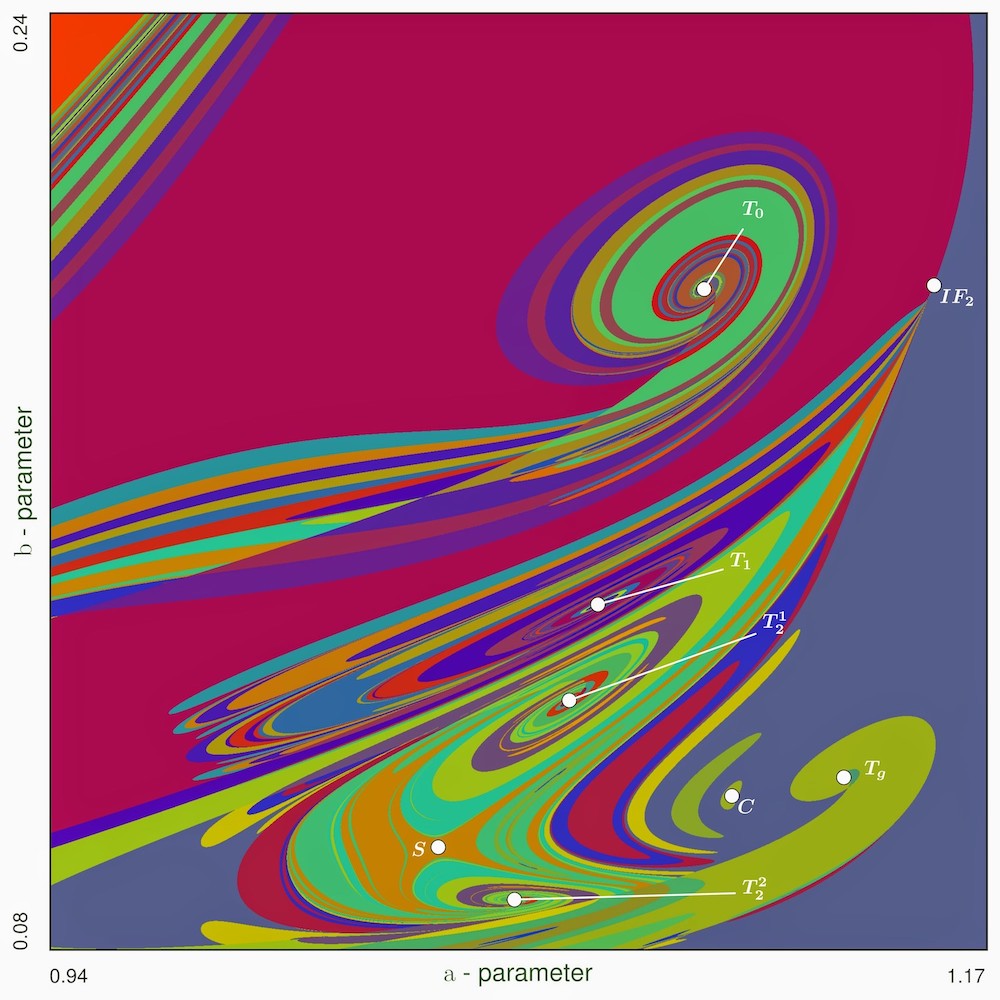}
\caption{The $(a,b)$-parameter sweep using symbolic sequences $\{k_i{\}}_{i=2}^{8}$ for $\sigma=10$. White circles in it denote selected cod-2 bifurcation points including Inclination Flip -- $IF_2$ (see Fig.~\ref{fig:sigma100_IF2_short_long}); T-points -- $T_0$ ($\{1\overline{0}\}$), $T_1$ ($\{10\overline{1}\}$), $S$ -- the bridging saddle between a pair of symmetric  T-points $T_2^1$ and $T_2^2$ below $S$ with the  identical symbolic coding $\{11\overline{0}\}$ (see Figs.\ref{fig:saddleMerge} and \ref{fig:saddle3D}). Here, $T_g$ marks the ghost of the T-point ($\{\overline{1}\}$) after its spiral structure is devoured by the periodic orbit ($\{\overline{01}\}$); $T_g$ and the semi-annular isolas near $C$ are detailed in Fig.~\ref{fig:ghostTpoint} below.}\label{fig:sigma100}
\bigskip
\includegraphics[width=0.45\textwidth]{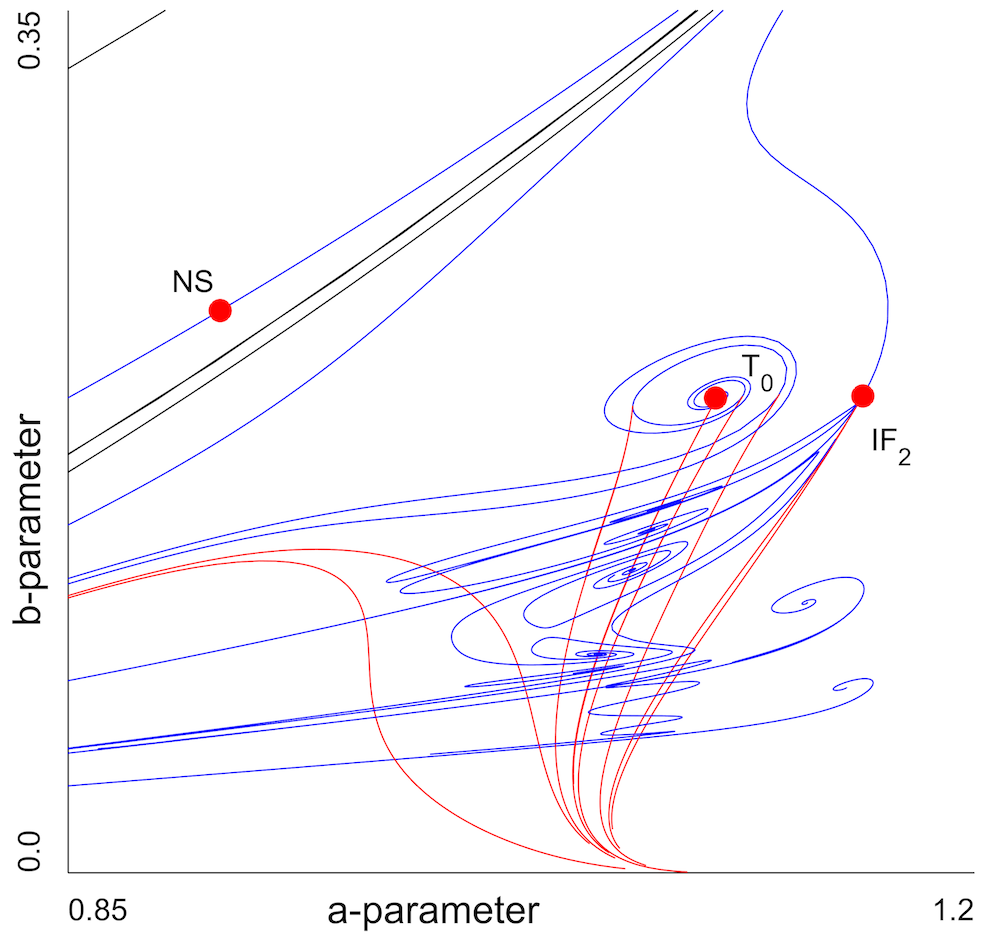}
\caption{Matching fragment of the $(a,b)$-bifurcation diagram done with the MatCont parameter continuation for $\sigma=10$. Blue and red curves correspond to the key homoclinic and heteroclinic bifurcations of the saddle and saddle-foci, respectively. Here, $IF_2$ stands for the Inclination Flip, and $NS$ for Neutral Saddle with zero saddle value.}\label{fig:sigma100Cont}
\end{figure*}
 
Moving on to $\sigma=10.0$, the bifurcation diagram using symbolic sequences $\{k_i{\}}_{i=2}^{8}$ is shown in Fig.~\ref{fig:sigma100}. The corresponding diagram using parametric continuation is shown in Fig.~\ref{fig:sigma100Cont}. The inclination flip $IF_2$ on the primary homoclinic is of a different type (type B), see Table \ref{tab:IF} for the eigenvalues. This involves small regions with stable limit cycles. Indeed, we find a limit point of cycles bifurcation (LPC) indicating a small region with stable periodic orbits. These LPCs are only connected to the primary homoclinic bifurcation. This may not be apparent from Figure \ref{fig:sigma100Cont} as the primary curve does not seem to make a loop towards the T-point. Detailed calculations though suggest the primary curve makes a turn near $\nu_1=0.3$, but accurate continuation did not succeed in that parameter region. All other inclination flip points are of type C. This implies that only the primary homoclinic bifurcation results in the emergence of additional stable periodic orbits nearby. The label $NS$ stands for the neutral homoclinic saddle with a zero saddle value. Here, $T_0$ ($\{1\overline{0}\}$), $T_1$ ($\{10\overline{1}\}$), $T_2^1$ ($\{11\overline{0}\}$) and $T_2^2$ ($\{11\overline{0}\}$) represent various T-points. Note that the T-points $T_2^1$ and $T_2^2$, above and below the saddle $S$, have identical construction and the same heteroclinic connection $\{11\overline{0}\}$. $T_g$ with the heteroclinic connection ($\{\overline{1}\}$) is a special case. Although $T_g$ resembles a T-point, it falls in the region of the periodic orbit ($\{\overline{01}\}$) which devours the spiral structure of $T_g$, leaving behind a ghost of the T-point. $T_g$ and the semi-annular isolas near $C$ are detailed in Fig.~\ref{fig:ghostTpoint}.
Further details of the special organizing structures are provided in the following sections. 

\subsubsection{Summary: $\sigma$ variations}

\begin{figure*}[t!]
\centering
\includegraphics[width=.99\textwidth]{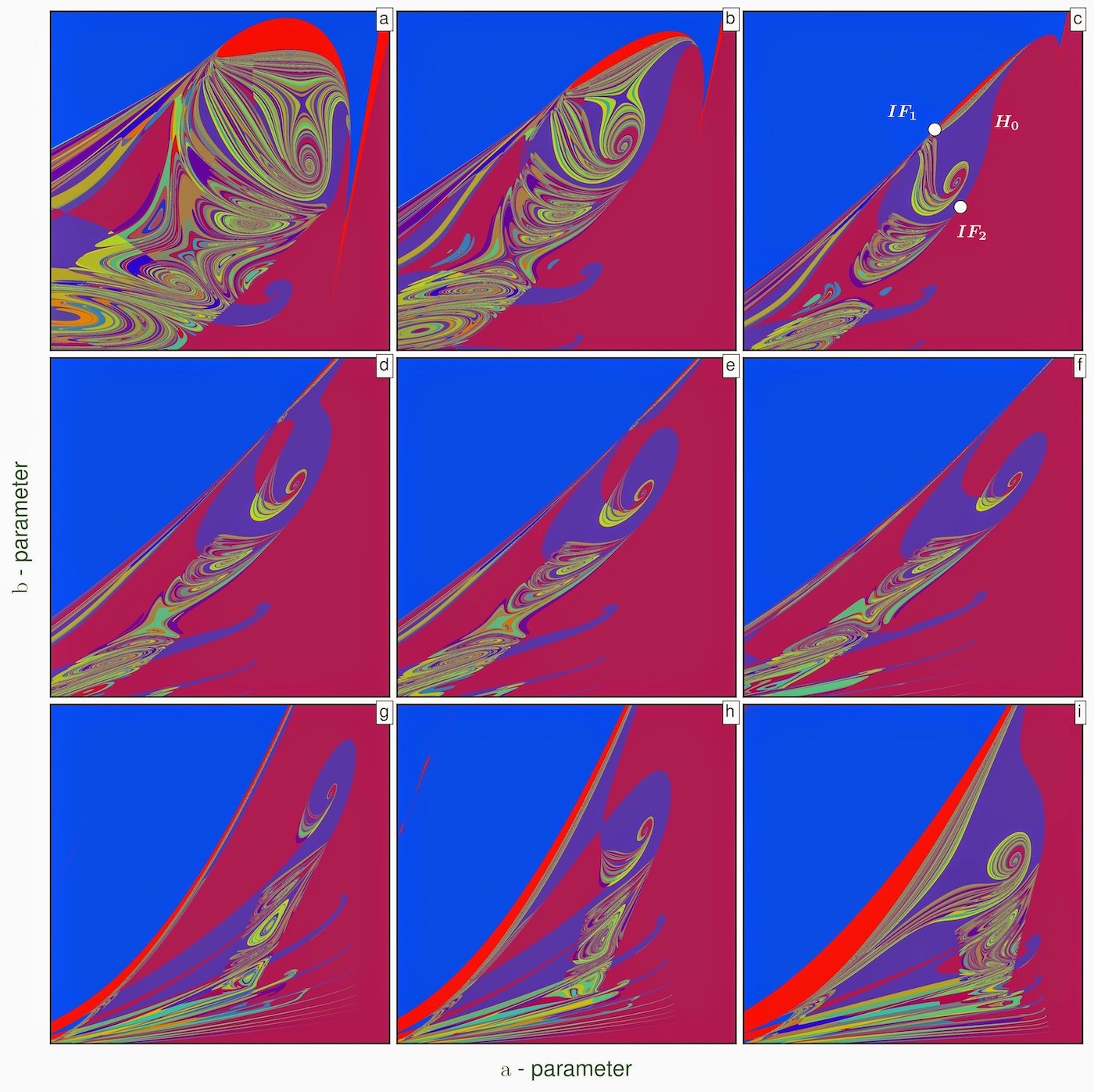}
\caption{Metamorphoses of global bifurcation structures in the the (a,b)-parameter plane as $\sigma$ is increased from  $1.2$ \textbf{(a)},  $1.32$ \textbf{(b)},  $1.5$ \textbf{(c)},  $1.72$ \textbf{(d)},  $1.76$ \textbf{(e)}, \ $2.0$ textbf{(f)},  $6.0$ \textbf{(g)},  $7.375$ \textbf{(h)} through \ $10.0$ textbf{(i)}, using $\{k_i{\}}_{i=4}^{11}$. Note that between (d) and (e), the primary  homoclinic bifurcation curve $H_0$ (which is the borderline between the red and purple solid regions) switches to spiraling onto $T_0$, while between (g) and (h), the reverse phenomena occur so that $H_0$ no longer ends up at $T_0$. (see also supplementary Movie~M1)
}\label{fig:sigmaVariations_shortTerm}
\end{figure*}

We conclude this section by presenting a summary of the bifurcation diagrams for varying $\sigma$-values. This also serves as a precursor for the construction of 3D parametric sweeps of Sec.\ref{section_Saddle_CL}. Fig.~\ref{fig:sigmaVariations_shortTerm}(a-h) reveal variations in the bifurcation structure using $\{k_i{\}}_{i=4}^{11}$ for $\sigma$-values $1.2$, $1.32$, $1.5$, $1.72$, $1.76$, $2.0$, $6.0$, $7.375$ and $10.0$. See also supplementary Movie~M1 that reveals these variations in further detail. Between $\sigma=1.72$ (Fig.~\ref{fig:sigmaVariations_shortTerm}d) and $\sigma=1.76$ (Fig.~\ref{fig:sigmaVariations_shortTerm}e), the primary homoclinic bifurcation curve branches and starts spiraling on to the primary T-point $T_0$. As we increase $\sigma$ further, the inclination flip $IF_1$ and the homoclinic bifurcation curves from $IF_1$ to $T_0$ change their orientation. Between $\sigma=6$ (Fig.~\ref{fig:sigmaVariations_shortTerm}g) and $\sigma=7.375$ (Fig.~\ref{fig:sigmaVariations_shortTerm}h), $H_0$ branches again to separate from $T_0$. 

\subsection{Global long term dynamics}

\begin{figure}[ht!]
\centering
\includegraphics[width=0.75\textwidth]{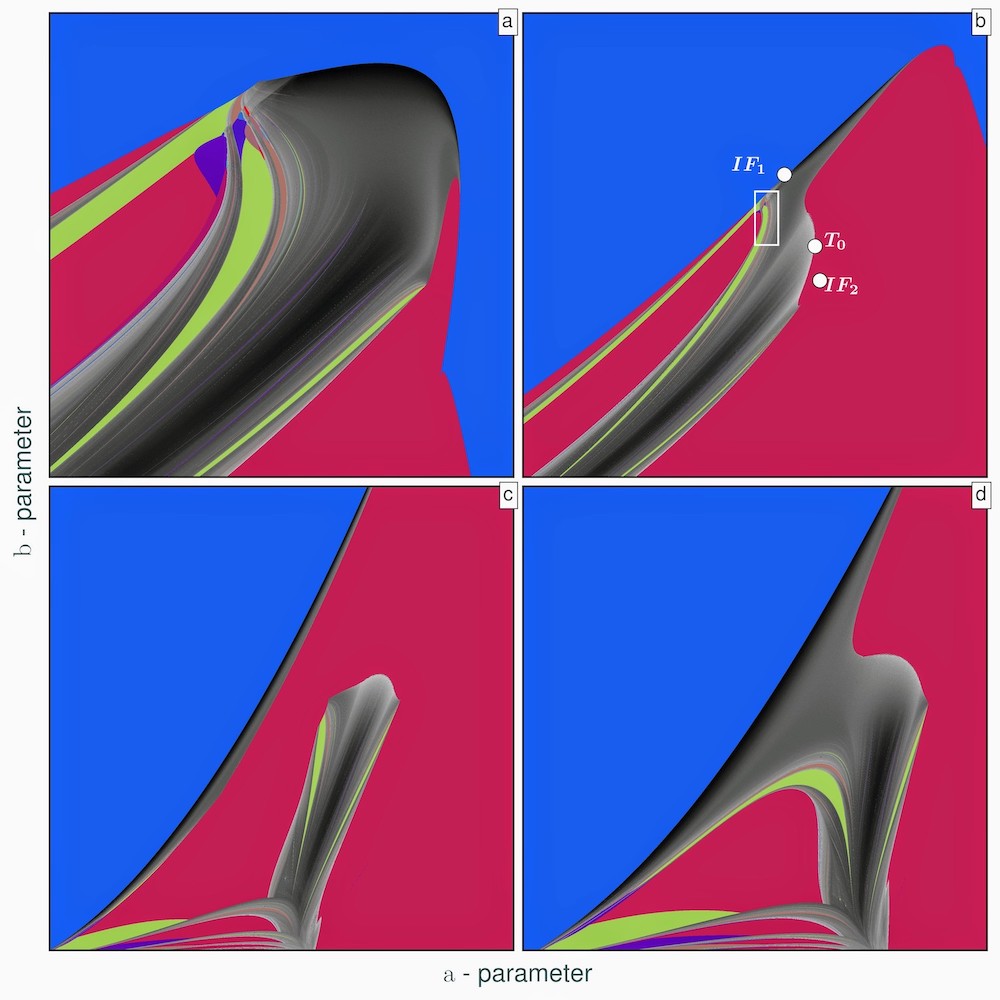}
\caption{Long term behavior of the OPL-model at different $\sigma$-values, revealed using the DCP-algorithm with $\{k_i{\}}_{i=999}^{1999}$. Solid colors represent regions of simple dynamics with stable equilibria or periodic orbits, computed using the periodicity correction (PC). Gray-sh noisy regions represent regions of structurally unstable, chaotic dynamics: the darker areas correspond to the greater LZ-complexity values for the corresponding binary sequences. Primary T-point $T_0$ and inclination flip points $IF_1$ and $IF_2$ for $\sigma=1.5$ are marked with white dots in (b) for reference, along with the white box enclosing a vicinity  of $IF_1$ which is expanded in Fig.~\ref{fig:sigma15_IF1_short_long}b. Observe the narrow Shilnikov flames emerging below secondary $IF$-points (located on homoclinic bifurcation spirals such as ones shown in Fig.~\ref{fig:sigma12_short_long}) in the chaos-land. Parameters: \textbf{(a)} $\sigma=1.2$; \textbf{(b)} $\sigma=1.5$; \textbf{(c)} $\sigma=7.375$ and \textbf{(d)} $\sigma=10$.}
\label{fig:sigmaVariations_longTerm} 
\end{figure}

To study the long term behavior of the system with DCP, we use long symbolic sequences $\{k_i{\}}_{i=999}^{1999}$, omitting the first 999 symbols as transients. Fig.~\ref{fig:sigmaVariations_longTerm} shows such long term behavior at $\sigma$ values: $1.2$, $1.5$, $7.375$ and $10$ in Panels~\textbf{(a)}-\textbf{(d)}, respectively. Simple, Morse-Smale dynamics of stable equilibria or periodic orbits are identified using Periodicity Correction and shown in solid colors. Distinct colors represent sequences with distinct periods. Structurally unstable, chaotic dynamics are identified by their lack of periodic structure, processed using LZ-complexity, and colored such that darker gray regions indicate greater LZ-complexity of the sequences, and therefore, increased structural instability and chaos. Between Fig.~\ref{fig:sigmaVariations_longTerm}(a,b) vs. Fig.~\ref{fig:sigmaVariations_longTerm}(c,)d note the change in orientation of the inclination flip $IF_1$ and the stability windows emerging from it. This is also in agreement with the change in orientation of the transient dynamics as seen in Fig.~\ref{fig:sigmaVariations_shortTerm} for the corresponding $\sigma$-values.

\subsection{Transient and long term behavior near the organizing centers}\label{section_T_IF}

In this section, we will present the detailed behavior of both short and long term dynamics near the organizing centers of T-points and inclination flips. 

\subsubsection{T-points}
For all three sigma values of $1.5$ (Fig.~\ref{fig:sigma15}), $2.0$ (Fig.~\ref{fig:sigma20}) and $10.0$ (Fig.~\ref{fig:sigma100}), the primary T-point $T_0$ is marked by the symbolic sequence $\{1\overline{0}\}$. The phase trajectory at $T_0$ is as shown by the light blue curve of Fig.~\ref{fig:sigma15_tracesHomoclinic}b, where $\Gamma_1$ makes one loop around $P^+$, followed by a heteroclinic connection to $P^-$. The trajectory effectively makes an infinite number of loops around $P^-$, before leaving the neighborhood of $P^{-}$. Fig.~\ref{fig:sigma15_IFs_T1T2}a shows a magnification near $T_0$ for $\sigma=1.5$ using $\{k_i{\}}_{i=5}^{12}$. It also shows multiple secondary T-points surrounding the primary T-point. Fig.~\ref{fig:sigma15_IFs_T1T2}b with $\{k_i{\}}_{i=3}^{10}$ shows a zoom of a small region with an analogous pair of T-points $T_1$ and $T_2$, above and below $H_0$, respectively, for $\sigma=1.5$ (see Fig.~\ref{fig:sigma15} white box). The heteroclinic connection at $T_1$ is shown in Fig.~\ref{fig:traces}d. Here $\Gamma_1$ makes one loop around $P^+$, followed by one loop around $P^-$, and then forms a heteroclinic connection to $P^+$. Hence, the symbolic representation of $\{10\overline{1}\}$. Similarly, the heteroclinic connection at $T_2$ is given by $\{11\overline{0}\}$. Such T-points always come in pairs on either side of a homoclinic bifurcation curve. Such a T-point pair can also be seen for $\sigma=10.0$ (Fig.~\ref{fig:sigma100}), with $T_1$ ($\{10\overline{1}\}$) above $H_0$, and $T_2^1$ ($\{11\overline{0}\}$ )below it. The T-point $T_2^2$ below the saddle $S$, also has the same structure in the parametric plane and the same heteroclinic connection ($\{11\overline{0}\}$) as $T_2^1$. This is explored further in Sec.\ref{section_Saddle_CL}. 

\subsubsection{Inclination flips}

\begin{figure}[ht!]
\centering
\includegraphics[width=0.35\textwidth]{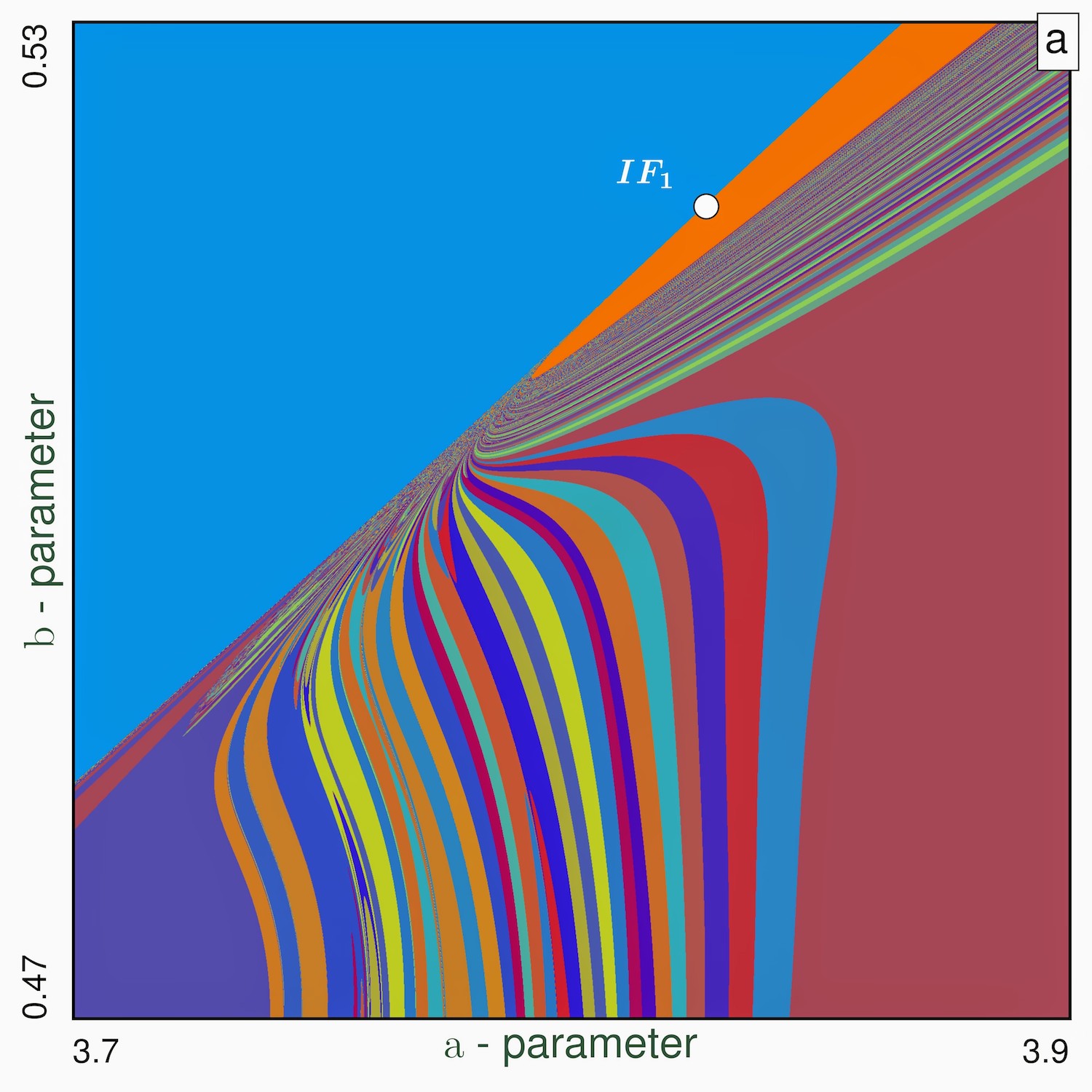}
\includegraphics[width=0.35\textwidth]{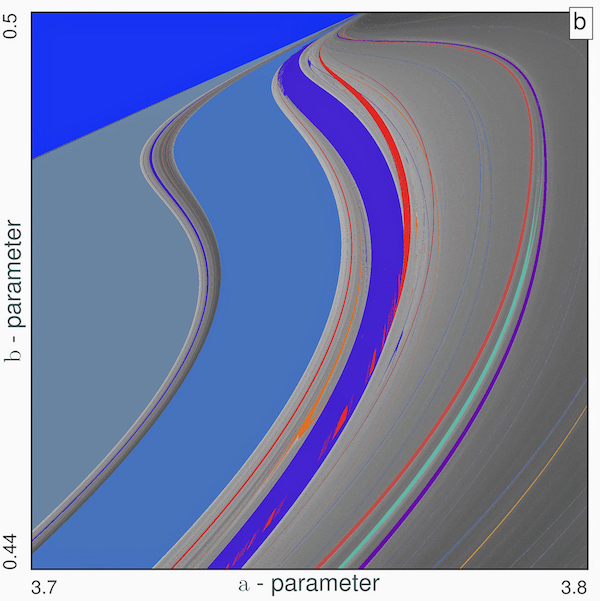}
\caption{Short-transient and long-term dynamics near the inclination flip point for $\sigma=1.5$. \textbf{(a)} $\{k_i{\}}_{i=5}^{12}$ sweep detects a locus of homoclinic bifurcation curves originating from $IF_1$ (compare Fig.~\ref{fig:sigma15}) \textbf{(b)} $\{k_i{\}}_{i=999}^{1999}$ sweep with DCP aimed for long-term behaviors reveals multiple stability windows (of solid colors) quickly  expanding away from $IF_1$ (not shown here, see near white box in Fig.~\ref{fig:sigmaVariations_longTerm}b).  Solid color regions correspond to the existence regions of stable periodic orbits detected by periodicity correction (PC) algorithm, while increasingly dark gray regions represent greater structural instability and chaos, obtained using LZ-complexity.} \label{fig:sigma15_IF1_short_long}
 \end{figure}

\begin{figure}[hb!]
\centering 
\includegraphics[width=0.35\textwidth]{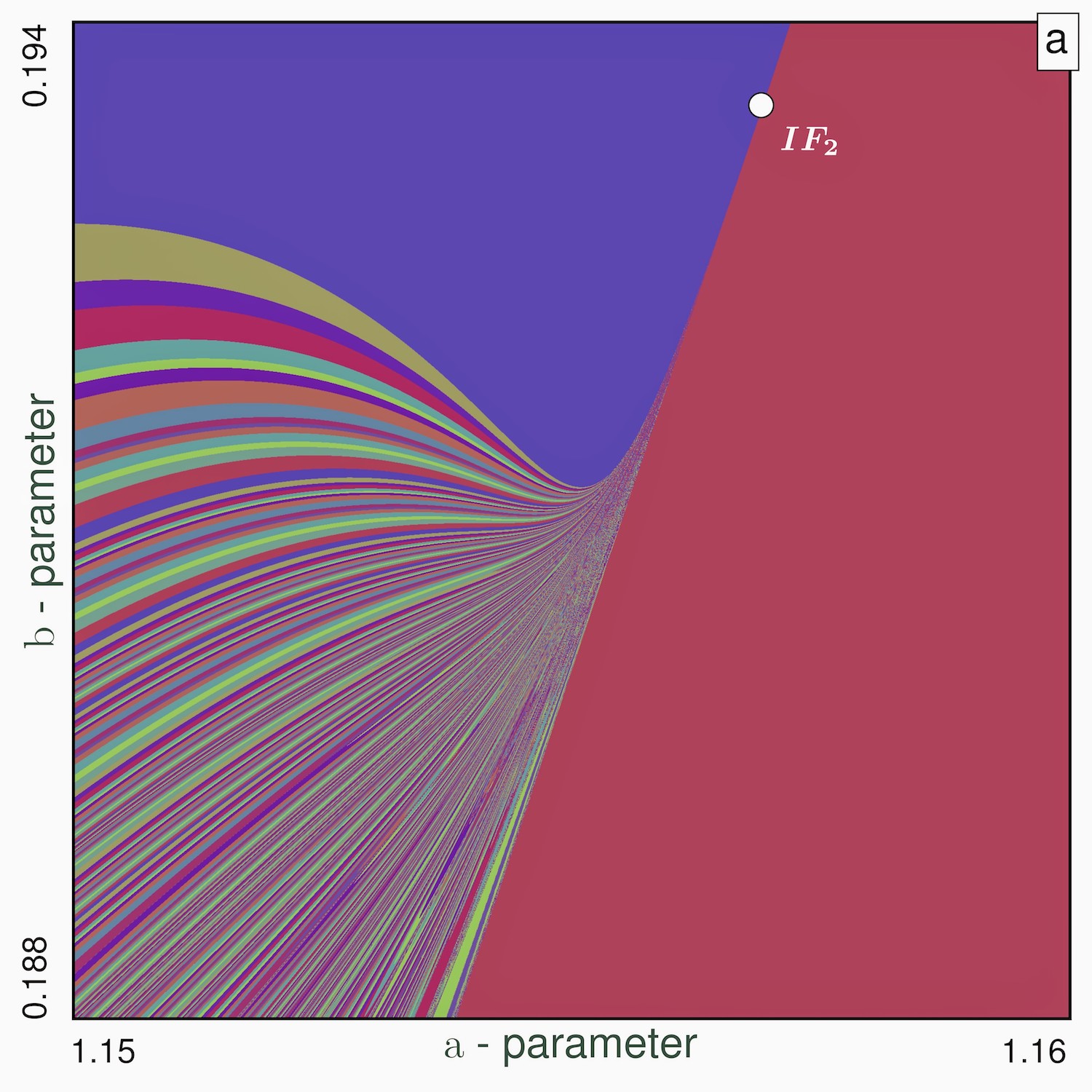}
\includegraphics[width=0.35\textwidth]{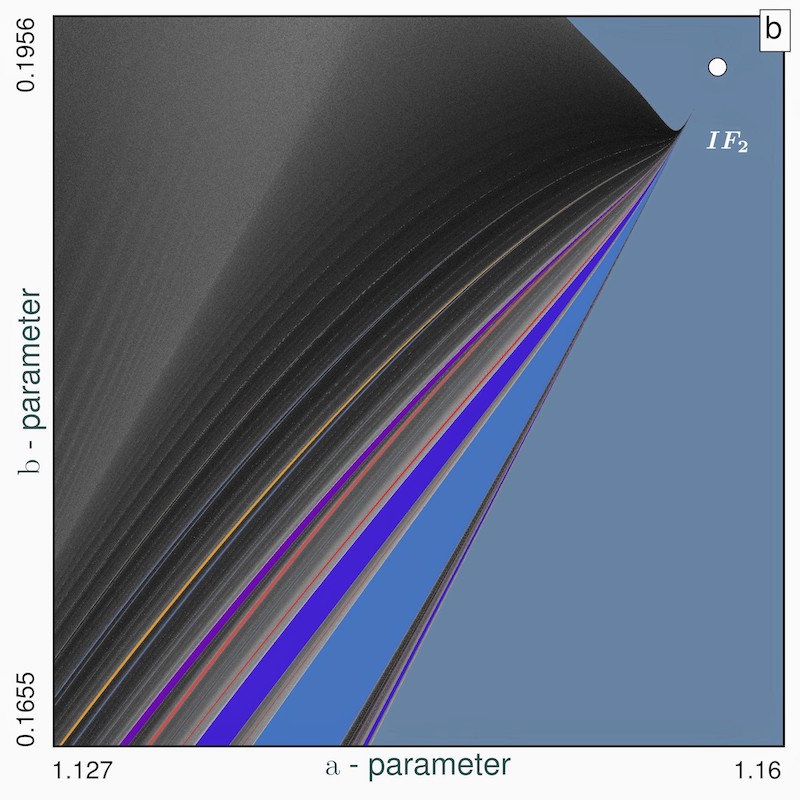}
\caption{Transient and long-term dynamics near the inclination-flip point $IF_2$ for $\sigma=10$. \textbf{(a)} $\{k_i{\}}_{i=16}^{23}$ sweep discloses an abundance of homoclinic bifurcation curves originating from $IF_2$.  \textbf{(b)} $\{k_i{\}}_{i=999}^{1999}$ DPC-sweep for long term behavior shows multiple stability windows converging towards $IF_2$; solid colors corresponds to stability windows of periodic orbits (PC) while gray areas correspond   structural unstable, chaotic dynamics generating poorly LZ-compressable binary sequences.}\label{fig:sigma100_IF2_short_long}
\end{figure}

Next, we show how the inclination flip points serve not only as the confluence of infinitely many homoclinic bifurcation curves, but also as the originators of multiple windows of stability with respect to long term dynamics in the system. Fig.~\ref{fig:sigma15_IFs_T1T2}a ($\sigma=1.5$) shows several inclination flips (white dots), starting from $IF_2$ and leading all the way beyond $T_0$, from which the homoclinic bifurcation curves arise leading up to the secondary T-points surrounding $T_0$. The specific values of IF bifurcation points are obtained with continuation. Fig.~\ref{fig:sigma15_IF1_short_long} shows the behavior near the inclination flip point $IF_1$ at $\sigma=1.5$ for both transient (compare Fig.~\ref{fig:sigma15}) and long term dynamics (compare Fig.~\ref{fig:sigmaVariations_longTerm}b). Fig.~\ref{fig:sigma15_IF1_short_long}a ($\{k_i{\}}_{i=5}^{12}$) reveals multiple homoclinic bifurcation curves that converge to $IF_1$, while Fig.~\ref{fig:sigma15_IF1_short_long}b ($\{k_i{\}}_{i=999}^{1999}$ with DCP) shows the convergence of multiple stability windows (solid colors) towards $IF_1$, interspersed within regions of structural instability and chaos (darker gray implies greater LZ-complexity and less compressible binary sequences thus generated from the model equations). Similar transient and long term behavior near $IF_2$ for $\sigma=10.0$ (compare Fig.~\ref{fig:sigma100},\ref{fig:sigmaVariations_longTerm}d) is summarized in Fig.~\ref{fig:sigma100_IF2_short_long}a ($\{k_i{\}}_{i=16}^{23}$) and Fig.~\ref{fig:sigma100_IF2_short_long}b ($\{k_i{\}}_{i=999}^{1999}$ with DCP), respectively.

\subsubsection{Summary at $\sigma=1.2$}
Finally, we magnify the behavior at $\sigma=1.2$ which nicely summarizes the behavior of the inclination flips and secondary T-points for short term (Fig.~\ref{fig:sigma12_short_long}a, $\{k_i{\}}_{i=2}^{9}$) and long term dynamics (Fig.~\ref{fig:sigma12_short_long}b, $\{k_i{\}}_{i=1000}^{1999}$ with DCP). Several inclination flips (white dots) close to the primary T-point $T_0$ are clearly seen as the congregation centers of homoclinic curves leading up to secondary T-points vis-à-vis transient dynamics, while those same inclination flip points also give rise to stability windows in the long term, amidst regions of structurally unstable and chaotic dynamics. See Fig.~\ref{fig:sigmaVariations_shortTerm}a,\ref{fig:sigmaVariations_longTerm}a for the corresponding global dynamics.

\subsection{Parametric saddles and isolas }\label{section_Saddle_CL}

In this last section, we present a discussion of more complicated organizing structures in the parameter space. Specifically, we describe the 3D structures of two types of parametric saddles and two types of isolas (isolated closed curves). Such structures have been theoretically described previously \cite{Shilnikov:1993, Wieczorek:2005, Algaba:2011,Algaba:2015}. To our knowledge, this is the first detailed three dimensional computational reconstruction of such bifurcation surfaces in parameter space, made possible by the fast parallel computation of trajectories on GPUs.

\subsubsection{Branching saddle for homoclinic bifurcation curves}

\begin{figure}[ht!]
\centering 
\includegraphics[width=0.35\textwidth]{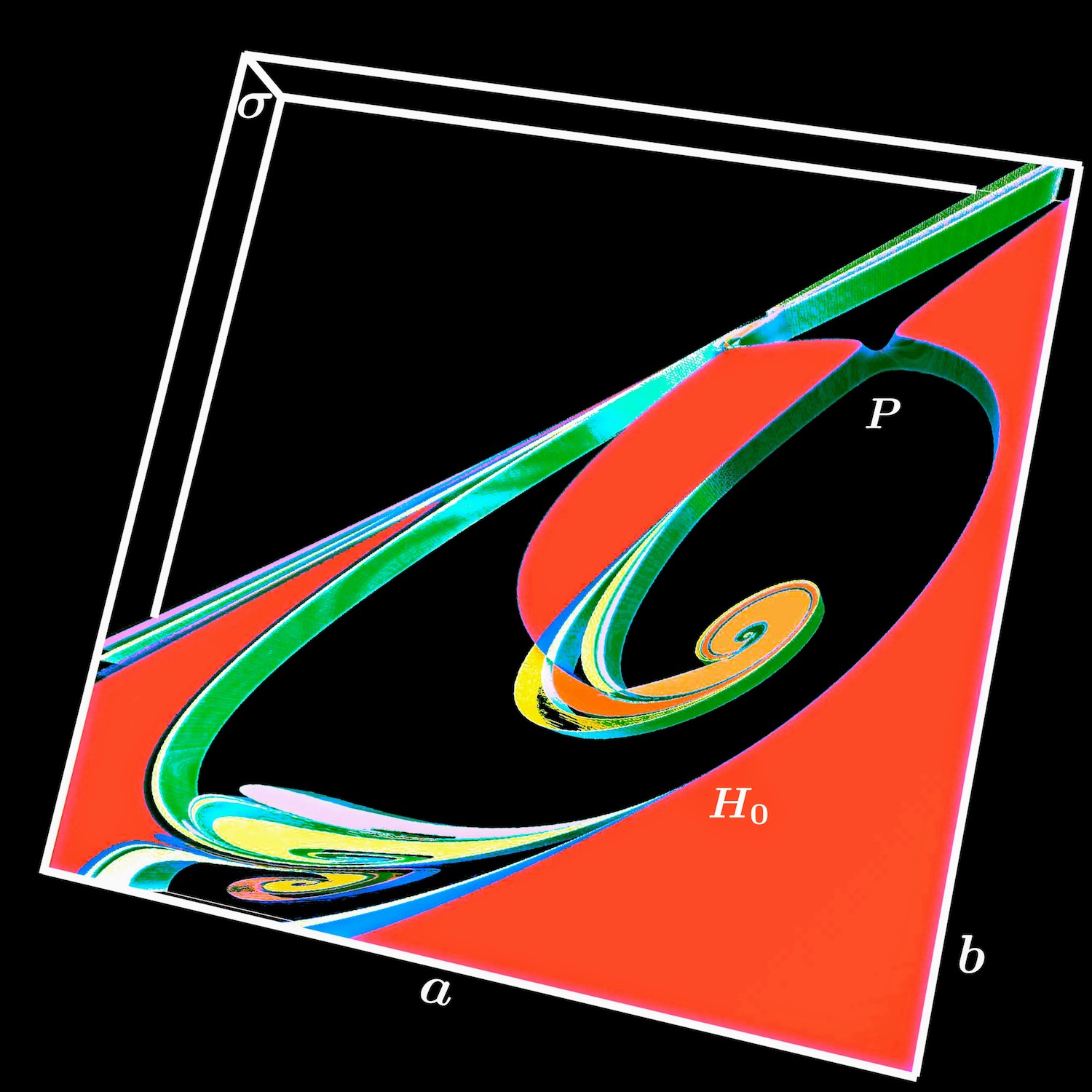}
\includegraphics[width=0.35\textwidth, height=0.4\textwidth]{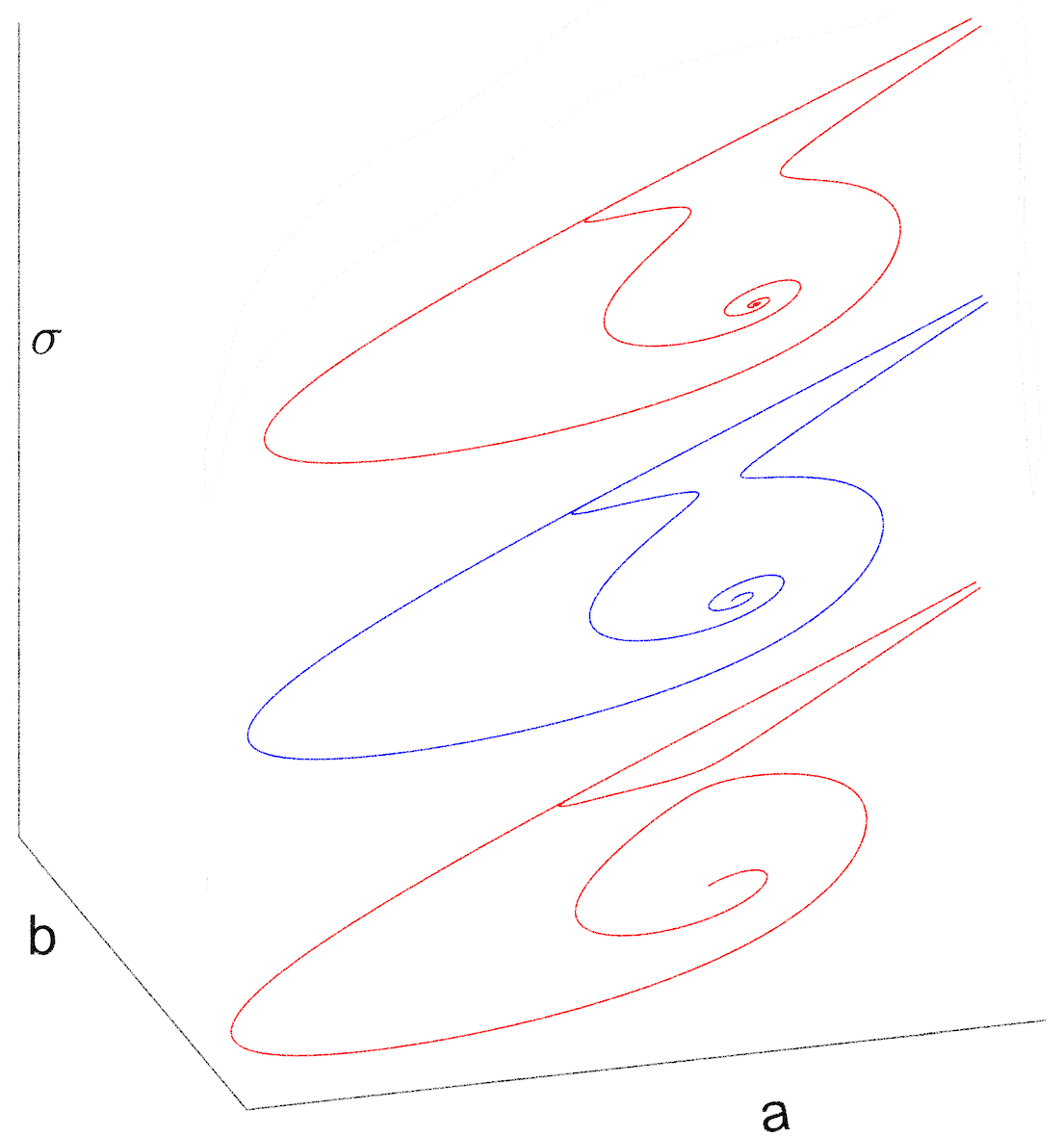}
\caption{Bifurcation diagram in $(a,b,\sigma)$-parameter space showing the transformation of the primary homoclinic bifurcation curve $H_0$, when it starts to spiral towards the primary T-point $T_0$ (see Fig.~\ref{fig:sigma20}) instead of going to $DRS$ (see Fig.~\ref{fig:sigma15}). This 3D reconstruction  (with $\sigma$ being on the vertical axes) is made of 100 sweeps with $\{k_i{\}}_{i=0}^{7}$ in the range $1.7418$ (top)$\leq\sigma\leq1.7439$ (bottom). The $P$-points marks the location of the branching saddle near $\sigma\approx1.7428$. MatCont-made bifurcation curves are shown on the right for three $\sigma$-values: $1.73$ (top), $1.74$ (middle) and $1.75$ (bottom).}\label{fig:sigma15_sigma20_transition3D}
\end{figure}

\begin{figure}[t!]
\centering 
\includegraphics[width=0.3\textwidth]{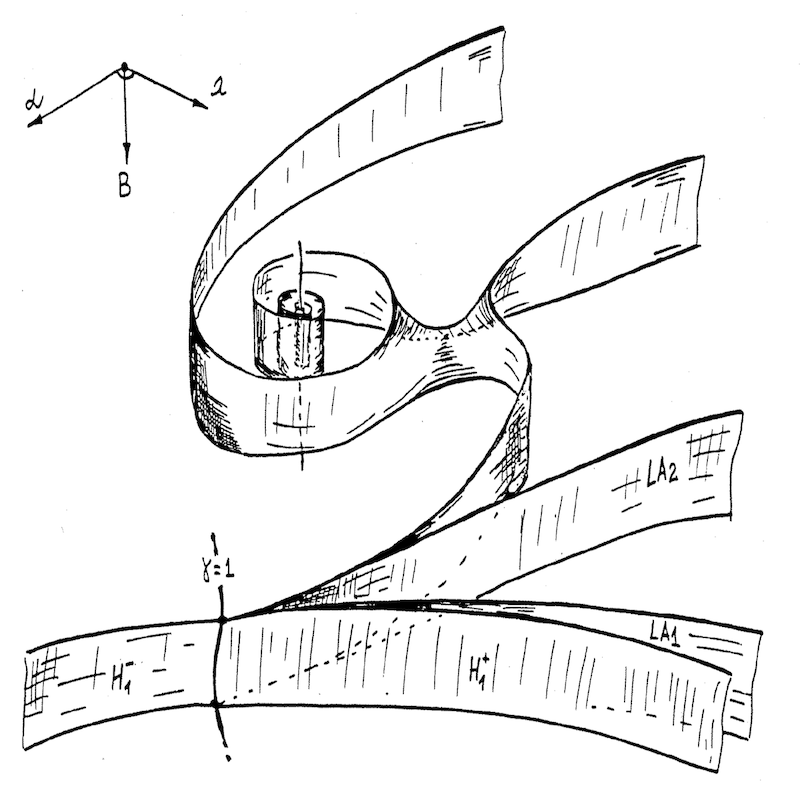}
\caption{Sketch of a bifurcation surface featuring a saddle causing the homoclinic bifurcation branching in the 3D parameter space of the Shimizu-Morioka system (courtesy A.L. Shilnikov et. al., 1993 \cite{ASHIL93}) }\label{fig:sigma15_sigma20_transition3D_old}

\bigskip

\includegraphics[width=0.3\textwidth]{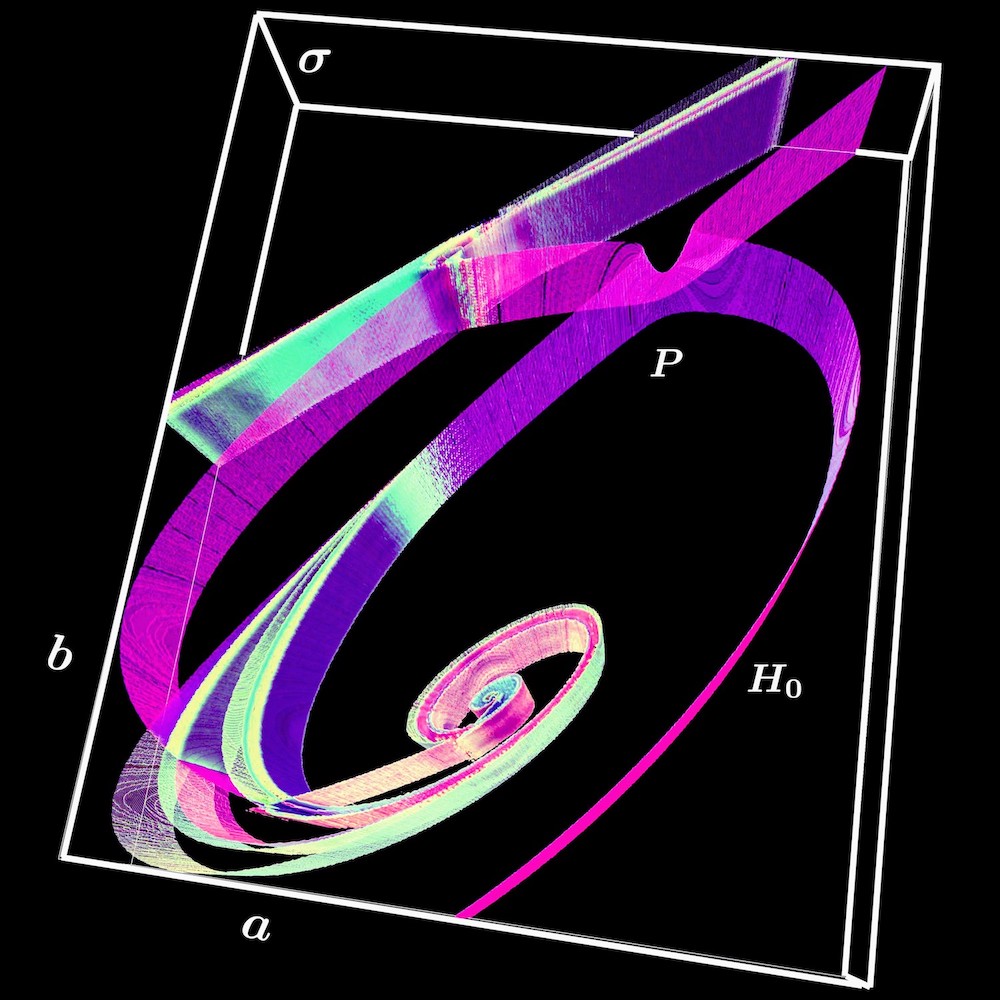}
\includegraphics[width=0.3\textwidth]{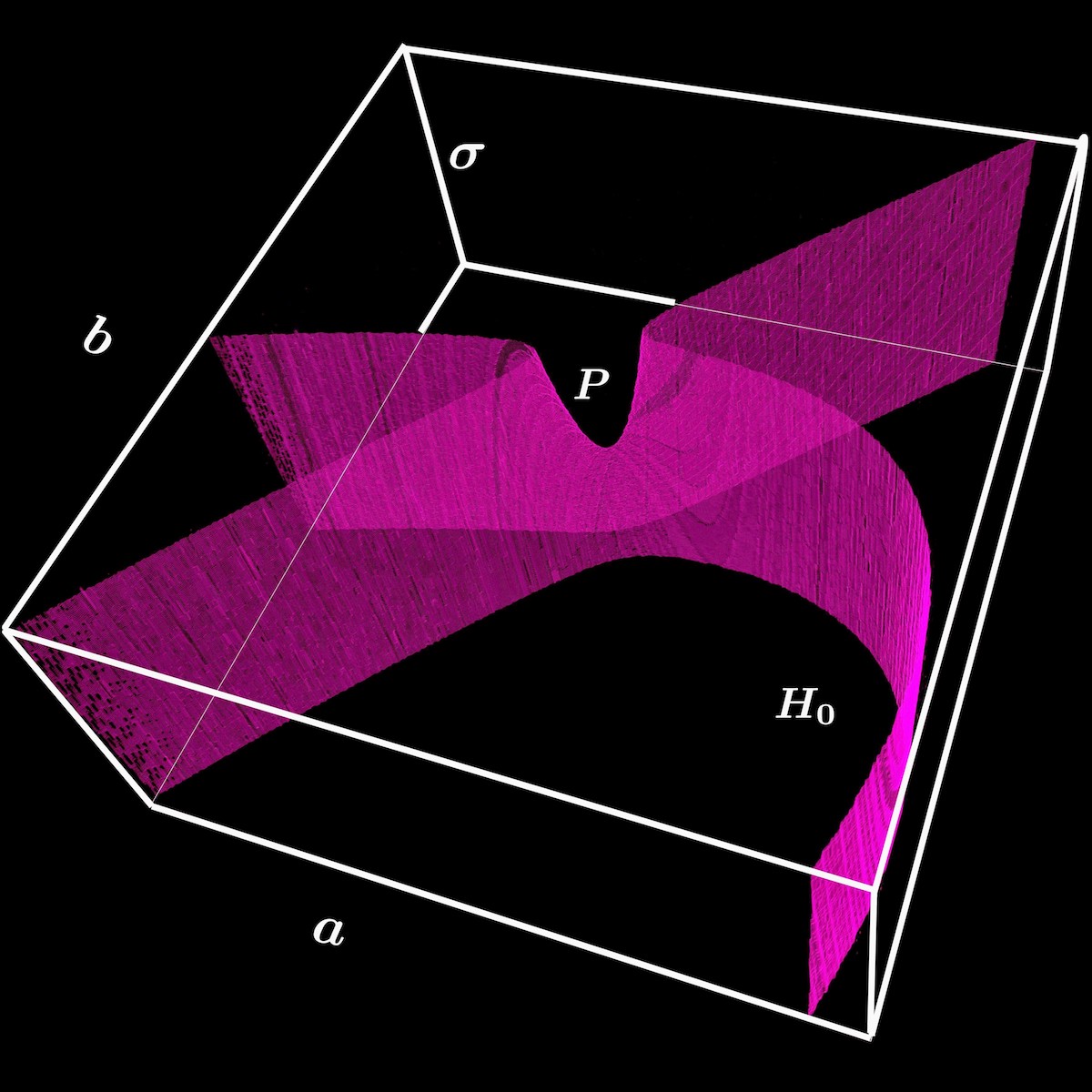}
\caption{Magnifications of the saddle $P$ in  Fig.~\ref{fig:sigma15_sigma20_transition3D} to better reveal the organization of the homoclinic bifurcation surface $H_0$ and how it branches to originate from 1D $IF_1$-curve and scroll onto the primary $T$-line in the $(a,b,\sigma)$-parameter space.}\label{fig:sigma15_sigma20_transition3D_zooms}

\end{figure}

In this section, we describe a saddle in the parameter space that causes branch switching between homoclinic bifurcation curves. As seen in Fig.~\ref{fig:sigma15Cont} and Fig.~\ref{fig:sigma20}, between $\sigma=1.5$ and $\sigma=2.0$, the primary homoclinic bifurcation curve shifts from a complete loop towards the codimension-2 point $DRS$, to a spiral organization around the primary T-point $T_0$. This implies the existence of a saddle in between the two $\sigma$ values that branches the primary homoclinic bifurcation curve $H_0$, as seen in Fig.~\ref{fig:sigmaVariations_shortTerm}d,e. Fig.~\ref{fig:sigma15_sigma20_transition3D}(left) shows a detailed 3D $(a,b,\sigma)$ bifurcation diagram close to this saddle. It is constructed using 100 bi-parametric sweeps in the $(a,b)$-plane with $1.7418$ (top surface) $\leq\sigma\leq1.7439$ (bottom surface), using $\{k_i{\}}_{i=0}^{7}$ and 3D volume is rendered with Drishti software \cite{Limaye:2012}. This branching of $H_0$ occurs at $\sigma\approx1.7428$ and is marked by $P$. Detailed zooms close to $P$, shown in Fig.~\ref{fig:sigma15_sigma20_transition3D_zooms}, further unravel this 3D organization of the homoclinic bifurcation surface at the homoclinic branching saddle $P$. A similar 3D bifurcation surface with a saddle in the Shimizu-Morioka system is presented in \cite{ASHIL93} and shown in Fig.~\ref{fig:sigma15_sigma20_transition3D_old}. Fig.~\ref{fig:sigma15_sigma20_transition3D}(right) shows the homoclinic bifurcation curves close to the branching saddle $P$ obtained with continuation at $\sigma$ values of $1.73$ (top), $1.74$ (middle) and $1.75$ (bottom).

\subsubsection{Bridging saddle between T-points}

\begin{figure}[ht]
\centering 
\includegraphics[width=0.65\textwidth]{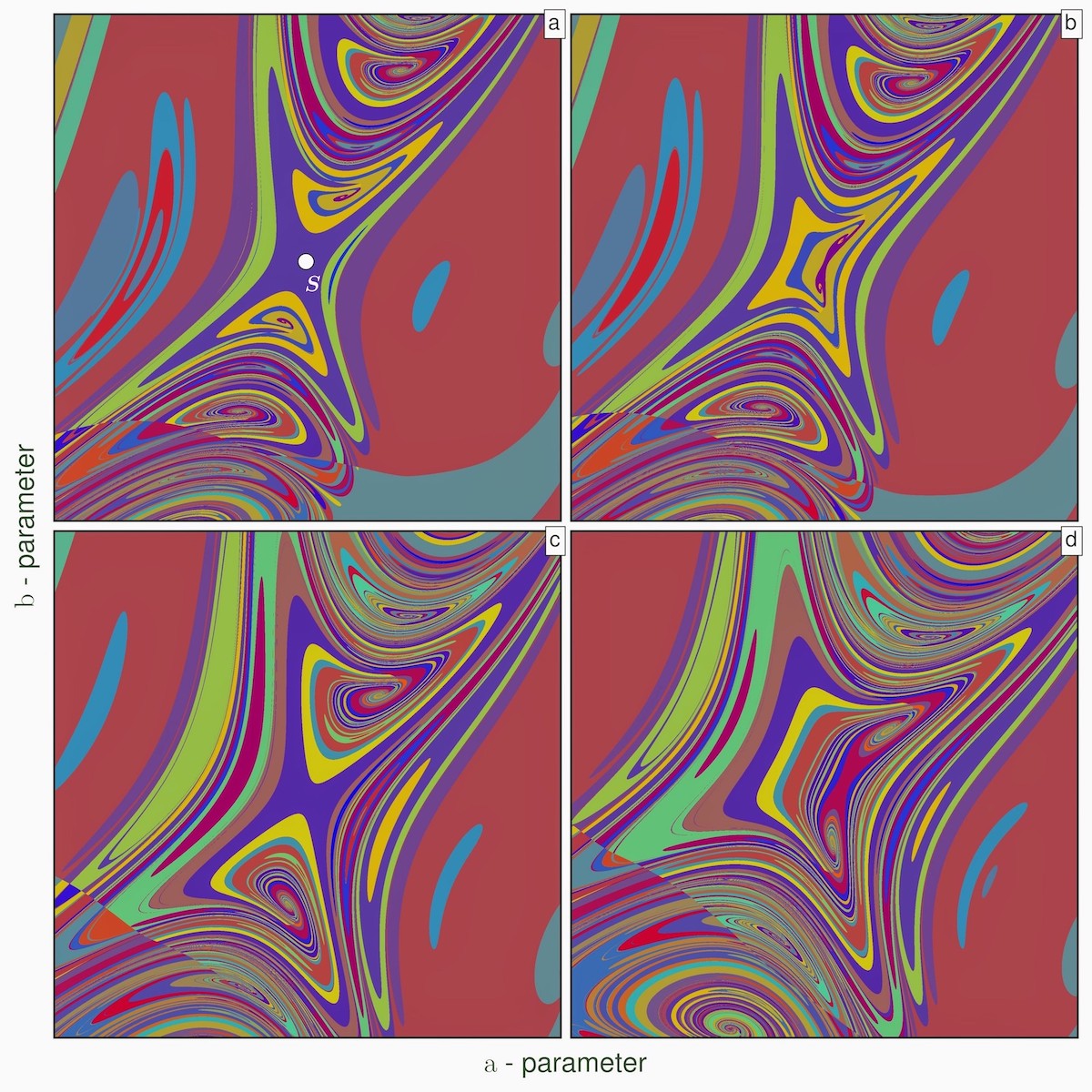}
\caption{Chaotic mixing near the bridging saddle $S$ (white dot in panel (a)) (see Fig.~\ref{fig:sigma15}) is revealed using four $\{k_i{\}}_{i=4}^{11}$-sweeps for varying $\sigma$ values: \textbf{(a)} $\sigma=1.372$,  \textbf{(b)} $\sigma=1.352$, \textbf{(c)} $\sigma=1.288$ and \textbf{(d)} $\sigma=1.264$. As $\sigma$ is changed, the symmetric 
 T-points (with an identical binary coding) above and below the saddle merge all together, giving rise to annular isolas. Compare with Fig.~\ref{fig:saddle3D} and watch the supplementary movie in the Appendix.}
\label{fig:saddleMerge}
\end{figure}

\begin{figure}[t!]
\centering 
\includegraphics[width=0.75\textwidth]{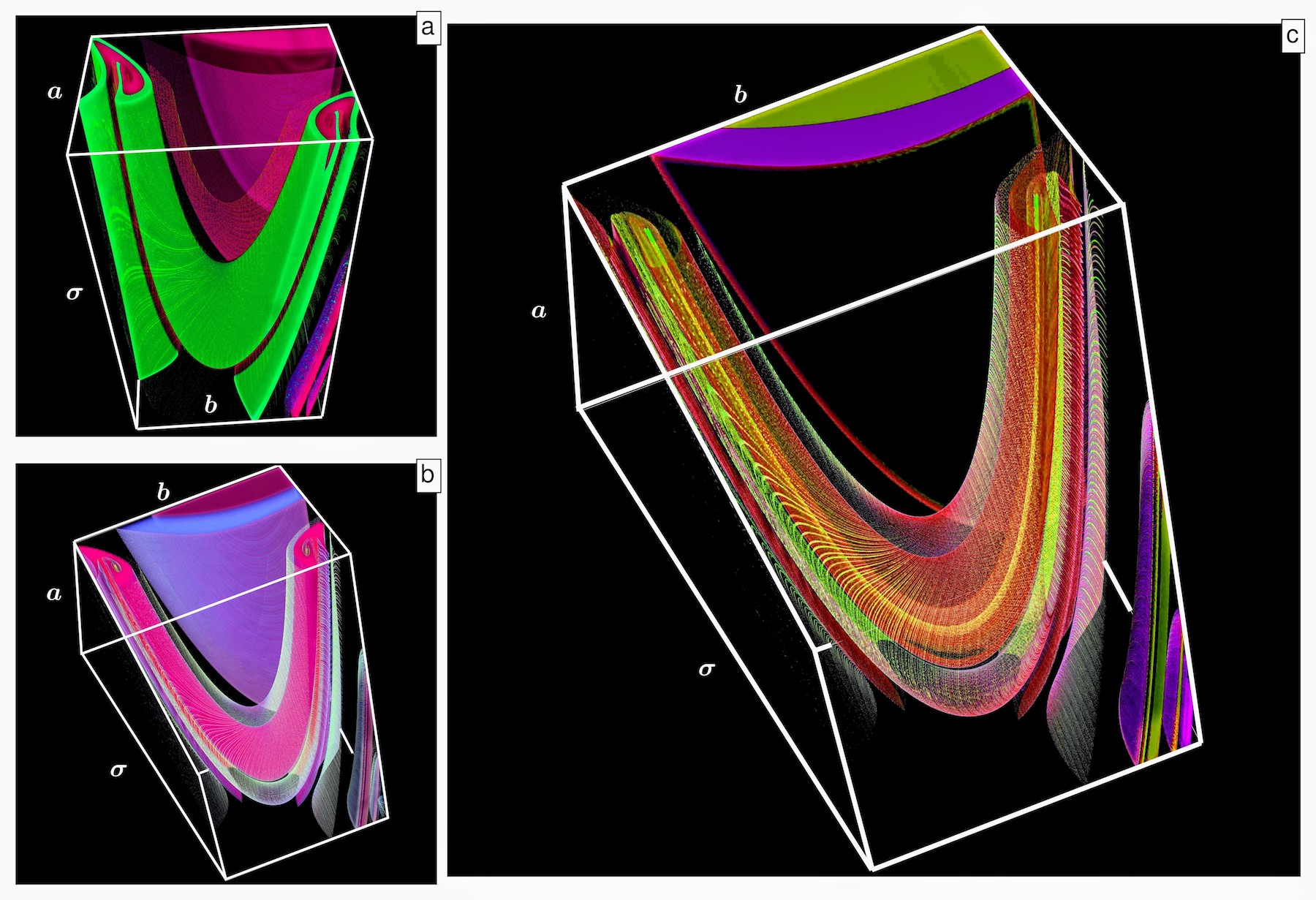}
\caption{3D bifurcation structure near the bridging saddle $S$ (see 2D bifurcation diagrams in Figs. \ref{fig:sigma15} and \ref{fig:saddleMerge}) in the $(a,b,\sigma)$-parameter space is rendered using 2000 $\{k_i{\}}_{i=4}^{11}$-sweeps (each of 2000x2000-resolution) in the $\sigma$-range: $1.344 \le \sigma \le 1.374985$.  Top and bottom surfaces are shown in Fig.~\ref{fig:saddle3D_top_bottom} (rotated through $90^\circ$). It reveals the connectivity between two identical T-points on either side of the saddle, with a gradually increasing depth, as a bending T-curve with the saddle $S$ in  the middle. Depending on how these strictures are sliced, they will look like spirals or concentric circles/isolas in the corresponding 2D parametric sweeps shown above and below.}
\label{fig:saddle3D}
\end{figure}

\begin{figure}[b!]
\centering 
\includegraphics[width=0.3\textwidth]{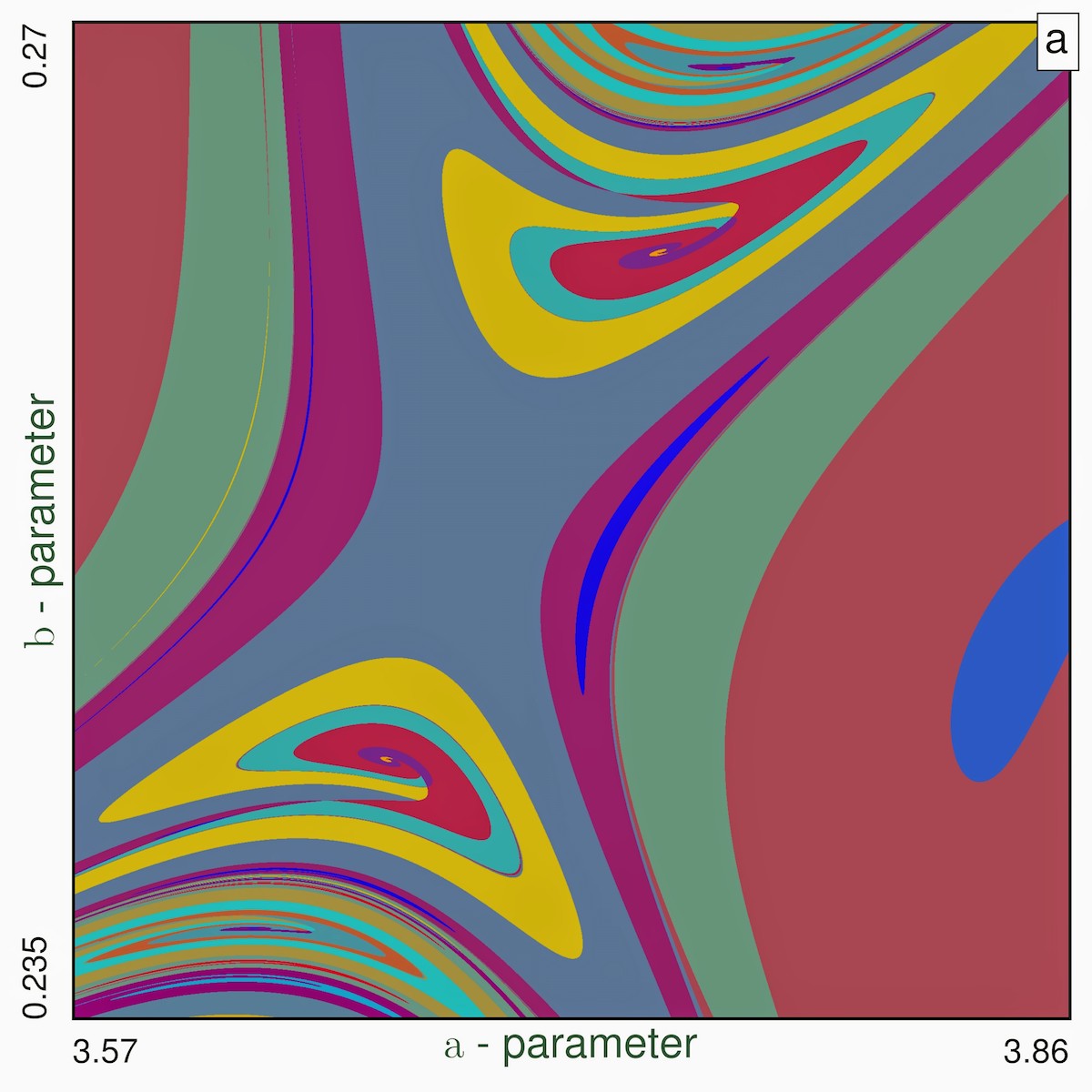}
\includegraphics[width=0.3\textwidth]{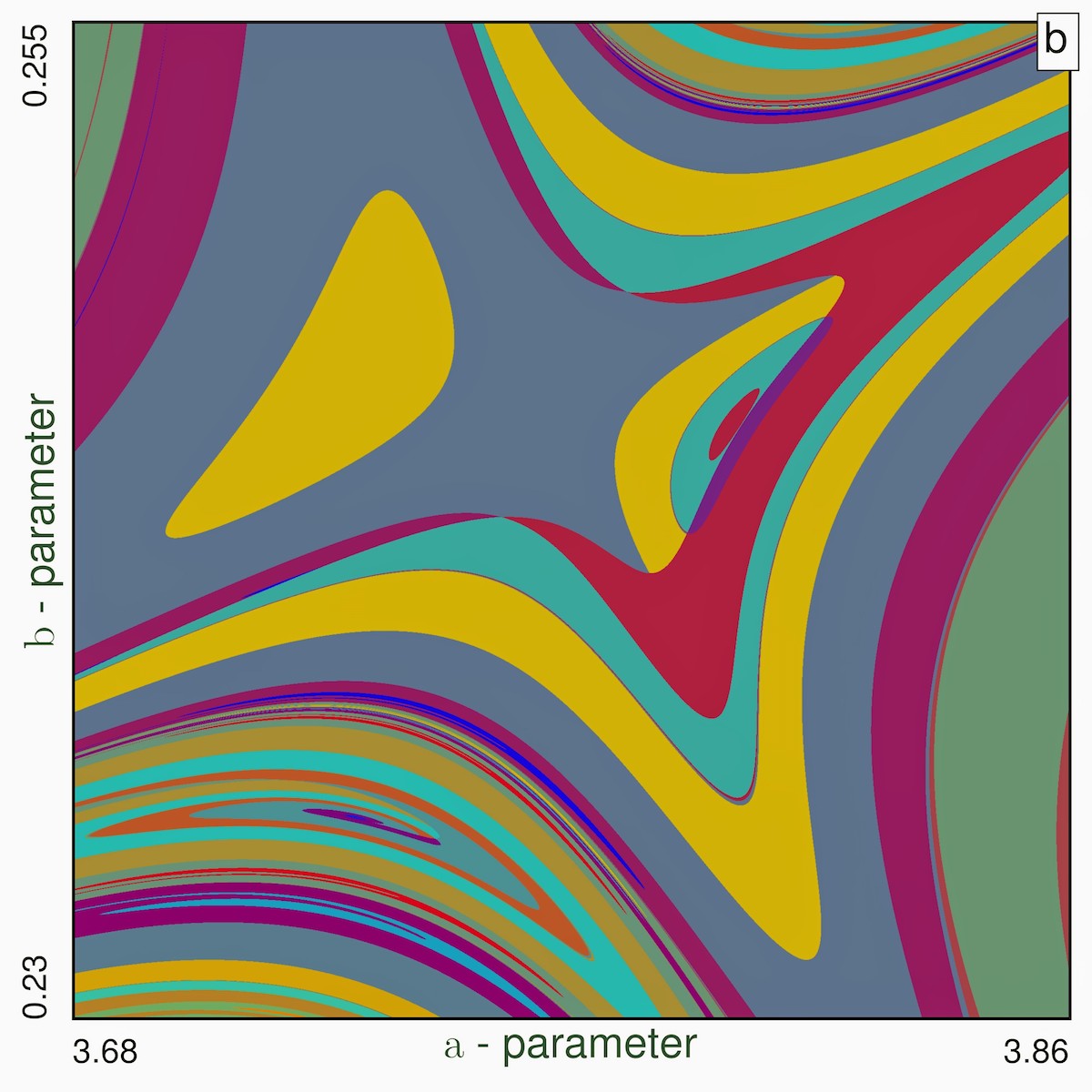}
\caption{ \textbf{(a)} Top sweep (before T-points merge) and the bottom sweep \textbf{(b)} (after T-points and the saddle merge) of Fig.~\ref{fig:saddle3D} are shown (rotated clockwise through  $90^\circ$). As the 3D structure in Fig.~\ref{fig:saddle3D} is sliced to obtain theses $(a,b)$-parametric sweeps, then the the orientation of both T-points  changes from horizontal to vertical so that they appear as they would have merged in the 2D sweeps and given rise to the onset of closed loops or annular isolas depicted in Figs.~\ref{fig:saddleMerge} and \ref{fig:saddle3D_top_bottom}b.}
\label{fig:saddle3D_top_bottom}
\end{figure}

We now focus on another kind of saddle ubiquitous to parametric sweeps, serving as a bridge between two identical T-points, marked $S$ in Figs.\ref{fig:sigma15},\ref{fig:sigma100}. The interesting feature of such a saddle is that the T-point spirals above and below have identical construction and the same heteroclinic connections, as described previously in Fig.~\ref{fig:sigma100} for $\sigma=10$, with $T_2^1$ above and $T_2^2$ below $S$. In order to see how such T-points on either side of a saddle are related to each other, we run multiple sweeps with closely varying $\sigma$ values and observe the structural changes in the spiral organization. Fig.~\ref{fig:saddleMerge} shows such chaotic mixing close to the saddle $S$ as $\sigma$ values are varied  from \textbf{(a)} $\sigma=1.372$, \textbf{(b)} $\sigma=1.352$, \textbf{(c)} $\sigma=1.288$, and through \textbf{(d)} $\sigma=1.264$. As $\sigma$ changes, the T-points above and below the saddle appear to be merging together, as depicted in the transitions: Fig.~\ref{fig:saddleMerge}a$\rightarrow$b and Fig.~\ref{fig:saddleMerge}c$\rightarrow$d.

The 3D scrolling structure of the bifurcation surface around the hyperbolic saddle $S$ for $1.344 \le \sigma \le 1.374985$ constructed using 2000 bi-parametric sweeps using $\{k_i{\}}_{i=4}^{11}$ and rendered with open-source scientific visualization software  {\em Drishti} \footnote{Drishti is available at https://github.com/nci/drishti} is shown in Fig.~\ref{fig:saddle3D}. Fig.~\ref{fig:saddle3D}a,b,c reveal with gradually increasing depth, the continuous structural connections between the T-points on either side of this bridging saddle $S$. As we move along the T-point curve by varying $\sigma$, the curve undergoes a change in orientation and re-enters the bifurcation planes of previous T-points, giving birth to the saddles in between them. Note that each sweep has the 2000x2000 resolution. All 2000 such produced slices imply a total computation of $2000^3=8 \times 10^9$ trajectories. Despite being computationally heavy, this was achieved in just about 8 hours on a single  Nvidia Tesla-K40 GPU. 

\subsubsection{Annular isolas from a bridging saddle}
As seen in Fig.~\ref{fig:saddleMerge}, as the T-points on either side of a bridging saddle merge together with varying $\sigma$, it results in the formation of isolas of homoclinic bifurcation curves resembling concentric annular structures. This is due to the fact that the T-point curve changes its orientation with respect to the bifurcation planes and the T-point becomes non-transverse to this plane, which was described as a codimension-two-plus-one event \cite{Wieczorek:2005,Algaba:2011}. This also becomes clear in Fig.~\ref{fig:saddle3D_top_bottom} which shows the top and bottom surfaces (rotated through $90^\circ$) of the 3D T-point curve in Fig.~\ref{fig:saddle3D}. As we slice several $(a,b)$ bi-parametric sweeps for changing $\sigma$-values, moving towards the bottom surface in Fig.~\ref{fig:saddle3D_top_bottom}b, the T-points appear to be merging as they become non-transverse to the bifurcation plane and we can only observe isolas made of homoclinic bifurcation curves.

\subsubsection{Semi-annular isolas from the ghost of a T-point}

\begin{figure}[t!]
\centering 
\includegraphics[width=0.65\textwidth]{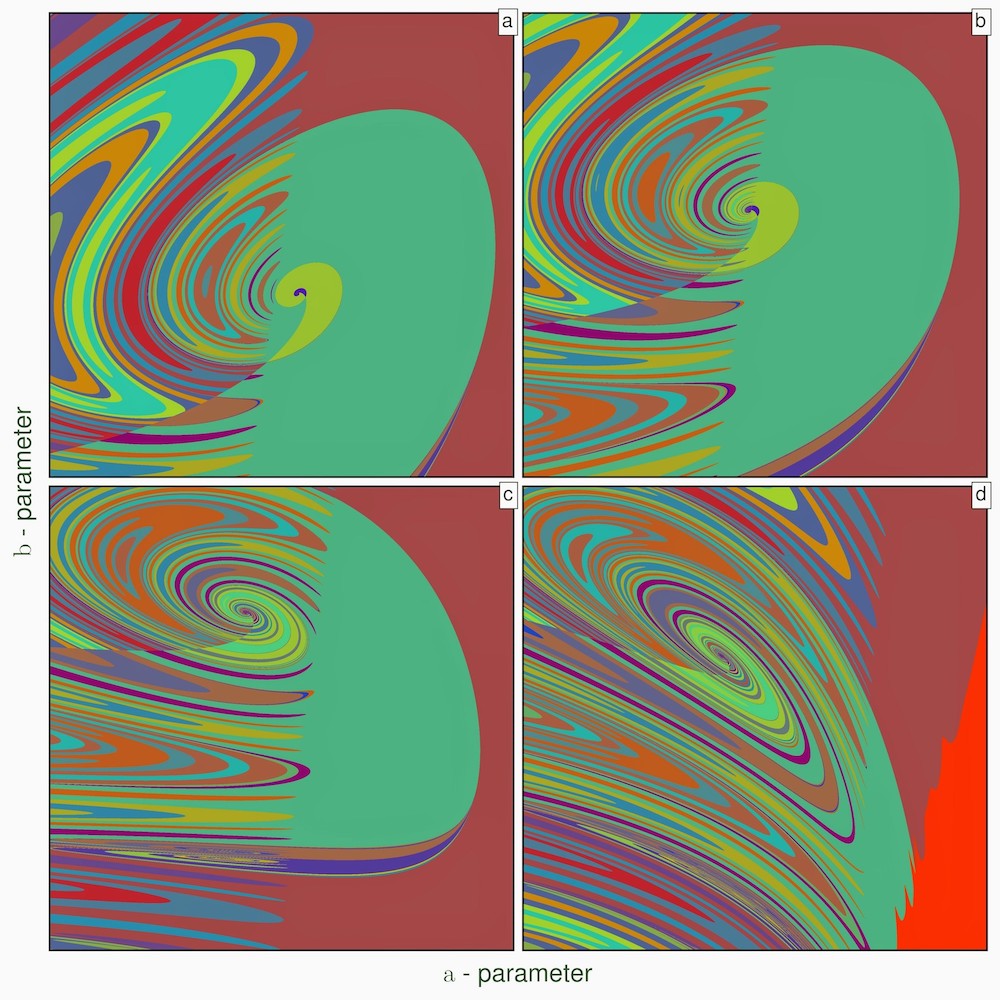}
\caption{Transformation of a T-point structure as it moves over the stability boundary of the periodic orbit $\{\overline{10}\}$ in the parameter plane, is visualized with four $\{k_i{\}}_{i=1}^{9}$-sweep as $\sigma$ changes from 11.1253 in \textbf{(a)}, next  11.2459  \textbf{(b)},  11.4501 \textbf{(c)} through 11.5522 in \textbf{(d)}. With increasing $\sigma$, the ghost of the T-point $T_g$ (see also Fig.~\ref{fig:sigma100}) crosses the boundary between the stable periodic orbit (green region) and structurally unstable chaotic regions in the parameter plane, the semi-annular isolas wraps and  merge to form the proper spiral organization of homoclinic bifurcation curves around the T-point in (d).}\label{fig:ghostTpoint}
\end{figure}

We end this section with a description of isolas of a different nature with semi-annular structure as seen near the point $C$ of Fig.~\ref{fig:sigma100}. The T-point $T_g$ (with the heteroclinic connection ) next to $C$ is of particular interest with respect to these isolas. To understand the relation between these isolas and $T_g$, we vary $\sigma$ in Fig.~\ref{fig:ghostTpoint} through \textbf{(a)} 11.1253 \textbf{(b)} 11.2459 \textbf{(c)} 11.4501 and \textbf{(d)} 11.5522. As $\sigma$ increases, $T_g$ starts crossing the boundary between the stable periodic orbit (green region) and structurally unstable chaotic regions in the parametric plane, the semi-annular isolas start merging around $T_g$ to gradually reveal the full spiral organization of homoclinic bifurcation curves around the T-point in Fig.~\ref{fig:ghostTpoint}(d). This explains how in Fig.~\ref{fig:sigma100}, when the T-point falls inside the region of the periodic orbit, the spiral structure of $T_g$ is devoured by the periodic orbit, leaving behind the remnants of the spirals surrounding the ghost of the T-point $T_g$, in the form of semi-annular isolas near the boundary. Note that when we say periodic orbit $\{\overline{01}\}$, for the sake of simplicity, we only mean an orbit that produces a periodic sequence in its symbolic representation. Its trajectory may still be chaotic but follow a periodic symbolic sequence as also seen in Fig.~\ref{fig:sigma15_IF_3D}d for a one sided chaotic trajectory, whose symbolic representation would still be periodic $\{\overline{1}\}$. In fact, the region surrounding $T_g$ seems to be the two sided chaotic trajectory following the periodic sequence $\{\overline{01}\}$. 

\section{Conclusions}

In this paper, we presented a novel framing combining applied dynamical systems, GPU-based parallel computation, and 2D and 3D parameter space visualization to extend the theory of non-local homoclinic bifurcations of lower codimensions. With this approach we explore and  determine     key universal rules of complex dynamics in diverse systems. Using the technique of symbolic dynamics with DCP, in corroboration with the parameter continuation methods powered by  MatCont, we could identify regions of simple and chaotic dynamics in the parameter space of a classic nonlinear laser model with a Lorenz-like attractor, as well as disclose key underlying bifurcation structures, including Bykov T-point spirals and inclination flips. We performed detailed computational reconstruction and visualization of the three dimensional embedding of bifurcation surfaces, parametric saddles and isolas. The knowledge and the methodology created in this study should further advance new ideas and approaches for a better understanding of cross-disciplinary nonlinear phenomena at large, while the generality of our modeling approaches are also applicable to other biological, medical, and engineering systems \cite{Pusuluri2020}.

\section*{Acknowledgements}

This work was partly funded by the NSF grant IOS-1455527 and the RSF grant 14-41-00044 at Lobachevsky University of Nizhny Novgorod. We thank the Brains and Behavior initiative of Georgia State University for providing pilot grant support and the doctoral fellowship of K. Pusuluri. We thank NVIDIA Corporation for supporting us with the Tesla K40 GPUs used in this study. Finally, we acknowledge the support of all the current and past members of the Shilnikov NeurDS lab for enriching discussions.


\begin{thebibliography}{10}
\expandafter\ifx\csname url\endcsname\relax
  \def\url#1{\texttt{#1}}\fi
\expandafter\ifx\csname urlprefix\endcsname\relax\def\urlprefix{URL }\fi
\expandafter\ifx\csname href\endcsname\relax
  \def\href#1#2{#2} \def\path#1{#1}\fi

\bibitem{ABS:1977}
V.~S. Afraimovich, V.~V. Bykov, L.~P. Shilnikov, The origin and structure of
  the {L}orenz attractor, Sov. Phys. Dokl. 22 (1977) 253--255.

\bibitem{AS:1983}
V.~S. Afraimovich, L.~P. Shilnikov, Strange attractors and quasiattractors, in:
  Nonlinear dynamics and turbulence, Interaction Mech. Math. Ser., Pitman,
  Boston, MA, 1983, pp. 1--34.

\bibitem{GW79}
J.~Guckenheimer, R.~F. Williams, Structural stability of {L}orenz attractors,
  Inst. Hautes \'Etudes Sci. Publ. Math. 50~(50) (1979) 59--72.

\bibitem{SSTC:1998}
L.~P. Shilnikov, A.~L. Shilnikov, D.~V. Turaev, L.~O. Chua, Methods of
  Qualitative Theory in Nonlinear Dynamics. Part I, World Scientific,
  Singapore, 1998.

\bibitem{Shilnikov:2001}
L.~P. Shilnikov, A.~L. Shilnikov, D.~Turaev, L.~O. Chua, Methods of Qualitative
  Theory in Nonlinear Dynamics, Vol. 1-2, World Sci. Publ., Singapore, 1998,
  2001.

\bibitem{Homburg:2010}
A.~J. Homburg, B.~Sandstede, Homoclinic and heteroclinic bifurcations in vector
  fields, in: H.~W. Broer, B.~Hasselblatt, F.~Takens (Eds.), Handbook of
  Dynamical Systems, Vol.~3, Elsevier Science, 2010, Ch.~8, pp. 379--524.
\newblock \href {https://doi.org/https://doi.org/10.1016/S1874-575X(10)00316-4}
  {\path{doi:https://doi.org/10.1016/S1874-575X(10)00316-4}}.

\bibitem{Moloney:1987}
J.~V. Moloney, J.~S. Uppal, R.~G. Harrison, Origin of chaotic relaxation
  oscillations in an optically pumped molecular laser, PRL 59~(25) (1987) 2868.
\newblock \href {https://doi.org/10.1103/PhysRevLett.59.2868}
  {\path{doi:10.1103/PhysRevLett.59.2868}}.

\bibitem{Forysiak:1991}
W.~Forysiak, J.~V. Moloney, R.~G. Harrison, Bifurcations of an optically pumped
  three-level laser model, Physica D 53~(1) (1991) 162 -- 186.
\newblock \href {https://doi.org/https://doi.org/10.1016/0167-2789(91)90170-E}
  {\path{doi:https://doi.org/10.1016/0167-2789(91)90170-E}}.

\bibitem{auto07p}
E.~J. Doedel, A.~R. Champneys, F.~Dercole, T.~F. Fairgrieve, Y.~A. Kuznetsov,
  B.~E. Oldeman, R.~C. Paffenroth, B.~Sandstede, X.~J. Wang, C.~H. Chang,
  Auto-07p: Continuation and bifurcation software for ordinary differential
  equations, ~, Concordia University, Canada (2007).

\bibitem{homcont}
A.~R. Champneys, Y.~A. Kuznetsov, B.~Sandstede, A numerical toolbox for
  homoclinic bifurcation analysis, Int. J. of Bifurcation and Chaos 6~(5)
  (1996) 867--887.
\newblock \href {https://doi.org/10.1142/S0218127496000485}
  {\path{doi:10.1142/S0218127496000485}}.

\bibitem{Matcont}
A.~Dhooge, W.~Govaerts, Y.~Kuznetsov, H.~G.~E. Meijer, B.~Sautois, New features
  of the software {M}at{C}ont for bifurcation analysis of dynamical systems,
  Math. Comput. Model. Dyn. Syst. 14~(2) (2008) 147--175.
\newblock \href {https://doi.org/10.1080/13873950701742754}
  {\path{doi:10.1080/13873950701742754}}.

\bibitem{ASHIL93}
A.~L. Shilnikov, On bifurcations of the {L}orenz attractor in the
  {S}himizu-{M}orioka model, Physica D 62~(1-4) (1993) 338--346.
\newblock \href {https://doi.org/10.1016/0167-2789(93)90292-9}
  {\path{doi:10.1016/0167-2789(93)90292-9}}.

\bibitem{DeWitte:2012}
V.~De~Witte, W.~Govaerts, Y.~A. Kuznetsov, M.~Friedman, Interactive
  initialization and continuation of homoclinic and heteroclinic orbits in
  {Matlab}, ACM Trans. Math. Softw. 38~(3) (2012) 18:1--18:34.
\newblock \href {https://doi.org/10.1145/2168773.2168776}
  {\path{doi:10.1145/2168773.2168776}}.

\bibitem{Wolf:1985}
A.~Wolf, J.~B. Swift, H.~L. Swinney, J.~A. Vastano, Determining {L}yapunov
  exponents from a time series, Physica D 16~(3) (1985) 285--317.
\newblock \href {https://doi.org/10.1016/0167-2789(85)90011-9}
  {\path{doi:10.1016/0167-2789(85)90011-9}}.

\bibitem{Gallas:2010}
J.~A.~C. Gallas, The structure of infinite periodic and chaotic hub cascades in
  phase diagrams of simple autonomous flows, Int. J. Bif. \& Chaos 20~(02)
  (2010) 197--211.
\newblock \href {https://doi.org/10.1142/S0218127410025636}
  {\path{doi:10.1142/S0218127410025636}}.

\bibitem{Barrio:2011}
R.~Barrio, F.~Blesa, S.~Serrano, A.~Shilnikov, Global organization of spiral
  structures in biparameter space of dissipative systems with {S}hilnikov
  saddle-foci, Physical Review E 84~(3) (2011) 035201.
\newblock \href {https://doi.org/10.1103/PhysRevE.84.035201}
  {\path{doi:10.1103/PhysRevE.84.035201}}.

\bibitem{Barrio:2012}
R.~Barrio, A.~Shilnikov, L.~Shilnikov, Kneadings, symbolic dynamics and
  painting {L}orenz chaos, Int. J. Bifurcation \& Chaos 22~(04) (2012) 1230016.
\newblock \href {https://doi.org/10.1142/S0218127412300169}
  {\path{doi:10.1142/S0218127412300169}}.

\bibitem{Xing:2014}
T.~Xing, R.~Barrio, A.~Shilnikov, Symbolic quest into homoclinic chaos, Int. J.
  Bifurcation \& Chaos 24~(8) (2014) 1440004.
\newblock \href {https://doi.org/https://doi.org/10.1142/S0218127414400045}
  {\path{doi:https://doi.org/10.1142/S0218127414400045}}.

\bibitem{Bykov:1980}
V.~V. Bykov, On the structure of bifurcations sets of synamical systems that
  are systems with a separatrix contour containing saddle-focus, Methods of
  Qualitative Theory of Differential Equations, Gorky University (in Russian).
  (1980) 44--72.

\bibitem{AGLST:2014}
V.~S. Afraimovich, S.~V. Gonchenko, L.~M. Lerman, A.~L. Shilnikov, D.~V.
  Turaev, Regular and chaotic dynamics, Scientific heritage of L.P. Shilnikov.
  Part 1 4~(19) (2014) 435--462.
\newblock \href {https://doi.org/10.1134/S1560354714040017}
  {\path{doi:10.1134/S1560354714040017}}.

\bibitem{AD02}
D.~Aubin, A.~D. Dalmedico, Writing the history of dynamical systems and chaos:
  longue dur\'ee and revolution, disciplines and cultures, Historia Math.
  29~(3) (2002) 273--339.
\newblock \href {https://doi.org/10.1006/hmat.2002.2351}
  {\path{doi:10.1006/hmat.2002.2351}}.

\bibitem{Belyakov1}
L.~A. Belyakov, Bifurcations of systems with a homoclinic curve of the
  saddle-focus with a zero saddle value, Mat. Zametki 36~(5) (1984) 681--689,
  798.

\bibitem{ALS91}
A.~L. Shilnikov, Bifurcation and chaos in the {M}arioka-{S}himizu system,
  Selecta Math. Soviet. 10(2) (1991) 105--117, originally published in Methods
  in qualitative theory and bifurcation theory (in Russian), Gorky State
  University, 1986, pp. 180--193.

\bibitem{Shilnikov:1965}
L.~P. Shilnikov, A case of the existence of a denumerable set of periodic
  motions, Dokl. Akad. Nauk SSSR 160 (1965) 558--561.

\bibitem{LP67}
L.~P. Shilnikov, The existence of a denumerable set of periodic motions in
  four-dimensional space in an extended neighborhood of a saddle-focus., Soviet
  Math. Dokl. 8(1) (1967) 54--58.

\bibitem{Shilnikov:2012}
A.~Shilnikov, Complete dynamical analysis of an interneuron model, J. Nonlinear
  Dynamics 68~(3) (2012) 305--328.
\newblock \href {https://doi.org/10.1007/s11071-011-0046-y}
  {\path{doi:10.1007/s11071-011-0046-y}}.

\bibitem{GOST05}
S.~Gonchenko, I.~Ovsyannikov, C.~Simo, D.~Turaev, Three-dimensional
  {H}{\'e}non-like maps and wild {L}orenz-like attractors, Int. J. Bif. Chaos
  15(11) (2005) 3493--3508.
\newblock \href {https://doi.org/10.1142/S0218127405014180}
  {\path{doi:10.1142/S0218127405014180}}.

\bibitem{Bykov:1998}
V.~V. Bykov, Bifurcations leading to chaos in {C}hua's circuit, Inter. J. Bif.
  Chaos 8 (1998) 685--699.
\newblock \href {https://doi.org/10.1142/s0218127498000486}
  {\path{doi:10.1142/s0218127498000486}}.

\bibitem{Pusuluri:2017}
K.~Pusuluri, A.~Pikovsky, A.~Shilnikov, Unraveling the chaos-land and its
  organization in the rabinovich system, in: Advances in Dynamics, Patterns,
  Cognition, Springer, Cham, 2017, pp. 41--60.
\newblock \href {https://doi.org/10.1007/978-3-319-53673-6_4}
  {\path{doi:10.1007/978-3-319-53673-6_4}}.

\bibitem{Pusuluri:2018}
K.~Pusuluri, A.~Shilnikov, Homoclinic chaos and its organization in a nonlinear
  optics model, Physics Review E 98~(4) (2018) 040202.
\newblock \href {https://doi.org/10.1103/PhysRevE.98.040202}
  {\path{doi:10.1103/PhysRevE.98.040202}}.

\bibitem{pusuluri2019symbolic}
K.~Pusuluri, A.~Shilnikov, Symbolic representation of neuronal dynamics, in:
  Advances on Nonlinear Dynamics of Electronic Systems, World Scientific, 2019,
  pp. 97--102.
\newblock \href {https://doi.org/10.1142/9789811201523_0018}
  {\path{doi:10.1142/9789811201523_0018}}.

\bibitem{Pusuluri2020}
K.~Pusuluri, H.~Ju, A.~Shilnikov, Chaotic dynamics in neural systems, in: R.~A.
  Meyers (Ed.), Encyclopedia of Complexity and Systems Science, Springer Berlin
  Heidelberg, Berlin, Heidelberg, 2020, pp. 1--13.
\newblock \href {https://doi.org/10.1007/978-3-642-27737-5_738-1}
  {\path{doi:10.1007/978-3-642-27737-5_738-1}}.

\bibitem{xwbs}
T.~Xing, J.~Wojcik, R.~Barrio, A.~Shilnikov, Symbolic toolkit for chaos
  explorations, in: Int. Conf. Theory and Application in Nonlinear Dynamics
  (ICAND 2012), Springer, 2014, pp. 129--140.
\newblock \href {https://doi.org/10.1007/978-3-319-02925-2_12}
  {\path{doi:10.1007/978-3-319-02925-2_12}}.

\bibitem{Shilnikov:1968}
L.~P. Shilnikov, The generation of a periodic motion from a trajectory which is
  doubly asymptotic to a saddle type equilibrium state, Mat. Sb. (N.S.) 77
  (119) (1968) 461--472.

\bibitem{Barrio:2014}
R.~Barrio, M.~Angeles~Mart{\'\i}nez, S.~Serrano, A.~L. Shilnikov, Macro- and
  micro-chaotic structures in the {H}indmarsh-{R}ose model of bursting neurons,
  Chaos 4~(2) (2014) 023128.
\newblock \href {https://doi.org/10.1063/1.4882171}
  {\path{doi:10.1063/1.4882171}}.

\bibitem{xwzs2015}
T.~Xing, J.~Wojcik, M.~Zaks, A.~L. Shilnikov, Fractal parameter space of
  lorenz-like attractors: A hierarchical approach, in: G.~Nicolis, V.~Basios
  (Eds.), Chaos, Information Processing and Paradoxical Games: The legacy of
  J.S. Nicolis, World Sci. Publ., Singapore, 2015, Ch.~5, pp. 87--104.
\newblock \href {https://doi.org/10.1142/9789814602136_0005}
  {\path{doi:10.1142/9789814602136_0005}}.

\bibitem{Lempel:1976}
A.~Lempel, J.~Ziv, On the complexity of finite sequences, IEEE Transactions on
  information theory 22~(1) (1976) 75--81.
\newblock \href {https://doi.org/10.1109/TIT.1976.1055501}
  {\path{doi:10.1109/TIT.1976.1055501}}.

\bibitem{Haken:1975}
H.~Haken, Analogy between higher instabilities in fluids and lasers, Physics
  Letters A 53~(1) (1975) 77--78.
\newblock \href {https://doi.org/10.1016/0375-9601(75)90353-9}
  {\path{doi:10.1016/0375-9601(75)90353-9}}.

\bibitem{Haken:1985}
H.~Haken, Laser Light Dynamics, North-Holland, Amsterdam, 1985.

\bibitem{Casperson:1978}
L.~Casperson, Spontaneous coherent pulsations in laser oscillators, IEEE
  Journal of Quantum Electronics 14~(10) (1978) 756--761.
\newblock \href {https://doi.org/10.1109/JQE.1978.1069683}
  {\path{doi:10.1109/JQE.1978.1069683}}.

\bibitem{Weiss:1982}
C.~Weiss, H.~King, Oscillation period doubling chaos in a laser, Optics
  Communications 44~(1) (1982) 59--61.
\newblock \href {https://doi.org/10.1016/0030-4018(82)90016-5}
  {\path{doi:10.1016/0030-4018(82)90016-5}}.

\bibitem{Weiss:1985}
C.~Weiss, W.~Klische, P.~Ering, M.~Cooper, Instabilities and chaos of a single
  mode {NH3} ring laser, Optics communications 52~(6) (1985) 405--408.
\newblock \href {https://doi.org/10.1016/0030-4018(86)90339-1}
  {\path{doi:10.1016/0030-4018(86)90339-1}}.

\bibitem{Weiss:1986}
C.~Weiss, J.~Brock, Evidence for {L}orenz-type chaos in a laser, Physical
  Review Letters 57~(22) (1986) 2804.
\newblock \href {https://doi.org/10.1103/PhysRevLett.57.2804}
  {\path{doi:10.1103/PhysRevLett.57.2804}}.

\bibitem{Weiss:1995}
C.~Weiss, U.~H{\"u}bner, N.~Abraham, D.~Tang, Lorenz-like chaos in {NH3-FIR}
  lasers, Infrared Physics \& Technology 36~(1) (1995) 489--512.
\newblock \href {https://doi.org/https://doi.org/10.1016/1350-4495(94)00088-3}
  {\path{doi:https://doi.org/10.1016/1350-4495(94)00088-3}}.

\bibitem{Moloney:1989}
J.~V. Moloney, W.~Forysiak, J.~S. Uppal, R.~G. Harrison, Regular and chaotic
  dynamics of optically pumped molecular lasers, Phys. Rev. A 39 (1989)
  1277--1285.
\newblock \href {https://doi.org/10.1103/PhysRevA.39.1277}
  {\path{doi:10.1103/PhysRevA.39.1277}}.

\bibitem{Bykov:1993}
V.~Bykov, The bifurcations of separatrix contours and chaos, Physica D
  62~(1--4) (1993) 290--299.
\newblock \href {https://doi.org/https://doi.org/10.1016/0167-2789(93)90288-C}
  {\path{doi:https://doi.org/10.1016/0167-2789(93)90288-C}}.

\bibitem{Vladimirov:1993}
A.~Vladimirov, D.~Volkov, Low-intensity chaotic operations of a laser with a
  saturable absorber, Optics Communications 100~(1) (1993) 351--360.
\newblock \href {https://doi.org/https://doi.org/10.1016/0030-4018(93)90597-X}
  {\path{doi:https://doi.org/10.1016/0030-4018(93)90597-X}}.

\bibitem{Shilnikov:1993}
A.~L. Shil'nikov, L.~P. Shil'nikov, D.~V. Turaev, Normal forms and {L}orenz
  attractors, Int. J. Bifurcation \& Chaos 3 (1993) 1123--1123.
\newblock \href {https://doi.org/10.1142/S0218127493000933}
  {\path{doi:10.1142/S0218127493000933}}.

\bibitem{LP62}
L.~P. Shilnikov, Some instances of generation of periodic motions in
  {$n$}-space, Dokl. Akad. Nauk SSSR 143 (1962) 289--292.

\bibitem{LP63}
L.~P. Shilnikov, Some cases of generation of period motions from singular
  trajectories, Mat. Sb. 61~(103) (1963) 443--466.

\bibitem{LP81}
L.~Shilnikov, The theory of bifurcations and quasiattractors, Uspeh. Math. Nauk
  36(4) (1981) 240--242.

\bibitem{Rob89}
C.~Robinson, Homoclinic bifurcation to a transitive attractor of {L}orenz type,
  Nonlinearity 2 (1989) 495--518.
\newblock \href {https://doi.org/10.1088/0951-7715/2/4/001}
  {\path{doi:10.1088/0951-7715/2/4/001}}.

\bibitem{Ry90}
M.~Rychlic, Lorenz attractor through {S}hil'nikov type bifurcation. {P}art {I},
  Ergodic Theory and Dynamical Systems 10~(4) (1990) 793--821.
\newblock \href {https://doi.org/10.1017/S0143385700005915}
  {\path{doi:10.1017/S0143385700005915}}.

\bibitem{LP70}
L.~P. Shilnikov, A contribution to the problem of the structure of an extended
  neighborhood of a rough equilibrium state of saddle-focus type, Math. USSR-Sb
  10~(1) (1970) 91--102.
\newblock \href {https://doi.org/10.1070/sm1970v010n01abeh001588}
  {\path{doi:10.1070/sm1970v010n01abeh001588}}.

\bibitem{LP69}
L.~P. Shilnikov, \href{http://mi.mathnet.ru/eng/dan34997}{A certain new type of
  bifurcation of multidimensional dynamic systems}, Dokl. Akad. Nauk SSSR
  189~(1) (1969) 59--62.
\newline\urlprefix\url{http://mi.mathnet.ru/eng/dan34997}

\bibitem{Shilnikov:2007}
L.~P. Shilnikov, A.~Shilnikov, {S}hilnikov bifurcation, Scholarpedia 2~(8)
  (2007) 1891, revision \#153014.
\newblock \href {https://doi.org/10.4249/scholarpedia.1891}
  {\path{doi:10.4249/scholarpedia.1891}}.

\bibitem{ABS:1983}
V.~S. Afraimovich, V.~V. Bykov, L.~P. Shilnikov, On structurally unstable
  attracting limit sets of {L}orenz attractor type, Trans. Moscow Math. Soc.
  44~(2) (1983) 153--216.

\bibitem{Bykov:1992}
V.~V. Bykov, A.~L. Shilnikov, On the boundaries of the domain of existence of
  the {L}orenz attractor, Selecta Math. Soviet. 11~(4) (1992) 375--382.

\bibitem{Tucker:1999}
W.~Tucker, The lorenz attractor exists, C.R. Acad. I - Math. 328~(12) (1999)
  1197--1202.
\newblock \href {https://doi.org/https://doi.org/10.1016/S0764-4442(99)80439-X}
  {\path{doi:https://doi.org/10.1016/S0764-4442(99)80439-X}}.

\bibitem{Viana:2000}
M.~Viana, What's new on {L}orenz strange attractors?, Math. Intelligencer
  22~(3) (2000) 6--19.
\newblock \href {https://doi.org/10.1007/BF03025276}
  {\path{doi:10.1007/BF03025276}}.

\bibitem{Glendinning:1986}
P.~Glendinning, C.~Sparrow, T-points: A codimension two heteroclinic
  bifurcation, Journal of Statistical Physics 43~(3) (1986) 479--488.
\newblock \href {https://doi.org/10.1007/BF01020649}
  {\path{doi:10.1007/BF01020649}}.

\bibitem{Milnor:1988}
J.~Milnor, W.~Thurston, On iterated maps of the interval, in: J.~Alexander
  (Ed.), Dynamical Systems: Proceedings of the Special Year held at the
  University of Maryland, College Park, 1986--87, Springer Berlin Heidelberg,
  Berlin, Heidelberg, 1988, pp. 465--563.
\newblock \href {https://doi.org/10.1007/BFb0082847}
  {\path{doi:10.1007/BFb0082847}}.

\bibitem{GS93}
P.~Glendinning, C.~Sparrow, Prime and renormalisable kneading invariants and
  the dynamics of expanding {L}orenz maps, Physica D 62~(1--4) (1993) 22--50.
\newblock \href {https://doi.org/10.1016/0167-2789(93)90270-B}
  {\path{doi:10.1016/0167-2789(93)90270-B}}.

\bibitem{GH96}
P.~Glendinning, T.~Hall, Zeros of the kneading invariant and topological
  entropy for {L}orenz maps, Nonlinearity 9~(4) (1996) 999--1014.
\newblock \href {https://doi.org/10.1088/0951-7715/9/4/010}
  {\path{doi:10.1088/0951-7715/9/4/010}}.

\bibitem{R78}
D.~Rand, The topological classification of {L}orenz attractors, Mathematical
  Proceedings of the Cambridge Philosophical Society 83~(3) (1978) 451--460.
\newblock \href {https://doi.org/10.1017/S0305004100054736}
  {\path{doi:10.1017/S0305004100054736}}.

\bibitem{Malkin91}
M.~I. Malkin, Rotation intervals and dynamics of {L}orenz type mappings,
  Selecta Math. Sovietica 10 (1991) 265--275.

\bibitem{TW93}
C.~Tresser, R.~F. Williams, Splitting words and {L}orenz braids, Physica D
  62~(1--4) (1993) 15--21.
\newblock \href {https://doi.org/10.1016/0167-2789(93)90269-7}
  {\path{doi:10.1016/0167-2789(93)90269-7}}.

\bibitem{GH90}
J.~Guckenheimer, P.~Holmes, Nonlinear oscillations, dynamical systems, and
  bifurcations of vector fields, Vol.~42 of Applied Mathematical Sciences,
  Springer-Verlag, New York, 1990.

\bibitem{Limaye:2012}
A.~Limaye, Drishti: a volume exploration and presentation tool, in: S.~Stock
  (Ed.), Developments in X-Ray Tomography Viii, Vol. 8506 of SPIE Proceedings,
  International Society for Optics and Photonics, 2012, p. 85060X.
\newblock \href {https://doi.org/10.1117/12.935640}
  {\path{doi:10.1117/12.935640}}.

\bibitem{Wieczorek:2005}
S.~Wieczorek, B.~Krauskopf, Bifurcations of {$n$}-homoclinic orbits in
  optically injected lasers, Nonlinearity 18~(3) (2005) 1095--1120.
\newblock \href {https://doi.org/10.1088/0951-7715/18/3/010}
  {\path{doi:10.1088/0951-7715/18/3/010}}.

\bibitem{Algaba:2011}
A.~Algaba, F.~Fern{\'a}ndez-S{\'a}nchez, M.~Merino, A.~Rodr{\'\i}guez-Luis,
  Structure of saddle-node and cusp bifurcations of periodic orbits near a
  non-transversal {T}-point, Nonlinear Dynamics 63~(3) (2011) 455--476.
\newblock \href {https://doi.org/10.1007/s11071-010-9815-2}
  {\path{doi:10.1007/s11071-010-9815-2}}.

\bibitem{Algaba:2015}
A.~Algaba, F.~Fern{\'a}ndez-S{\'a}nchez, M.~Merino, A.~Rodr{\'\i}guez-Luis,
  Analysis of the {T}-point-{H}opf bifurcation in the {L}orenz system,
  Communications in Nonlinear Science and Numerical Simulation 22~(1-3) (2015)
  676--691.
\newblock \href {https://doi.org/10.1016/j.cnsns.2014.09.025}
  {\path{doi:10.1016/j.cnsns.2014.09.025}}.

\end{thebibliography}

\end{document}